\documentclass[review]{elsarticle}

\usepackage{tabularx}
\usepackage{multirow}
\usepackage{xspace}
\usepackage{subfigure}
\usepackage{algorithm}
\usepackage{algorithmic}
\usepackage{ifthen}
\usepackage{clrscode}
\usepackage{url}
\usepackage{paralist}
\usepackage{array}
\usepackage{amsfonts}
\usepackage{amsmath}
\usepackage{amssymb}
\usepackage{color}

\usepackage{lineno,hyperref}
\modulolinenumbers[5]

\usepackage{setspace}
\journal{Journal Of Computational Physics}

\newcommand{\alertblu}[1]{{\color{black}#1}}
\newcommand{\figpath}{./}

\newcommand{\ldbracket}{[\hspace{-1.5pt}[}
\newcommand{\rdbracket}{]\hspace{-1.5pt}]}
\newcommand{\lgbracket}{\{\hspace{-3.5pt}\{}
\newcommand{\rgbracket}{\}\hspace{-3.5pt}\}}

\newcommand{\ud}{~\mathrm{d}}
\newcommand{\NABLA}{\boldsymbol{\nabla}}

\newcommand{\normal}{\mathbf{n}}
\newcommand{\F}{\sigma}

\newcommand{\faceref}{\widehat{\sigma}}

\newcommand{\jump}[1]{\ldbracket #1 \rdbracket}
\newcommand{\averg}[1]{\lgbracket #1 \rgbracket}

\newcommand{\xvec}{{\bf x}}

\newcommand{\ie}{\textit{i.e.}\@\xspace}
\newcommand{\eg}{\textit{e.g.}\@\xspace}

\newcommand{\st}{\textit{s.t.}\@\xspace}
\newcommand{\cf}{\textit{cf.}\@\xspace}
\newcommand{\ea}{\textit{et al.}\@\xspace}

\newcommand{\xivec}{{\boldsymbol \xi}}

\newcommand{\eqbydef}{\stackrel{\mathrm{def}}{=}}

\newcommand{\Sys}[1]{A_\mathrm{#1}}
\newcommand{\SysM}[1]{\mathbf{A}_{#1}}

\newcommand{\Tf}{\mathcal{T}_0}
\newcommand{\TLc}{\mathcal{T}_L}

\newcommand{\Tl}{\mathcal{T}_{\ell}}
\newcommand{\Tlpo}{\mathcal{T}_{\ell+1}}
\newcommand{\Tlmo}{\mathcal{T}_{\ell-1}}

\newcommand{\Tlone}{\mathcal{T}_{1}}
\newcommand{\Tltwo}{\mathcal{T}_{2}}
\newcommand{\Th}{\{\mathcal{T}_{\ell}\}_{\ell=0,...,L}}
\newcommand{\Thagglo}{\{\mathcal{T}_{\ell}\}_{\ell=1}^L}
\newcommand{\Thr}{\mathcal{T}_{h}}

\newcommand{\TF}{\mathcal{T}_{\F}}
\newcommand{\TFl}{\mathcal{T}_{\F_\ell}}

\newcommand{\T}{\kappa}

\newcommand{\Tpr}{\kappa'}
\newcommand{\Tprl}{\kappa'_{\ell}}

\newcommand{\Tprlpo}{\kappa'_{\ell+1}}
\newcommand{\Tref}{\widehat{\kappa}}

\newcommand{\Flpo}{\mathcal{F}_{\ell+1}}
\newcommand{\Fl}{\mathcal{F}_{\ell}}
\newcommand{\Fln}[1]{\mathcal{F}_{\mathrm{#1}}}
\newcommand{\FT}{\mathcal{F}_{\T}}
\newcommand{\Ff}{\mathcal{F}_{0}}

\newcommand{\It}{\mathcal{I}}

\newcommand{\Itvhlp}{(\mathcal{I}_{\ell}^0 v_{\ell})}
\newcommand{\Itwhlp}{(\mathcal{I}_{\ell}^0 w_{\ell})}

\newcommand{\Itvhl}{\mathcal{I}_{\ell}^0 v_{\ell}}
\newcommand{\Itwhl}{\mathcal{I}_{\ell}^0 w_{\ell}}
\newcommand{\Itpr}{\mathcal{I}^{\ell}_{\ell+1}}
\newcommand{\Itre}{\mathcal{I}^{\ell+1}_{\ell}}
\newcommand{\Itmatrixre}{\mathbf{I}^{\ell+1}_{\ell}}
\newcommand{\Itmatrixpr}{\mathbf{I}^{\ell}_{\ell+1}}

\newcommand{\Itb}{\mathbf{I}}

\newcommand{\M}{\mathbf{M}}

\newcommand{\Ffi}{\Ff^{\rm i}}
\newcommand{\Ffb}{\Ff^{\rm b}}
\newcommand{\Fli}{\Fl^{\rm i}}

\newcommand{\Flb}{\Fl^{\rm b}}

\newcommand{\Poly}[2]{\mathbb{P}_{#1}^{#2}}

\newcommand{\ST}{\,|\,}
\newcommand{\closure}[1]{\overline{#1}}

\newcommand{\intB}[2]{\int_{#1} #2 } 

\newcommand{\disint}[1]{\displaystyle \int_{#1}}
\newcommand{\dissum}[1]{\displaystyle \sum_{#1}}

\DeclareMathOperator{\card}{card}

\newcommand{\GRADh}{\nabla_{\ell}}
\newcommand{\GRADhl}{\nabla_{\ell}}
\newcommand{\GRADhf}{\nabla_{0}}

\newcommand{\inv}{^{^{\mathrm{-1}}}}

\newcommand{\trn}{^{^{\mathrm{T}}}}

\newcommand{\nlev}{L}

\newcommand{\SCAL}{{\cdot}}

\newcommand{\DIV}{{\nabla\SCAL}}

\newcommand{\sumT}{\displaystyle\sum_{\kappa\in\Tf}}
\newcommand{\sumTl}{\displaystyle\sum_{\kappa\in\Tl}}

\newcommand{\sumF}{\displaystyle\sum_{{\sigma}\in\Ff}}
\newcommand{\sumFi}{\displaystyle\sum_{{\sigma}\in\Ff^i}}

\newcommand{\intO}[1]{\displaystyle \int_{\Omega} #1 }

\newcommand{\wvec}{{\bf w}}
\newcommand{\vvec}{{\bf v}}
\newcommand{\rvec}{{\bf r}}

\newcommand{\uvec}{{\bf u}}
\newcommand{\kvec}{{\bf k}}

\newcommand{\Amat}{{A}}
\newcommand{\Bmat}{{B}}
\newcommand{\Cmat}{{C}}
\newcommand{\Mmat}{{M}}
\newcommand{\Smat}{{S}}
\newcommand{\Jmat}{{J}}
\newcommand{\Idmat}{{I}}
\newcommand{\Imat}{{\bf Id}}
\newcommand{\Fvec}{{\bf F}}

\begin{document}

\begin{frontmatter}

\title{$h$-multigrid agglomeration based solution strategies 
       for discontinuous Galerkin discretizations of incompressible flow problems}

  \author[unibg]{L. Botti\corref{cor1}}\ead{lorenzo.botti@unibg.it}
  \author[unibg]{A. Colombo}\ead{alessandro.colombo@unibg.it}
  \author[unibg]{F. Bassi}\ead{francesco.bassi@unibg.it}
  \address[unibg]{Dipartimento di Ingegneria e Scienze Applicate,
    Universit\`a di Bergamo \\
    via Marconi 4, 24044 Dalmine (BG), Italy}
  \cortext[cor1]{Corresponding author}

\begin{abstract}
In this work we exploit agglomeration based $h$-multigrid preconditioners to speed-up the iterative solution 
of discontinuous Galerkin discretizations of the Stokes and Navier-Stokes equations.
As a distinctive feature $h$-coarsened mesh sequences are generated by recursive agglomeration of a fine grid,
admitting arbitrarily unstructured grids of complex domains, and agglomeration based discontinuous Galerkin 
discretizations are employed to deal with agglomerated elements of coarse levels.
Both the expense of building coarse grid operators
and the performance of the resulting multigrid iteration are investigated. 
For the sake of efficiency coarse grid operators are inherited through 
element-by-element $L^2$ projections, avoiding the cost of 
numerical integration over agglomerated elements. 
Specific care is devoted to the projection of viscous terms discretized by means of the BR2 dG method.
We demonstrate that enforcing the correct amount of stabilization on coarse grids levels
is mandatory for achieving uniform convergence with respect to the number of levels.
The numerical solution of steady and unsteady, linear and non-linear problems is considered 
tackling challenging 2D test cases and 3D real life computations on parallel architectures.
Significant execution time gains are documented.
\end{abstract}


\end{frontmatter}

\section{Introduction}
Discontinuous Galerkin (dG) methods have proved to be effective in the CFD field allowing to simulate 
complex physics in complex domains while guaranteeing accuracy and robustness. 
Although very popular for compressible fluid flow simulations,
their adoption by incompressible fluid flow practitioners is still limited due to 
difficulties involved in the numerical solution of the Incompressible Navier-Stokes (INS) equations. 
On the one hand explicit and decoupled time integration strategies (\eg Pressure Poisson Equation segregated methods)
complicate the achievement of high-order pressure accuracy 
reducing the appeal of high-order accurate spatial discretizations. 
On the other hand fully implicit fully coupled velocity-pressure spatial discretisations 
result in systems of Differential Algebraic Equations (DAEs)
that are very expensive to solve 
due to the indefiniteness of the resulting system matrices, their poor spectral properties, 
and the saddle point nature of the problem \cite{Benzi05numericalsolution}.

In this work, in order to speed up the numerical solutions of coupled variables dG discretizations  
of incompressible flow problems, we consider $h$-multigrid solution strategies
on $h$-coarsened mesh sequences generated by recursive agglomeration of a fine grid.
$h$-multigrid is very attractive from the efficiency viewpoint in the sense that the number of 
arithmetic operations needed to solve a discrete problem is proportional to the number of 
degrees of freedom. Convergence factors, that is the average residual decrease at each 
multigrid iteration, can be made $h$-independent and small.

In the context of dG discretizations $p$-multigrid has been fruitfully applied 
in practical applications see \eg \cite{Fidkowski05,Nastase06,BassiGhidoni09,ShahbaziMavriplis-MGAlgorithms:2009},
while the theoretical and practical investigation of $h$- and $hp$-multigrid is more recent.
In 2003 Gopalakrishnan and Kanschat \cite{KanschatMultiAdvDiff} analyzed a V-cycle preconditioner for diffusion and 
advection-diffusion problems. 
Multigrid algorithms for dG discretizations of elliptic problems were considered by 
Brenner \ea \cite{Brenner11}, who proved uniform convergence with respect to the number of levels for F-,V- and W-cycle on graded meshes,
and Antonietti \ea \cite{AntoniettiSartiVerani}, who provided similar results for W-cycle $h$-,$p$- and $hp$-multigrid.
While the previous works employed $h$-refined mesh sequences, 
Prill \ea \cite{PrillHartmann} considered smoothed aggregation to build coarse problems for $h$-multigrid dG solvers.
\alertblu{The issue of developing optimal solvers for Composite discontinuous Galerkin Methods, first developed and analyzed Antonietti \ea \cite{AntoniettiGianiHouston2013}
was considered by Antonietti \ea \cite{Antonietti2014,AntoniettiHoustonSmears2016}.}
More recently Antonietti \ea \cite{AntoniettiHoustonSartiVerani} analysed multigrid strategies for Interior Penalty dG discretizations 
over agglomerated elements meshes, while Wallraff and Leicht \cite{WallraffLeicht14} and Wallraff \ea \cite{WallraffHartmannLeicht15}
applied an agglomeration based $h$-multigrid solver to dG discretizations of the compressible Reynolds Averaged Navier-Stokes (RANS) equations.

$h$-coarsening by agglomeration leads to unprecedented 
flexibility in the definition of the coarse meshes.
Starting from a fine grid, a coarse mesh can be generated on the fly 
clustering together a number of mesh elements. The process can be repeated at will in 
a recursive manner resulting in a nested mesh sequence.
Note that the generation of a sequence of nested grids by recursive refinement of a coarse mesh, 
\eg by means of element subdivision techniques, might require to improve the rough approximation 
of the computational domain provided by the coarse mesh. 
While coarsening by agglomeration is flexible enough to account for complex 3D domains,
physical frame dG discretizations allows to handle 
polyhedral elements of very general shape \cite{BassiFlexy12,BassiFree14,Cangiani14,Giani14,Cangiani16}. 
Nevertheless, as a consequence of the lack 
of efficient quadrature rules for agglomerated elements, 
numerical integration of bilinear and trilinear forms
might lead to excessive matrix assembly costs, see Bassi \ea \cite{BassiFree14}.

The present investigation focuses on efficiency of building coarse grid operators 
for dG discretizations of incompressible flow problems and effectiveness of the multigrid V-cycle iteration. 
In particular, we introduce a strategy for inheriting the BR2 dG formulation of \cite{Bassi.Rebay.ea:1997}, 
which provides optimal convergence properties and does not require numerical integration during assembly of coarse grid operators.
Besides the BR2 formulation, here employed for the discretization of the viscous terms, inherited multigrid 
can be fruitfully employed for the discrete divergence and the discrete gradient operators, and also for the discretization 
of the non-linear convective flux terms appearing in the Navier-Stokes equations.

The material is organized as follows. In Section \ref{sec:agglodGd} we  
introduce agglomeration based dG discretization over $h$-coarsened mesh sequences.
Section \ref{sec:dg_flow} is dedicated to presenting dG discretizations of incompressible flow problems:
\begin{inparaenum}[i)]
\item
the incompressible Navier-Stokes equations spatial and temporal discretization in Section \ref{sec:dg_ns} and \ref{sec:dg_ns_t}, respectively;
\item
the discretization of the steady Stokes problem in Section \ref{sec:dg_stk};
\item 
the BR2 dG formulation in Section \ref{sec:LapBR2}.
\end{inparaenum}
The ingredients of the $h$-multigrid iteration are described in Section \ref{sec:multiALL}:
\begin{inparaenum}[i)]
\item 
the V-cycle in Section \ref{sec:multiVcicle}; 
\item 
intergrid transfer operators in Section \ref{sec:intergrid};
\item
inherited coarse grid operators in Section \ref{sec:coarseGridOp}.
\end{inparaenum}
Section \ref{sec:precINS} briefly comments on the use of the $h$-multigrid V-cycle iteration as a preconditioner
for iterative solvers and introduces block preconditioners for the Stokes problem.
Performance gain assessment as compared to state-of-the-art iterative and direct solvers 
is conducted in Section \ref{sec:numres}. We consider
\begin{inparaenum}[i)]
\item
elliptic problems in Section \ref{sec:numresBR2}; 
\item
linear Stokes problems in Section \ref{sec:numresStokes};
\item
non-linear incompressible flow problems in Section \ref{sec:numresNS}.
\end{inparaenum}

\section{Agglomeration based dG discretizations}
\label{sec:agglodGd}
\subsection{Coarsening by agglomeration}
\label{sec:aggloTopo}
Let $\Omega$ be a bounded connected open domain.
Consider a (possibly non conforming) mesh $\Tf$ of $\Omega$ 
composed of (possibly curved) elements $\T\in\Tf$ such that
\begin{inparaenum}[(i)]
\item for any $\T\in\Tf$, there exists a reference polygon $\Tref$ 
and a polynomial mapping $\Psi_{\T}:\Tref \to \T$ such that $\T=\Psi_{\T}(\Tref)$.
\item quadrature rules of arbitrary order are available on the reference polygon $\Tref$.
\end{inparaenum}
The set of reference polygons includes but is not limited to triangular and quadrilateral reference elements in 2D, 
tetrahedral, hexahedral, pyramidal and prismatic reference elements in 3D.

\begin{figure}[H]
\begin{tabular}{lr}
\includegraphics[width=0.53 \textwidth]{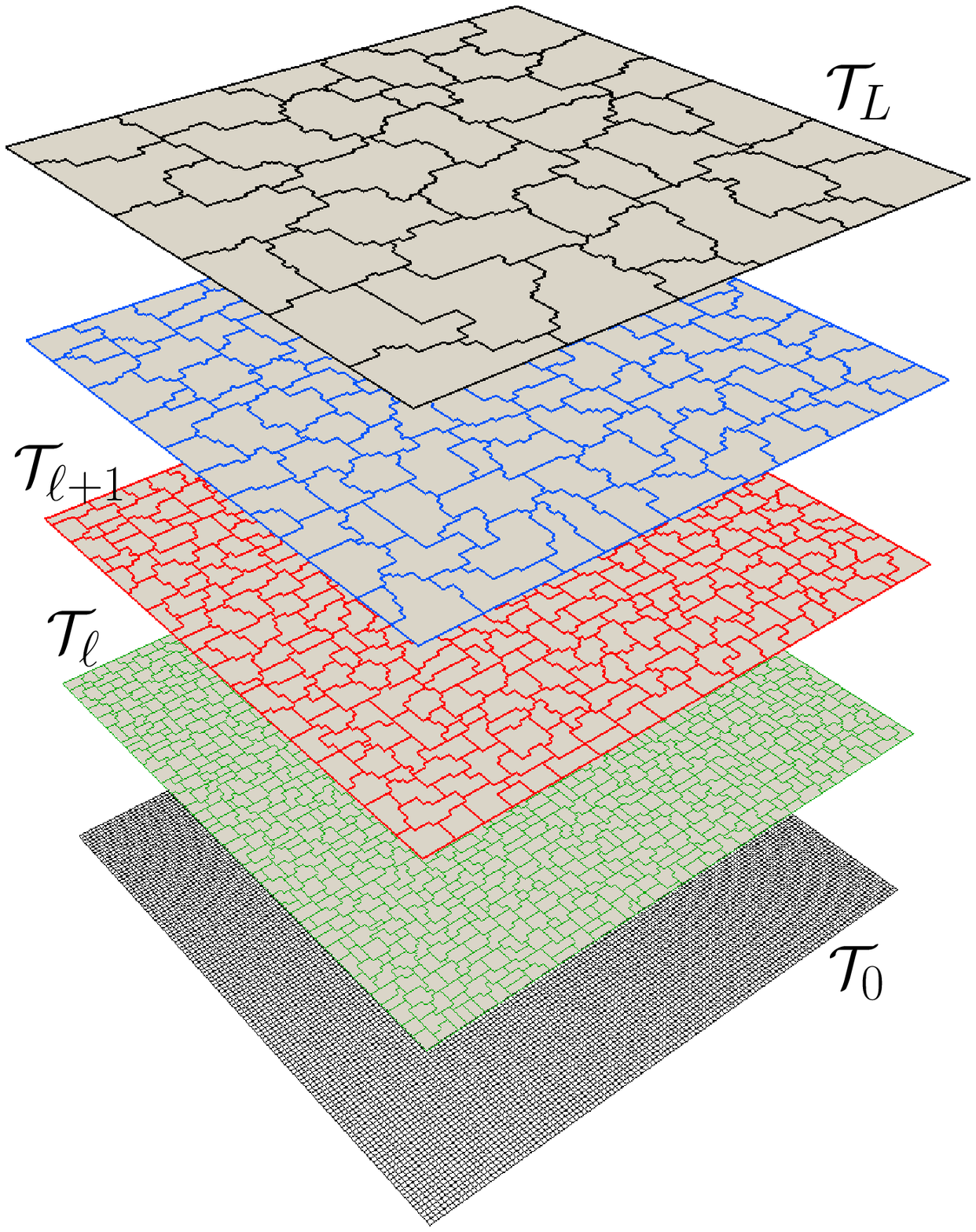} &
\includegraphics[width=0.43 \textwidth]{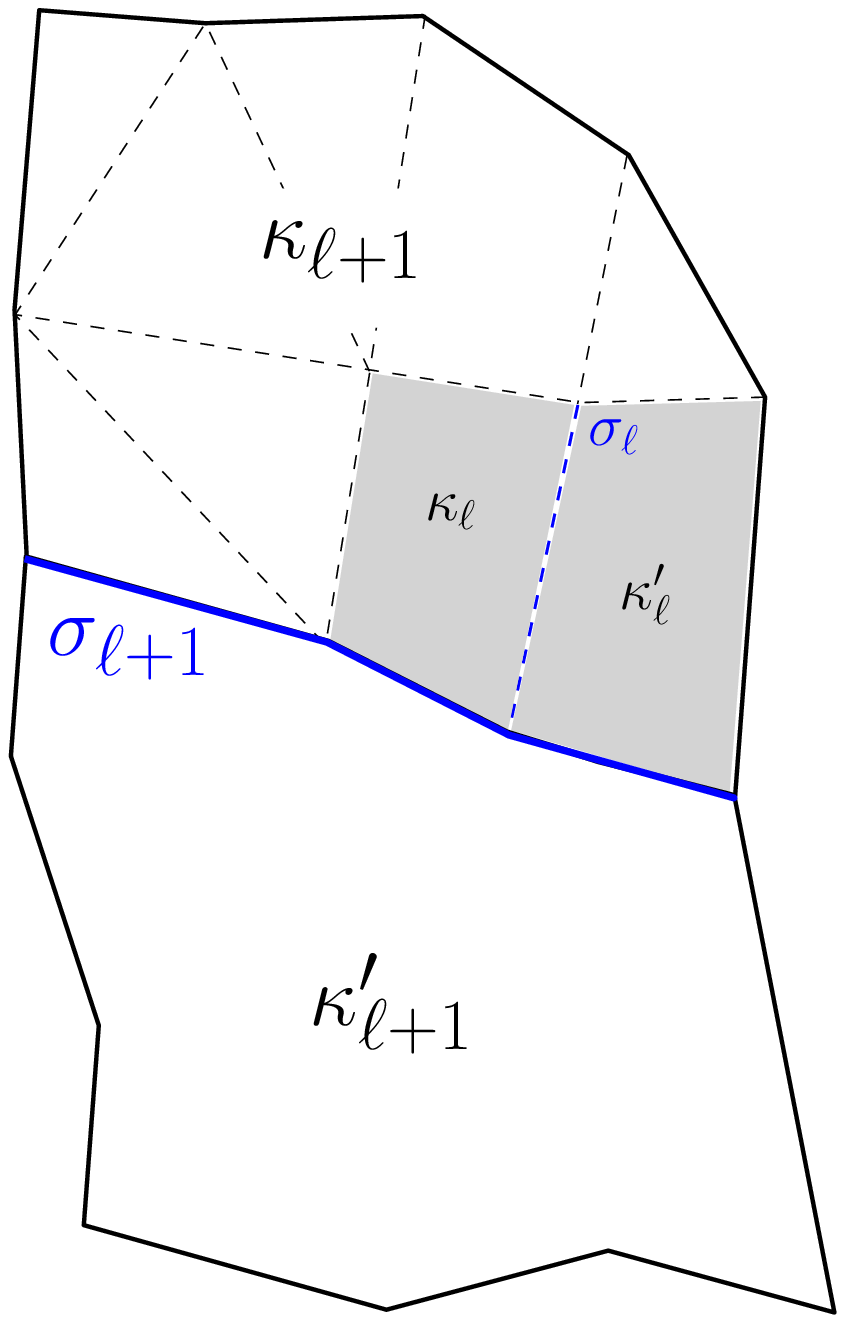} \\
\end{tabular}
\caption{\emph{Left}, example of a five levels ($L=4$) $h$-coarsened mesh sequence. 
         \emph{Right}, two mesh elements $\kappa_{\ell+1},\kappa'_{\ell+1} \in \Tlpo$ sharing a face $\sigma_{l+1}$ and 
	               two mesh elements $\kappa_{\ell},\kappa'_{\ell} \in K_{\ell}^{\ell+1} \subset \Tl$ sharing a face $\sigma_{l}$.
		       \label{fig:grid_seq}}
\end{figure}

Starting from $\Tf$ we can define a sequence of coarsened meshes $\Th$ by
agglomeration, see Figure \ref{fig:grid_seq}. 
For the sake of notation we denote by $\T_{\ell}$ any element $\T \in \Tl$ whose diameter is $h_\T$, and we denote the mesh 
size of $\Tl,\; \ell = 0,...,L$, by $h_{\ell} = \max_{\T \in \Tl} (h_\T)$.
Agglomeration generates a hierarchic sequence of nested grids, in particular for any $\Tl, \; \ell = 0,...,L-1$, we suppose that 
\begin{itemize}
\item $\Tlpo$ is a disjoint partition of $\Omega$ obtained clustering together the elements of $\Tl$;
\item every $\T\in\Tlpo$ is an open bounded connected subset of $\Omega$ and there exists $K_{\ell}^{\ell+1} \subset \Tl$ such that
  \begin{equation}
  \closure{\T}_{\ell+1}=\bigcup_{\T \in K_{\ell}^{\ell+1}}\closure{\T}_{\ell}. \label{eq:meshElement}
  \end{equation}
  The $\card{(K_{\ell}^{\ell+1})}$ cells clustered into the agglomerated element $\T_{\ell+1}$
  are referred to as sub-elements. 
\item for every $\T\in\Tl$ 
  there exists $K_0^{\ell} \subset \Tf$ such that
  \[
  \closure{\T}_{\ell}=\bigcup_{\T \in K_0^{\ell}}\closure{\T}_0.
  \]
  This can be obtained by applying \eqref{eq:meshElement} recursively and formalizes the fact that
  agglomerated elements on any mesh level $\ell$ can be expressed as a composition 
  of elements belonging to the finer mesh $\Tf$.
\end{itemize}

Clearly the coarsening steepness $h_{\ell+1}/h_{\ell}$ is influenced by 
the number of sub-elements composing aggregate elements as well as by the aspect ratio
of agglomerated elements, see Figure \ref{fig:grid_seq}.
In this work all the mesh sequences are generated setting $\overline{\card}{(K)} = 4,8$ in two and three space dimensions, respectively,
where the agglomeration rate $\overline{\card}{(K)}$ is a strict upper bound for the number of sub-elements,
so that $\card{(K)} \leq \overline{\card}{(K)}, \forall \T \in \Thagglo$.
The sequence of coarse meshes are generated by means of the library MGridGen \cite{Moulitsas.Karypis.ea:2001},
which allows to fix $\overline{\card}{(K)}$ and 
relies on optimization algorithms in order to ensure overall good quality of the agglomerated elements.
\alertblu{Note that, while the typical coarsening steepness $h_{\ell+1}/h_{\ell} = 2$ can be
obtained on regular Cartesian grids by regrouping 4 quadrilateral elements (2D case, 8 hexahedral elements in 3D) sharing a node,
on general unstructured grids leaving $\card{(K)}$ unbounded from below
gives room for more aggressive aspect ratio optimizations.}

To complete the definition of agglomerated grids we 
introduce inter-element boundaries 
where to define trace operators and fluxes of the dG discretization.
\begin{itemize}
\item Faces of an element $\T\in\Tf$ are defined 
as a portion of $\partial \T_0$ such that there exists
a (hyperplanar) face $\faceref$ of the corresponding reference
element $\Tref$ such that ${\F}$ is the image of $\faceref$
through the mapping $\Psi_{\T}$.
\item Faces of an agglomerated element $\T \in \Tl, \; \ell = 1,...,L$,
are defined as a portion of $\partial \T$ such that either $\F = \partial \T \cap
\partial \Omega$ or there exists $\Tpr\in\Tl$, $\Tpr\neq \T$, such that
$\F = \partial \T \cap \partial \Tpr$.
\end{itemize}
Mesh faces are collected in the sets $\Fl,\ell = 0,...,L$.
As mesh elements are composed by sub-elements,
every face $\F \in \Flpo$ is composed by sub-faces, also called facets, 
which belong to the set $\Fl$.
Moreover, for every face $\F$ we introduce the set
$\Sigma_{\ell}^{\ell+1} \subset \Fl$ collecting the facets partitioning
$\F_{\ell+1}$, \ie, 
\[
\closure{\F}_{\ell+1} = \bigcup_{\F \in \Sigma_{\ell}^{\ell+1}} \closure{\F}_{\ell},
\]
and applying the definition recursively we get
\[
\closure{\F}_{\ell} = \bigcup_{\F \in \Sigma_{0}^{\ell}} \closure{\F}_{0},
\]
where $\Sigma_{0}^{\ell} \subset \Fln{0}$.

We introduce the set of boundary mesh faces $\F\in\Flb$ such that
$\F\subset\partial\Omega$ and let $\Fli\eqbydef\Fl\setminus\Flb$ denote
the set of internal faces.
Moreover, for any mesh element $\T \in \Tl$, the set 
\begin{equation}
\FT \eqbydef \{ \F \in \Fl \ST \F \subset \partial \T  \},
\end{equation}
collects the mesh faces composing the boundary of $\T$.
The maximum number of mesh faces composing the boundary of mesh
elements is denoted by 
\begin{equation}
N_{{\partial}_\ell} \eqbydef \max_{\T \in \Tl} \left( \card{(\FT)} \right).
\end{equation}
For any mesh face $\F \in \Fl$ we define the set 
\begin{equation}
\TF \eqbydef \{ \T \in \Tl \ST \F \subset \partial \T  \}.
\end{equation}
$\TF$ regroups the two mesh elements $\T,\Tpr$ sharing $\F$ if $\F \in \Fli$
while it consists of a single mesh element if $\F \in \Flb$.

\subsection{Physical frame dG discretizations}
\label{sec:PFdGd}
For each mesh level $\ell = 0,...,L$ we consider the following broken polynomial spaces
\begin{equation}
  \label{eq:PTh}
  \Poly{d}{k}(\Tl)\eqbydef \left \{v_{\ell}\in L^2(\Omega) : 
  v_{\ell} |_{\T}\in\Poly{d}{k}(\T),\,\forall \T\in\Tl
  \right\},
\end{equation}
where $\Poly{d}{k}(\T)$ is the restriction to a mesh element $\T$ 
of the polynomials functions of $d$ variables and total degree at most $k$, 
such that $N_\mathrm{dof}^{\T} = {\rm dim}(\Poly{d}{k}) = {k+d \choose k}$.
Since in this work $d=\{2,3\}$ and no confusion is possible,
we drop the subscript and simply use the notation $\Poly{}{k}$ in place of 
$\Poly{d}{k}$.
Due to the nestedness of mesh elements we have $ \Poly{}{k}(\Tf) \supset \Poly{}{k}(\Tlone) \supset \Poly{}{k}(\Tltwo) ... \supset \Poly{}{k}(\TLc)$.

It is interesting to remark that  
physical frame discretizations are defined so to inherently span 
the space $\Poly{}{k}(\Tl)$ and provide optimal approximation properties on regular
$h$-refined mesh sequences $(\Thr)_{h>0}$, see \eg Botti~\cite{appPoly}.
Accordingly, for all $\T \in \Thr$ and for each polynomial degree $k$, the $L^2$-orthogonal projection operator 
${\pi}_{\T}^k: L^2(\T) \rightarrow \Poly{}{k}(\T)$ is such that 
for all $v \in H^{k+1}(\T)$, there holds
\begin{equation}
\label{eq:approxOPT}
\| v - \pi_{\T}^k v \|_{L^2(\T)} \leq C_{\rm app} h_{\T}^{k+1} |v |_{H^{k+1}(\T)} 
\end{equation}
where $C_{\rm app}$ is independent of $h$ and $k$.
The optimal approximation estimate \eqref{eq:approxOPT} holds true
over mesh sequences composed of agglomerated elements of very general shape,
in particular agglomerated elements meshes built on top of a curved elements mesh $\Tf$ 
are eligible to provide optimal approximation properties, see \eg \cite{Polygons-cf:2011}.
While the mesh regularity assumption implies star-shapedness of agglomerated elements,
see \cite{BrennerScott:2008} or \cite{DiPiErn11} for additional details,
the numerical convergence rates assessed in \cite{BassiFlexy12} and \cite{BassiFree14}
allow to claim that optimal approximation properties are achieved 
over mesh sequences obtained by means of the MGridGen library
(for instance using $\Th$ in reversed order as a $h$-refined mesh sequence).

Sharp approximation properties estimates valid in the general framework of $hp$-discontinuous Galerkin discretizations 
have been obtained introducing the concept of shape regular $d$-simplexes coverings of polygonal/polyhedral meshes, see \cite{Cangiani14}.

\subsection{Basis functions choice}
\label{sec:basisFunction}
For a given $\T\in\Tl, \; \ell = 0,...,L$, 
let $\Phi^k_{\Poly{}{}(\T)}=\{\varphi_i^{\T},i = 1,...,{\rm dim}(\Poly{}{k})\}$ denote a basis for $\Poly{}{k}(\T)$.
A basis for the space $\Poly{}{k}(\Tl)$ is given by
\begin{equation}
  \label{def:basis}
  \Phi^k \eqbydef \{ \Phi^k_{\Poly{}{}(\T)} \}_{\T \in \Tl}.
\end{equation}
where each basis functions $\varphi_i^{\T}$ is extended to
$\Omega$ by simply setting $\varphi_i^{\T}=0$ on $\Omega\setminus \T$.

From a practical viewpoint, in order to find a numerically
satisfactory physical frame basis function
we rely on the procedure proposed by
Bassi, Botti, Colombo, Di Pietro and Tesini~\cite{BassiFlexy12}. 
Starting from a monomial basis for each elementary space $\Poly{}{k}(\T)$
defined according to a reference frame whose 
axes are aligned with the principal axes of inertia of $\T$, 
an $L^2$-orthonormal basis is inferred by means of the Modified 
Gram-Schmidt (MGS) orthogonalization procedure.
The resulting basis functions $\Phi^k=\{\varphi_i^\T\}$ are hierarchical, orthogonal with respect 
to the $L^2$ inner product and provide well conditioned local matrices at high polynomial degrees.
In particular the elementary mass matrices
are unit diagonal, for any element shape.

The sole requirement to apply the orthogonalization strategy is the capability to
compute the integrals of polynomial functions on each
element $\T$. In the case of agglomerated elements this is achieved by exploiting the partition
$K_0$ into standard-shaped sub-elements.
The integral of any $v\in\Poly{d}{k}(\Tl)$ is computed as follows
\begin{equation} 
\label{eq:intRefAgglo}
  \int_{\T_{\ell}} v(\xvec)\ud\xvec
  =\sum_{\T\in K_0^{\ell}} \int_{\T_0} v(\xvec)\ud\xvec
  =\sum_{\T\in K_0^{\ell},\,\T_0=\Psi_{\T}(\Tref)}
      \int_{\Tref} (v\circ\Psi_{\T})(\xivec)|J_{\Psi_{\T}}(\xivec)|\ud\xivec,
\end{equation}
where $\xvec$ and $\xivec$ are physical and reference space
coordinates, respectively, and $J_{\Psi_{\T}}$ is the Jacobian of the
mapping function $\Psi_{\T}$.
The order of exactness required for exact integration over each sub-element
rapidly increases when considering high order polynomials on curved elements.
Moreover, the use of Gaussian quadrature rules defined on the reference frame 
polygon $\Tref$ might lead to an excessive growth of the number of quadrature points 
if the agglomerated elements are composed of many sub-elements. 
\subsection{Average, jump and lifting operators}
\label{sec:dGoperators}
For all $\F\in\Fli$ and all $v_{\ell}\in \Poly{}{k}(\Tl)$ we introduce the jump and
average operators defined as follows:
$$
\jump{v_{\ell}}_\F \eqbydef{v_{\ell}}{|_{\T}}-{v_{\ell}}{|_{\Tpr}},\qquad
\averg{v_{\ell}}_\F \eqbydef\frac12({v_{\ell}} |_{\T}+{v_{\ell}} |_{\Tpr}).
$$
Whenever no confusion can arise we drop the subscript $\F$.
On boundary faces, we conventionally set
$\jump{v_{\ell}}=\averg{v_{\ell}}=v_{\ell}$.
When $v$ is vector-valued, the weighted average operator acts componentwise on the function $v$.

For all $\F\in\Flb$,  $\normal_{\F}$ denotes the unit outward
normal to $\Omega$, whereas, for all $\F\in\Fli$ such that
$\F\subset\partial \T \cap \partial \Tpr$,
$\normal_{\F}$ is defined as the unit normal pointing out of $\T$ (the
order of the elements sharing $\F$ is arbitrary but fixed).
For all $\F\in\Fl$ we define the (local) lifting
operator \mbox{$\mathbf{r}_{\F}: L^2(\F) \rightarrow
  [\mathbb{P}_d^k(\Tl)]^d$}, such that, for all $\phi\in
L^2(\F)$,
\begin{equation}
  \label{eq:rF}
  \int_{\Omega} \mathbf{r}_{\F}(\phi) \SCAL \boldsymbol{\tau}_{\ell} = \int_{\F}
  \phi \averg{\boldsymbol{\tau}_{\ell}}\SCAL \normal_{\F}\qquad
  \forall\boldsymbol{\tau}_{\ell} \in [\mathbb{P}^k(\Tl)]^d.
\end{equation}
Note that the support of $\mathbf{r}_{\F}$ consists of one and two mesh elements 
if $\F \in \Fl^{\mathrm{b}}$ and $\F \in \Fl^{\mathrm{i}}$, respectively, that is
$$
\mathrm{supp}(\mathbf{r}_{\F}) = \bigcup_{\T \in \TF} \overline{\T}.
$$
For any function $v \in H^1(\Tl)$, we also introduce the global lifting
\begin{equation}
  \label{eq:RF}
 \displaystyle \mathbf{R}_{l}(v) := 
     \sum_{\F\in\mathcal{F}_l} \mathbf{r}_{\F}(\jump{v}),
\end{equation}
which collects the local lifting contributions, note that $\jump{v}_\F \in L^2(\sigma)$.

\section{Incompressible flow problems}
\label{sec:dg_flow}
\subsection{Incompressible Navier-Stokes equations dG discretization}
\label{sec:dg_ns}
We consider the unsteady INS equations with Dirichlet
boundary conditions,
\begin{subequations}
  \label{eq:ns}
  \begin{alignat}{2}
    \label{eq:ns.momentum}
    \partial_t \mathbf{u}
    + \mathbf{u} \cdot \NABLA \mathbf{u} 
    -\DIV (\nu\NABLA\mathbf{u})+\NABLA p= 0
    &\qquad& \text{in $\Omega\times (0,t_F)$}, \\
    \label{eq:ns.mass}
    \DIV\mathbf{u} = 0
    &\qquad& \text{in $\Omega\times (0,t_F)$}, \\
    \label{eq:ns.bc}
    \mathbf{u}= \mathbf{f} 
    &\qquad& \text{on $\partial\Omega\times (0,t_F)$}, \\
    \label{eq:ns.ic}
    \mathbf{u}(\cdot,t=0) = \mathbf{u}^0,
    &\qquad& \text{in $\Omega$}, \\
    \label{eq:ns.zero-mean}
    \langle p\rangle_\Omega = 0,
  \end{alignat}
\end{subequations}
where $\mathbf{u} \in \mathbb{R}^{d}$ is the velocity vector, $p$ is the pressure, $\nu>0$ denotes the (constant) viscosity, 
$\mathbf{f}$ is the boundary datum, $\mathbf{u}^0$ is the initial condition, and
$\langle\cdot\rangle_\Omega$ denotes the average value over $\Omega$.
The density has been assumed to be uniform and equal to one.

Letting $\Fvec^{\nu} = -\nu \nabla \otimes \uvec$ and
$\Fvec^c = \mathbf{u} \otimes \mathbf{u} + p \Imat$ 
be the viscous and convective flux functions,
Eqs.~\eqref{eq:ns.momentum}-\eqref{eq:ns.mass}
can be written in conservation form as
\begin{equation}\label{eq:NS}
{\partial_t} \uvec+ 
\nabla \cdot \Fvec = {\bf 0},
\end{equation}
where $ \Fvec \eqbydef \left[ \Fvec^c + \Fvec^{\nu}, \uvec \right] \in \mathbb{R}^d \otimes \mathbb{R}^{d+1}$. 
For $d=3$ we get 
\begin{equation}
\Fvec = 
\left[\begin{array}{l l l c}
uu + \nu \frac{\partial u}{\partial x}+p & uv + \nu \frac{\partial v}{\partial x}   & uw + \nu \frac{\partial w}{\partial x}   & u \\
vu + \nu \frac{\partial u}{\partial y}   & vv + \nu \frac{\partial v}{\partial y}+p & vw + \nu \frac{\partial w}{\partial y}   & v \\
wu + \nu \frac{\partial u}{\partial z}   & wv + \nu \frac{\partial v}{\partial z}   & ww + \nu \frac{\partial w}{\partial z}+p & w \\
\end{array} \right]
\end{equation}

The dG discretization of the Navier-Stokes equations we rely upon consists in seeking 
$(\mathbf{u}_0,p_0) \in [\Poly{}{k}(\Tf)]^{d+1}$  such that
\begin{subequations}
\label{discr:ns}
\begin{align}
& \intO{ \vvec_0 \cdot \partial_t \uvec_0}
-  \intO{ \nabla_0 \vvec_0 : \left[ \Fvec^c(\uvec_0,p_0) + \widetilde{\Fvec}^{\nu}\left(\nabla_0 \uvec_0, \mathbf{R}_{0}(\uvec_0) \right)\right]} \nonumber \\
&+\sumF \intB{\sigma}{\normal_{\sigma}\otimes \jump{\vvec_0}  :
\left[\widehat{\Fvec}^c\left( \uvec_0^{\T,\Tpr}, p_0^{\T,\Tpr}\right) + 
      \widehat{\Fvec}^{\nu}\left(\nabla_0 \uvec_0^{\T,\Tpr}, \eta_\F \rvec_\sigma^{\T,\Tpr} \left({ \jump{\uvec_0} }\right) \right)\right]} = 0, \\
- &\intO{ \nabla_0 q_0 \cdot \uvec_0} + \sumF \intB{\sigma}{\jump{q_0} \,\normal_{\sigma} \; \cdot \widehat{\uvec}(\uvec_0^{\T,\Tpr},p_0^{\T,\Tpr})} = 0, \\
&\intO{p_0} = 0 \nonumber
\end{align}
\end{subequations}
for all $(\mathbf{v}_0,q_0) \in [\Poly{}{k}(\Tf)]^{d+1}$.

According to the BR2 scheme, proposed in~\cite{Bassi.Rebay.ea:1997} and theoretically 
analyzed in \cite{Brezzi.Manzini.ea:2000} and \cite{Arnold.Brezzi.ea:2002}, 
the viscous numerical fluxes read
\begin{eqnarray}
\label{eq:vflux}
&\widetilde\Fvec^{\nu}\left( \nabla_0 \uvec_0,\mathbf{R}_{0} \left(\uvec_0\right)\right)
\eqbydef -\nu \nabla_0 \uvec_0 + \mathbf{R}_{0} \left(\uvec_0\right), \\
&\widehat\Fvec^{\nu}\left(\nabla_0 \uvec_0^{\T,\Tpr}, \eta_\F \rvec_\sigma^{\T,\Tpr} \left({ \jump{\uvec_0} }\right)\right)
\eqbydef -\nu \averg{\nabla_0 \uvec_0} + \eta_\F \averg{\rvec_\F \left(\jump{\uvec_0}\right)}.
\end{eqnarray}
$\widetilde\Fvec^{\nu}$ is the consistent discrete gradient while 
$\widehat\Fvec^{\nu}$ is the consistent diffusive flux ensuring symmetry and stability of the scheme.
In particular coercivity holds provided that $\eta_\F$ is greater 
than the maximum number of faces of the elements sharing $\F$.
The inviscid physical and numerical fluxes of the dG discretization reads
\begin{equation}
 {\Fvec}^{!\nu}(\wvec_0) \eqbydef \left[ \Fvec^c(\wvec_0), \uvec_0 \right]  \;\; \mbox{and} \;\;
 \widehat{\Fvec}^{!\nu}(\wvec_0) \eqbydef \left[ \widehat{\Fvec}^c(\wvec_0), \widehat{\uvec}_0(\wvec_0) \right],
\end{equation}
respectively.
The inviscid numerical fluxes $\widehat{\Fvec}^{!\nu}$ 
result from the exact solution of local Riemann problems 
based on an artificial compressibility perturbation of the Euler equations, as proposed in~\cite{Bassi.Crivellini.ea:2006}.

Boundary conditions are enforced weakly by properly defining for each $\F \in \Ffb$ a boundary state $(\uvec^{\Tpr^b}, p^{\Tpr^b})$ 
having support on the interface of a ghost neighboring elements $\Tpr^b$.
The ghost boundary state is defined based on the method of characteristics 
exploiting the hyperbolic nature of the artificial compressibility perturbation of the Euler equation.
Accordingly the ghost state depend on the Dirichlet datum $f$ but also on the internal state $(\uvec^{\T}, p^{\T})$.
Once ghost states are computed, handling of internal and boundary faces is similar:
for each $\F \in \Ff$ two neighboring elements $\T,\Tpr$ 
concur to the computation of numerical fluxes and lifting operators.

\subsection{Navier-Stokes equations temporal discretization}
\label{sec:dg_ns_t}
For the sake of notation we collect the vector velocity and the pressure polynomial expansions
in the vector $\wvec \eqbydef (u_{0,1},...,u_{0,d},p_0) \in [\Poly{}{k}(\Tf)]^{d+1}$
and identify the unknown vector at time $t_n$ with $\wvec_0^n$, that is 
$\wvec_0^n = [\uvec_0(t_n),p_0(t_n)]$.
For all $\mathbf{\delta} \wvec_0, \wvec_0, \kvec_0 \in [\Poly{}{k}(\Tf)]^{d+1}$ we introduce the following bilinear and trilinear forms 
\begin{align}
 m_{i}(\delta w_i,k_i) &= +\sumT\intB{\T}{ k_{i} \, \delta w_{i}} \nonumber \\
 j^{!\nu}_{i,j}(\wvec, {\delta} w_j, k_i) &= 
   - \sumT \sum_{l=1}^d \intB{\T}{ \frac{\partial k_{i}}{\partial x_l} \frac{\partial {F}_{l,i}^{!\nu}(\wvec)}{\partial{w_j}} \delta w_j} 
   + \sumF \sum_{l=1}^d\intB{\sigma}{\jump{k_{i}} n_{\sigma,l} \frac{\partial \widehat{F}^{!\nu}_{l,i}(\wvec)}{\partial{w_j}} \delta w_j}  \label{eq:jacTrilF}\\
 j^{\nu}_{i}({\delta} w_i, k_i) &= 
   - \sumT \sum_{l=1}^d \intB{\T}{ \frac{\partial k_{i}}{\partial x_l} \frac{\partial \widetilde{F}^{\nu}_{l,i}(w_i)}{\partial{w_i}} \delta w_i} 
   + \sumF \sum_{l=1}^d\intB{\sigma}{\jump{k_{i}} n_{\sigma,l} \frac{\partial \widehat{F}^{\nu}_{l,i}(w_i)}{\partial{w_i}} \delta w_i}  \label{eq:jacTrilFv}\\
 f_i^m(w_i, k_i) &= -\sumT\intB{\T}{ k_{i} \, w_{i}} \label{eq:rhsNStime} \\ 
 f^{!\nu}_i(\wvec, k_i) &= +  \sumT\sum_{l=1}^d\intB{\T}{ \frac{\partial k_{i}}{\partial x_l} {F}^{!\nu}_{l,i}(\wvec)} 
                 - \sumF \sum_{l=1}^d\intB{\sigma}{\jump{k_{i}} n_{\sigma,l} \widehat{F}^{!\nu}_{l,i}(\wvec)}, \nonumber \\
 f^{\nu}_i(w_i, k_i) &= +  \sumT\sum_{l=1}^d\intB{\T}{ \frac{\partial k_{i}}{\partial x_l} \widetilde{F}^{\nu}_{l,i}(w_i)} 
                 - \sumF \sum_{l=1}^d\intB{\sigma}{\jump{k_{i}} n_{\sigma,l} \widehat{F}^{\nu}_{l,i}(w_i)}. \nonumber 
\end{align}
In the above definitions we dropped the mesh sequence subscript for notation convenience. 
Note that, by abuse of notation, \eqref{eq:jacTrilF} is a 
bilinear (resp. trilinear) when $F_{l,i}(\wvec)$ is a linear (resp. non-linear) function of $w_j$. 

Given the initial condition $\wvec_0^0=\wvec_0(t=0) \in [\Poly{}{k}(\Tf)]^{d+1}$ we define the sequence $\wvec_0^{n+1} $
iteratively by means of the backward Euler method: 
\begin{algorithm}[H]
\caption{Backward Euler \label{algo:be}}
\begin{algorithmic}[1]
  \STATE set ${\wvec}_0^{n} = \wvec_0^0$, $n_F = \displaystyle\frac{t_F}{\delta t}$
  \FOR{$n = 0,1,...,n_F$}
    \STATE set ${\wvec}_0^{n+1} \leftarrow \wvec_0^n$
    \WHILE{$\mathbf{\delta} \wvec_0$ is too large \label{bewhile:convdw}}
      \STATE{find $\mathbf{\delta} \wvec_0 \in [\Poly{}{k}(\Tf)]^{d+1}$ such that, for all $\kvec_0 \in [\Poly{}{k}(\Tf)]^{d+1}$
        \begin{align}
        &\displaystyle \frac{1}{\delta t} \sum_{i=1}^{d} m_i(\mathbf{\delta} w_{0,i}, k_{0,i}) + 
	                                  \sum_{i=1}^{d+1} \sum_{j=1}^{d+1} j_{i,j}^{!\nu}({\wvec}_0^{n+1}, \delta w_{0,j}, k_{0,i}) +
	                                  \sum_{i=1}^{d} j_{i}^{\nu}(\delta w_{0,i}, k_{0,i}) = \nonumber\\
        &\frac{1}{\delta t} \sum_{i=1}^{d} f^m_i({w}_{0,i}^{n+1}-w_{0,i}^{n},k_i) + \sum_{i=1}^{d+1}f^{!\nu}_i({\wvec}_0^{n+1},k_i) 
                 	+ \sum_{i=1}^{d}f^{\nu}_i({w}_{0,i}^{n+1},k_i) \label{eq:nsbe} \\
        &\langle \delta w_{0,d+1} \rangle_\Omega = 0,\label{eq:pabe}
        \end{align}}
        \vspace{-0.7cm}
      \STATE set ${\wvec}_0^{n+1} \mathrel{+}= \delta \wvec_0$
    \ENDWHILE
  \ENDFOR
\end{algorithmic}
\end{algorithm}
Note that the continuation condition at line \ref{bewhile:convdw} can be replaced by checking that a proper norm of the right hand side 
of Equation \eqref{eq:nsbe} is too large.
Equation \eqref{eq:pabe} is needed since the average value 
of the pressure increment is left undefined in Equation $\eqref{eq:nsbe}$.

To recast Problem \eqref{eq:nsbe} in operator form
we let $X^k(\mathcal{T}_0)= \Poly{}{k}(\mathcal{T}_0)\times[\Poly{}{k}(\mathcal{T}_0)]^{d}$ 
and introduce the linear operators such that,
$\forall \wvec_0 \in [\Poly{}{k}(\mathcal{T}_0)]^{d+1}$
\begin{align} 
(J^{A}_{0} \delta \uvec_0, \mathbf{v}_0)_{[{L^2}(\Omega)]^{d}} &= \sum_{i=1}^d j^{\nu}_{i}(\mathbf{\delta} w_{0,i}, k_{0,i}),
& \forall \mathbf{\delta} \uvec_0,\mathbf{v}_0 \in [\Poly{}{k}(\mathcal{T}_0)]^{d}, \nonumber \\
(J^B_{0}(\wvec_0) \delta \uvec_0, q_0)_{{L^2}(\Omega)} &= \sum_{j=1}^d j^{!\nu}_{d+1,j}({\wvec}_0, \mathbf{\delta} w_{0,j}, k_{0,d+1}), 
& \forall \mathbf{\delta} (q_0,\uvec_0) \in X^k(\mathcal{T}_0), \nonumber\\
(J^{B^t}_{0}(\wvec_0) \delta p_0, \mathbf{v}_0)_{[{L^2}(\Omega)]^{d}} &= \sum_{i=1}^d j^{!\nu}_{i,d+1}({\wvec}_0, \mathbf{\delta} w_{0,d+1}, k_{0,i}), 
& \forall (\mathbf{\delta} p_0,\mathbf{v}_0) \in X^k(\mathcal{T}_0), \nonumber\\
(J^C_{0}(\wvec_0) \delta p_0, q_0)_{{L^2}(\Omega)} &= j^{!\nu}_{d+1,d+1}({\wvec}_0, \mathbf{\delta} w_{0,d+1}, k_{0,d+1}), 
& \forall \mathbf{\delta} p_0,q_0 \in \Poly{}{k}(\mathcal{T}_0), \nonumber\\
(J^{D}_{0}(\wvec_0) \delta \uvec_0, \mathbf{v}_0)_{[{L^2}(\Omega)]^{d}} &= \sum_{i=1}^d \sum_{j=1}^d j^{!\nu}_{i,j}({\wvec}_0, \mathbf{\delta} w_{0,j}, k_{0,i}),
& \forall \mathbf{\delta} \uvec_0,\mathbf{v}_0 \in [\Poly{}{k}(\mathcal{T}_0)]^{d}, \nonumber \\
(M_{0} \delta \uvec_{0}, \vvec_{0})_{[{L^2}(\Omega)]^d} &= \sum_{i=1}^d m_{0}(\delta w_{0,i},k_{0,i}),                    
& \forall \mathbf{\delta} \uvec_0,\mathbf{v}_0 \in [\Poly{}{k}(\mathcal{T}_0)]^{d}. \nonumber 
\end{align}
Moveover we introduce the residuals of momentum and continuity equations
\begin{align} 
\mathbf{f}^M_0(\wvec_0^{n,n+1},\vvec_0) &= \sum_{i=1}^{d}f_i^m(u_{0,i}^{n+1}-u_{0,i}^n, v_{0,i}) + \sum_{i=1}^{d}f^{v,nv}_i(\wvec_0^{n+1}, v_{0,i}), 
                                                                 & \forall \vvec_0 \in[\Poly{}{k}(\mathcal{T}_0)]^{d}, \nonumber\\
f^C_0(\wvec_0^{n+1},q_0) &= f_{d+1}(\wvec_0^{n+1}, q_{0}), 
                           & \forall q_0 \in \Poly{}{k}(\mathcal{T}_0). \nonumber
\end{align}
Problem \eqref{eq:nsbe} amounts at solving a linear system in the form:
\begin{equation}
\label{discr:navierStokesBlock} 
\Amat_0^{\mathrm{INS}} \; 
\left[ \begin{array}{c}
{\delta \uvec}_0 \\ {\delta p}_0
\end{array} \right]
= 
\left[ \begin{array}{c}
{\mathbf{f}}_0^M \\ {f}_{0}^C
\end{array} \right]
,\;
\mbox{with}
\;
\Amat_0^{\mathrm{INS}} =
\left[
\begin{array}{cr}
\Mmat_0 + \Jmat^{A}_0 + \Jmat^{D}_0  & \Jmat_0^{B^t} \\ 
\Jmat_0^B  & \Jmat_0^C      
\end{array} \right].
\end{equation}

\subsection{Stokes equations dG discretization}
\label{sec:dg_stk}
The steady Stokes equation problem can be obtained dropping the time derivative 
and the convective term, that is the first two terms in equation \eqref{eq:ns.momentum}.
In case of a steady Stokes flow the inviscid interface fluxes $\widehat{\Fvec}^c$ and $\widehat{\uvec}$ can be explicitly 
computed as the solution of a linear hyperbolic system, see ~\cite{Bassi.Crivellini.ea:2006} for details.
The resulting dG discretization reads:
find $(\mathbf{u}_0,p_0) \in [\Poly{}{k}(\Tf)]^{d+1}$ such that
\begin{subequations}
\label{discr:stokes} 
\begin{align}
\nu a_0(\uvec_0, \vvec_0) + b_0(\vvec_0, p_0) &= \mathbf{f}_0^M(\mathbf{f},\vvec_0), && \forall \vvec_0 \in [\Poly{}{k}(\Tf)]^d, \\
-   b_0(\uvec_0, q_0)     + c_0(p_0, q_0)     &= f_0^C(\mathbf{f},q_0),     && \forall q_0 \in \Poly{}{k}(\Tf), \\ 
   \langle p_0 \rangle_\Omega &= 0,
\end{align}
\end{subequations}
where 
\begin{align}
b_0(\vvec_0,q_0) &\eqbydef - \intO q_0 \nabla_0 \cdot \vvec_0 + \sumF \intB{\F}{\jump{\vvec_0}} \cdot \normal_{\F} \averg{q_0}, \label{eq:Bdiv} \\
                 & = \intO \vvec_0 \cdot \nabla_0 q_0 - \sumFi \intB{\F}{\averg{\vvec_0}} \cdot \normal_{\F} \jump{q_0}, \label{eq:Bgrad} \\
c_0(q_0,t_0) &\eqbydef \sumFi h_\F \intB{\F}{\jump{q_0}} \jump{t_0}, \label{eq:stabStokes}  \\
a_0(\vvec_0,\wvec_0) &\eqbydef \sum_{i=1}^d j_0^{\nu}(v_{0,i}, w_{0,i}) = \sum_{i=1}^d a_0^{\mathrm{BR2}}(v_{0,i}, w_{0,i}), \label{eq:BR2Stokes}
\end{align}
and the terms on right hand side defined below accounts for the week imposition of Dirichlet boundary conditions
\begin{align}
\mathbf{f}_0^M(\mathbf{f}; \vvec_0) &= - \sum_{i=0}^{d} \intO \mathbf{R}_{0}({f}_i) \cdot \nabla_0 v_{0,i} + 
                         \eta_{\sigma} \sum_{\F \in \Ffb} \sum_{i=0}^{d} \intO \mathbf{r}_{\F}({f}_i) \cdot \mathbf{r}_{\F}(v_{0,i}), \\
f_0^C(\mathbf{f}; q_0) &= -\sum_{\F \in \Ffb} \intB{\F} \mathbf{f} \cdot \normal_\F q_0.
\end{align}
According to \eqref{eq:BR2Stokes} the discretization of the viscous term can be obtained applying the BR2 method to each velocity component, 
\cf definitions \eqref{eq:jacTrilFv} and \eqref{eq:ahBR2}. 
The dG discretization in \eqref{discr:stokes} was analysed by Di Pietro in \cite{DiPiStokes07}, see also \cite[Chapter~6]{DiPiErn11}.

Problem \eqref{discr:stokes} has a block structure which we can take advantage for devising effective preconditioners.
To this end we define the operators $A_0$, $C_0$ and $B_0$ 
\begin{align} 
(A_{0} \vvec_{0},\wvec_{0})_{[{L^2}(\Omega)]^d} &= \nu a_{0}(\vvec_0,\wvec_{0}), \;  &\forall \vvec_{0},\wvec_{0} \in [\Poly{}{k}(\mathcal{T}_0)]^d, \label{discr:opA}\\
(C_{0} q_{0},r_{0})_{{L^2}(\Omega)} &= c_{0}(q_0,r_{0}), \;                      &\forall q_{0},r_{0} \in \Poly{}{k}(\mathcal{T}_0), \label{discr:opC}\\
(B_{0} \vvec_{0},q_{0})_{{L^2}(\Omega)} &= -b_{0}(\vvec_0,q_{0}), \;              &\forall (\vvec_{0},q_{0}) \in [\Poly{}{k}(\mathcal{T}_0)]^d \times \Poly{}{k}(\mathcal{T}_0)\label{discr:opB}.
\end{align}
Note that according to \eqref{eq:Bgrad} we are able to infer $(\vvec_{0}, B_{0}^t q_{0})_{[{L^2}(\Omega)]^d} = b_{0}(\vvec_0,q_{0})$.

Problem \eqref{discr:stokes} amounts at solving a linear system in the form:
\begin{equation}
\label{discr:stokesBlock} 
\Amat_0^{\mathrm{Stk}} \;
\left[ \begin{array}{c}
{\uvec}_0 \\ {p}_0
\end{array} \right]
= 
\left[ \begin{array}{c}
{{\mathbf{f}}}_0^M \\ {f}_0^C 
\end{array} \right]
,\;
\mbox{with}
\;
\Amat_0^{\mathrm{Stk}} =
\left[
\begin{array}{rr}
\Amat_0  & \Bmat_0^t \\ 
\Bmat_0 & \Cmat_0      
\end{array} \right].
\end{equation}

\subsection{Viscous terms dG discretization}
\label{sec:LapBR2}
The BR2 dG formulation is employed for the discretization of the viscous terms of Equation \eqref{eq:ns.momentum} being an 
important building block of both the Stokes and the Navier-Stokes dG discretizations.
In this work we focus on the performance of solving the BR2 dG 
discretization of the following model Poisson problem
\begin{equation}
  \label{eq:lap}
  \begin{cases}
    -\triangle u = f & \text{in $\Omega$}, \\
    u = 0 & \text{on $\partial\Omega$}.
  \end{cases}
\end{equation}
Assessing and improving the performance of $h$-multigrid for elliptic problems 
has been a critical step for achieving satisfactory performances on incompressible flow problems.

The BR2 method can be written into a consistent and symmetric bilinear form plus stabilization term as follows: For all $ v_0,w_0 \in \Poly{}{k}(\Tf)$,
\begin{equation}
  \label{eq:ahBR2}
  a_{0}^{\mathrm{BR2}}(v_{0}, w_{0}) \eqbydef a^{\mathrm{BR2,CS}}_{0}(v_{0}, w_{0}) +s^{\mathrm{BR2}}_{0}(v_{0},w_{0}), 
\end{equation}
where  
\begin{align}
  a^{\mathrm{BR2,CS}}_{0}(v_{0}, w_{0}) & \eqbydef \int_\Omega\left(\nabla_0 v_{0}-\mathbf{R}_{0}(v_{0})\right)\SCAL \left(\nabla_0 w_{0}-\mathbf{R}_{0}(w_{0})\right) 
                                              -\int_\Omega \mathbf{R}_{0}(v_{0})\SCAL\mathbf{R}_{0}(w_{0}), \label{eq:ahcs} \\
  s^{\mathrm{BR2}}_{0}(v_{0},w_{0}) & \eqbydef \sum_{\F \in \Fl} \eta_{\F}\int_{\Omega}\mathbf{r}_{\F}(\jump{v_{0}})\SCAL\mathbf{r}_{\F}(\jump{w_{0}}). \label{eq:stabBR2}
\end{align}
Using the definition of the BR2 bilinear form in \eqref{eq:ahBR2} the discretization of~\eqref{eq:lap} reads:
\begin{equation}
  \label{eq:lapdiscr}
  \text{Find $u_{0}\in W_{0}$ \st $a^{\mathrm{BR2}}_{0}(u_{0},v_{0})=\int_{\Omega} f v_{0}$
    for all $v_{0}\in W_{0}$.}
\end{equation}
Well-posedness of problem \eqref{eq:lapdiscr}
was proved by Brezzi et al.~\cite{Brezzi.Manzini.ea:2000}.

The BR2 method in \eqref{eq:lapdiscr} can be reformulated as follows: Given $f \in L^2(\Omega)$,
\begin{equation}
  \label{eq:lapdiscrMatrix}
  \text{find $u_0 \in \Poly{}{k}(\Tf)$ \st $\Sys{0}^{\mathrm{BR2}} u_{0} = {\pi}_{\Tf}^k f$},
\end{equation}
where ${\pi}_{\Tf}^k$ is the $L^2(\Omega)$-orthogonal projection operator into $\Poly{}{k}(\Tf)$ and 
$A_0^{\mathrm{BR2}}$ is the fine grid operator, such that
\begin{equation} 
(A_{0}^{\mathrm{BR2}} u_{0},v_{0})_{{L^2}(\Omega)} = a^{\mathrm{BR2}}_{0}(u_0,v_{0}), \;  \forall u_{0},v_{0} \in \Poly{}{k}(\mathcal{T}_0). \label{discr:opBR2}
\end{equation}
Solving \eqref{eq:lapdiscrMatrix} amounts to solving a linear system in the form
\begin{equation}
  \Amat_0^{\mathrm{BR2}} {u}_0 = {f}_0. \label{discr:Lap}
\end{equation}

\section{$h$-multigrid V-cycle}
\label{sec:multiALL}
The ability to define $h$-coarsened mesh sequences by agglomeration and 
to perform high-order accurate dG discretizations on general polygonal
grids allows to exploit \emph{h}-multigrid solvers to improve the efficiency of the solution strategy.
In this work we consider $h$-multigrid preconditioners for the dG discretization of the Stokes problem \eqref{discr:stokes}
and the linearized Navier-Stokes problem \eqref{eq:nsbe}.
Besides incompressible flow problems we will also focus on the performance of $h$-multigrid applied to purely elliptic scalar problems.
The interest is twofold: firstly the discretization of the viscous terms relies on a BR2 dG discretization and secondly block preconditioners for the
Stokes problem require effective preconditioners for the discrete Laplace operator, see Section \ref{sec:precINS}. 

The linear (or linearized) systems arising from dG discretizations of the Navier-Stokes, Stokes and Laplace equation,
see Eqs. \eqref{discr:navierStokesBlock}, \eqref{discr:stokesBlock} and \eqref{discr:Lap}, are in the form
\begin{equation}
  A_0 {w}_0 = {f}_0. \label{discr:general}
\end{equation}
Solving \eqref{discr:general} allows to compute the degrees of freedom $\underline{w}_0$ of $w_0 \in [\Poly{}{k}(\Tl)]^\mathsf{v}$,
where $\mathsf{v} = 4$ in case of incompressible flow problems. As opposite we deal with a scalar function in case of the Laplace equation, \ie $\mathsf{v} = 1$.
In order to accelerate convergence towards $w_0$ the multigrid iteration relies on several coarse grid problems 
in the form 
\begin{equation}
  A_{\ell} w_\ell = f_\ell, \;\; \ell = 1,...,L. \label{discr:generalCoarse}
\end{equation}
Coarse grids $\Tl$ are explicitly built and 
coarse grid solutions belong to piecewise polynomial spaces defined over them, $w_\ell \in [\Poly{}{k}(\Tl)]^{\mathsf{v}}$.
The purpose of this section is to provide some insight on how coarse grid 
operators $A_\ell$ are built and how coarse grid solutions $w_\ell$ can be effectively employed to 
speed up the achievement of the fine grid solution $w_0$. 
A comprehensive review 
can be found in Refs.~\cite{BriggsMultigridBook:2000,SmithBookDPM:2004,TrottenbergMultigridBook:2001},
while the analysis of multigrid as a preconditioner for Krylov solvers, is analyzed in detail
by Smith, Bj{\o}rstad and Gropp~\cite{SmithBookDPM:2004}.

\subsection{Multigrid V-cycle iteration}
\label{sec:multiVcicle}
In this work we consider the multigrid V-cycle iteration, that is the simplest 
way of traversing the mesh sequence generated by agglomeration coarsening of the fine grid, see Section \ref{sec:aggloTopo}.
The recursive multigrid V-cycle for the problem $A_{\ell} w_{\ell} = f_{\ell}$ on level $\ell$ reads:
\begin{algorithm}[H]
\caption{$\mathrm{MG}_{\mathcal{V}}(l,f_{\ell},w_{\ell})$ \label{eq:vcic}}
 \begin{algorithmic}[0]
 \IF {$(\ell = L)$}
 \STATE{$\overline{w_{\ell}} = A_{\ell}^{-1} f_{\ell}$}
 \ENDIF
 \IF {$(\ell < L)$}
   \STATE{\underline{\emph{Pre-smoothing:}}}
   \STATE{$\overline{w}_{\ell} = \mathrm{SMOOTH}(w_{\ell}, f_{\ell})$} 
   \vspace{0.2cm}
   \STATE{\underline{\emph{Coarse grid correction:}}}
   \STATE $r_{\ell} = f_{\ell} - A_{\ell} \overline{w}_{\ell}$
   \STATE $r_{\ell+1} = \mathcal{I}_{\ell}^{\ell+1} r_{\ell}$
   \STATE ${e}_{\ell+1} = \mathrm{MG}_{\mathcal{V}}(\ell+1,r_{\ell+1},0)$ 
   \STATE $\widehat{w}_{\ell} = \overline{w}_{\ell} + \mathcal{I}_{\ell+1}^{\ell} e_{\ell+1}$
   \vspace{0.2cm} 
   \STATE{\underline{\emph{Post-smoothing:}}}
   \STATE{$\overline{w}_{\ell} = \mathrm{SMOOTH}(\widehat{w}_{\ell}, f_{\ell})$} 
 \ENDIF
 \STATE{return $\overline{w}_{\ell}$}
\end{algorithmic}
\end{algorithm}
As a result of invoking $\mathrm{MG}_{\mathcal{V}}(0,f_0,u_0)$ the grid sequence is traversed
moving towards coarser levels, one grid at a time, until the coarsest level $L$ is reached.
Note that the coarsest level can be thought to be located at bottom of the V-shaped cycle. 
The descending phase is followed by ascension towards finer levels, until
a new approximation $\overline{u}_0$ of the exact solution over the fine grid is available.
This marks the completion of one V-cycle iteration.

On each level $\ell$ except the coarsest one, three distinct phases 
take place: pre-smoothing, coarse grid correction and post-smoothing, see Algorithm \ref{eq:vcic}.
In the pre-smoothing phase a few iterations (one or two, in this work) of a standard preconditioned iterative solver
are performed in order to damp high-frequency modes of the error $e_{\ell} = \overline{w}_{\ell} - w_{\ell}$.
Since the convergence of iterative solvers deteriorates when trying to damp low-frequency modes resulting in an inefficient solution process,
the error equations $A_{\ell+1} e_{\ell+1} = r_{\ell+1}$ are solved on a coarser grid (level $\ell+1$), where low-frequency modes appears 
more oscillatory. Once the error is computed it is transferred back to level $\ell$ and used to correct the solution: $\widehat{w}_{\ell} = \widehat{w}_{\ell} + e_{\ell}$.
Before doing so post-smoothing ensures that only smooth components of the error have survived. 
Note that correction takes place only after the
residual equations have been accurately solved on the coarsest level, usually with a direct solver.

It is interesting to remark that, due to the linearity of the original problem, solving $A_{\ell} w_{\ell} = f_{\ell}$ with 
an arbitrary initial guess is equivalent to solving the residual equations $A_{\ell} e_{\ell} = r_{\ell}$ with a zero initial guess.
As a consequence the coarse grid correction requires to compute the residual $r_{\ell+1}$ 
but the computation of an initial guess $e_{\ell+1}$ is not needed.
The residual is approximated by projecting its fine counterpart $r_{\ell} = f_{\ell} - A_{\ell} \overline{w}_{\ell}$ to level $\ell+1$
which requires the definition of the so called restriction operator $\mathcal{I}_{\ell}^{\ell+1}$: $\Poly{}{k}(\Tl) \rightarrow \Poly{}{k}(\Tlpo)$. 
Similarly the error $e_{\ell+1}$ needs to be prolongated to the coarse mesh by means 
of the prolongation operator $\mathcal{I}_{\ell+1}^{\ell}$: $\Poly{}{k}(\Tlpo) \rightarrow \Poly{}{k}(\Tl)$.
The implementation of the intergrid transfer operators $\mathcal{I}_{\ell+1}^{\ell}$ and $\mathcal{I}_{\ell}^{\ell+1}$ is described in detail in the next section.

\subsection{Intergrid transfer operators}
\label{sec:intergrid}
In any geometric multigrid strategy, transfer operators are required to map functions between 
two subsequent spaces in the set $\{ \Poly{}{k}(\Tl) \}_{\ell=0,...,L}$.
Since nested grids are generated by recursive coarsening of a fine grid,
also the polynomial spaces are nested, that is $ \Poly{}{k}(\Tf) \supset \Poly{}{k}(\Tlone) ... \supset \Poly{}{k}(\TLc)$.
Accordingly, \textit{prolongation} is the natural injection $\Itpr: \Poly{}{k}(\Tlpo) \rightarrow \Poly{}{k}(\Tl)$ such that 
\begin{equation}
\sumTl \intB{\T} (\Itpr u_{\ell+1} - u_{\ell+1})\; = 0, \;\; \forall u_{\ell+1} \in \Poly{}{k}(\Tlpo),
\end{equation}
while \textit{restriction} is the $L^2$ projection $\Itre: \Poly{}{k}(\Tl) \rightarrow \Poly{}{k}(\Tlpo)$ such that
\begin{equation}
\sumTl \intB{\T} (\Itre u_{\ell} - u_{\ell})\; v_{\ell+1} = 0, \;\; \forall (u_{\ell},v_{\ell+1}) \in \Poly{}{k}(\Tl) \times \Poly{}{k}(\Tlpo)
\end{equation}

The matrix counterpart $\Itmatrixre \in \mathbb{R}^{m,n}, m = \card(\Tlpo)\, N^{\T}_{\mathrm{dof}}, n = \card(\Tl)\, N^{\T}_{\mathrm{dof}}$,
of the restriction operator $\Itre$ is a sparse block matrix composed of  
$\card(\Tl)$ blocks $\Itb_{\T_{\ell+1},\T_{\ell}}\in \mathbb{R}^{N^{\T}_{\mathrm{dof}},N^{\T}_{\mathrm{dof}}}$, each defined as
\begin{equation}
\label{eq:restrictme}
\Itb_{\T_{\ell+1},\T_{\ell}} \eqbydef \mathbf{M}_{\T_{\ell+1}}\inv\;\M_{\T_{\ell+1},\T_{\ell}},
\end{equation} 
where
\begin{align}
\left( \M_{\T_{\ell}} \right)_{i,j}          &\eqbydef \intB{\T_{\ell}} \varphi_i^{\T_{\ell}} \;  \varphi_j^{\T_{\ell}},  &i,j\in{1,...,N^{\T}_\mathrm{dof}},& \\
\left( \M_{\T_{\ell+1},\T_{\ell}} \right)_{i,j} &\eqbydef \intB{\T_{\ell}} \varphi_i^{\T_{\ell+1}} \; \varphi_j^{\T_{\ell}}, &i,j\in{1,...,N^{\T}_\mathrm{dof}}.&
\end{align}
In particular each row of the matrix $\Itmatrixre$ is associated to an element $\T_{\ell+1} \in \Tlpo$ and consist of $\card(K_{\ell}^{\ell+1})$ blocks.
Similarly the prolongation matrix $\Itmatrixpr \in \mathbb{R}^{n,m}$ consists of  
$\card(\Tlmo)$ blocks $\Itb_{\T_{\ell},\T_{\ell+1}} \in \mathbb{R}^{N^{\T}_{\mathrm{dof}},N^{\T}_{\mathrm{dof}}}$, each defined as
\begin{equation}
\label{eq:prolongme}
\Itb_{\T_{\ell},\T_{\ell+1}} \eqbydef \M_{\T_{\ell}}\inv\;\left(\M_{\T_{\ell+1},\T_{\ell}}\right)\trn.
\end{equation} 
Thanks to the use of orthonormal basis functions, the elemental mass matrices reduce to the identity 
matrix, that is $\M_{\T}=\M_{\T_{\ell+1}}=\mathbf{Id}$, reducing the computational cost of computing transfer operators.
Moreover, since $\Itmatrixre = (\Itmatrixpr)^T$,
storing transfer operators requires to store 
only $\displaystyle \sum_{\ell=0}^{L-1} \card(\Tl)$ blocks of size $(N^{\T}_{\mathrm{dof}})^2$.

Interestingly the same holds true when considering intergrid transfer operators for a vector function $w_{\ell} \in [\Poly{}{k}(\Tl)]^{\mathsf{v}}$.
Restriction and prolongation can be efficiently performed componentwise, that is 
\begin{align}
&\mbox{Restriction:}  &w_{\ell+1,i} = \It_{\ell}^{\ell+1} w_{\ell,i}, && i=1,...,\mathsf{v}, \ell=0,...,\mathrel{L}-1 \nonumber \\
&\mbox{Prolongation:} &w_{\ell,i} = \It_{\ell+1}^{\ell} w_{\ell+1,i}, && i=1,...,\mathsf{v}, \ell=\mathrel{L}-1,...,0 \nonumber
\end{align}
without explicitly building the matrix $\Itmatrixre$ associated to the restriction operator
$\Itre: [\Poly{}{k}(\Tl)]^{\mathsf{v}}\rightarrow [\Poly{}{k}(\Tlpo)]^{\mathsf{v}}$. 
Matrix-free restriction and prolongation algorithms are implemented as follows. 
\begin{algorithm}[H]
\caption{Restriction of a vector function $w_{\ell} \in [\Poly{}{k}(\Tl)]^{\mathsf{v}}$}
 \begin{algorithmic}[0]
    \label{algo:f2c_v}
    \FOR {$\T_{\ell+1} \in \mathcal{T}_{\ell+1}$}
      \FOR {$\T_{\ell} \in K_{\ell}^{\ell+1}$}
        \FOR {$i \in \{1,...,\mathsf{v}\}$}
          \STATE $ \underline{w}_{i,\T_{\ell+1}} \mathrel{+}= \M_{\T_{\ell+1},\T_{\ell}} \underline{w}_{i,\T_{\ell}}$
	\ENDFOR
      \ENDFOR
    \ENDFOR
 \end{algorithmic}
\end{algorithm}
\noindent
\begin{algorithm}[H]
\caption{Prolongation of a vector function $w_{\ell} \in [\Poly{}{k}(\Tl)]^{\mathsf{v}}$}
 \begin{algorithmic}[0]
   \label{algo:c2f_v}
   \FOR {$\T_{\ell+1} \in \Tlpo$}
      \FOR {$\T_{\ell} \in K_{\ell}^{\ell+1}$}
        \FOR {$i \in \{1,...,\mathsf{v}\}$}
          \STATE $ \underline{w}_{i,\T_{\ell}} = \M_{\T_{\ell+1},\T_{\ell}}^T \underline{w}_{i,\T_{\ell+1}}$
	\ENDFOR
      \ENDFOR	
   \ENDFOR
 \end{algorithmic}
\end{algorithm}
Note that $\underline{w}_{i,\T_{\ell+1}}$ and $\underline{w}_{i,\T_{\ell}}$ are the 
degrees of freedom associated with the i-th component of the function $w_{\ell+1}|_{\T_{\ell+1}} \in [\Poly{}{k}(\T_{\ell+1})]^{\mathsf{v}}$ and 
of the function $w_{\ell} |_{\T_{\ell}} \in [\Poly{}{k}(\T_{\ell})]^{\mathsf{v}}$, respectively.

\subsection{Coarse grid operators}
\label{sec:coarseGridOp}
Two possibilities are available for building coarse grid problems in the form of \eqref{discr:generalCoarse}, 
the so called \emph{non-inherited} multigrid, where 
discrete operators are assembled on each grid of the mesh sequence, and \emph{inherited} multigrid,
where coarse operators are recursively built by restricting the fine grid operators.

Recently, evidence emerged that non-inherited multigrid might be preferable from the convergence rates viewpoint, in particular
Antonietti \ea \cite{AntoniettiSartiVerani} have analyzed $h$-multigrid Interion Penalty dG discretization of the Laplace equation 
demonstrating that only non-inherited multigrid provides uniform convergence with respect to the number of levels.
Nevertheless, inherited coarse grid operators are significantly cheaper to compute since 
evaluation of numerical fluxes and assembly of bilinear forms over agglomerated elements grids is avoided.
In particular, from the implementation viewpoint
\begin{inparaenum}[i)]
\item numerical integration and basis function orthogonalization over agglomerated elements meshes 
      are required only for the computation of intergrid transfer operators;
\item the parallel implementation is simpler since flux computation on partition boundaries 
      requires to access data from ghost agglomerated elements
      (note that ghost agglomerated elements are composed by many layers of fine ghost cells). 
\end{inparaenum}
Accordingly choosing between inherited and non-inherited version of $h$-multigrid might involve a trade-off  
between efficiency of the solver strategy and computational cost of 
assembling coarse grid operators.
In order to avoid such an uncomfortable situation, in what follows we propose to heal the convergence degradation 
of inherited multigrid using a rescaled Galerkin projection of the stabilization 
terms of the BR2 dG discretization. 
\alertblu{The possibility to suitably rescale the stabilization terms of dG discretizations to improve the 
performance of coarse grid solvers was first proposed by Antonietti \ea \cite{AntoniettiDiosBrenner} 
in the context of two level Schwarz methods for overpenalized Interior Penalty formulations.}

\subsubsection{BR2 dG discretization}
\label{sec:coarseBR2}
Consider the BR2 bilinear form $a^{\mathrm{BR2}}_{0}(v_{0},w_{0}):\Poly{}{k}(\Tf) \times \Poly{}{k}(\Tf) \rightarrow \mathbb{R}$ 
defined in \eqref{eq:ahBR2} and the corresponding fine grid operator $\Sys{0}^{\mathrm{BR2}} : \Poly{}{k}(\Tf) \rightarrow \Poly{}{k}(\Tf)$, see definition \eqref{discr:opBR2}.
The coarse grid operators $\Sys{\ell}^{\mathrm{BR2}},\Sys{\ell}^{\It,\mathrm{BR2}}: \Poly{}{k}(\Tl) \rightarrow \Poly{}{k}(\Tl), \; \ell=1,...,L$, read
\begin{align}
\mbox{Non-Inherited:} && (\Sys{\ell}^{\mathrm{BR2}} v_{\ell},w_{\ell})_{L^2(\Omega)} 
                       &\eqbydef a_{\ell}^{\mathrm{BR2}}(v_{\ell},w_{\ell}), &\forall v_{\ell},w_{\ell} \in \Poly{}{k}(\Tl), \label{def:ninhM}\\
\mbox{Inherited:} && (\Sys{\ell}^{\It,\mathrm{BR2}} v_{\ell}, w_{\ell})_{L^2(\Omega)} 
                       &\eqbydef a_{0}^{\mathrm{BR2}}(\It_{\ell}^0 v_{\ell}, \It_{\ell}^0 w_{\ell}), &\forall v_{\ell},w_{\ell} \in \Poly{}{k}(\Tl), \label{def:inhM}
\end{align}
where $\It_{\ell}^0 = \It_1^0 \; \It_2^1 \; ... \; \It_{\ell}^{\ell-1}$
and $\Itpr: \Poly{}{k}(\Tlpo) \rightarrow \Poly{}{k}(\Tl), \, \ell=0,...,L\text{-}1$ are the prolongation operators introduced in Section \ref{sec:intergrid}.
Continuity and coercivity bounds for the 
$a^{\mathrm{BR2}}_{\ell}(u_{\ell},v_{\ell})$ bilinear form over agglomerated elements meshes 
were proven by Bassi \ea \cite{BassiFlexy12}, in particular on level $\ell$ stability holds provided that $\eta_{\F} > N_{{\partial}_\ell}$.
Accordingly the coarse grid problems $A_{\ell}^{\mathrm{BR2}} u_\ell = f_\ell$, $\ell=1,...,L$, arising in the non-inherited 
version of the multigrid V-cycle iteration, see Section \ref{sec:multiVcicle}, are well-posed.

In what follows we demonstrate that, given the BR2 bilinear form in \eqref{eq:ahBR2}, for all $v_{\ell},w_{\ell} \in \Poly{}{k}(\Tl)$
\begin{align}
 a^{\mathrm{BR2,CS}}_{0}(\It_{\ell}^0 v_{\ell}, \It_{\ell}^0 w_{\ell}) 
                                                       &= a^{\mathrm{BR2,CS}}_{\ell}(v_{\ell}, w_{\ell}), \label{eq:ahcs_inh} \\
 s^{\mathrm{BR2}}_{0}(\It_{\ell}^0 v_{\ell},\It_{\ell}^0 w_{\ell})                            
                                                       &\neq s^{\mathrm{BR2}}_{\ell}(v_{\ell},w_{\ell}). \label{eq:ahs_inh}
\end{align}
Accordingly the difference between $\Sys{\ell}^{\mathrm{BR2}}$ and $\Sys{\ell}^{\It,\mathrm{BR2}}$ hinges on the stabilization term.

Using the local and global lifting operator definitions \eqref{eq:rF} and \eqref{eq:RF}
the consistency and symmetry BR2 bilinear form in \eqref{eq:ahcs} can be rewritten as 
$$a^{\mathrm{BR2,CS}}_{0}(v_{0}, w_{0}) = a^{\mathrm{BR2,CS_\T}}_{0}(v_{0}, w_{0}) + a^{\mathrm{BR2,CS_\sigma}}_{0}(v_{0}, w_{0})$$
where
\begin{align}
  a^{\mathrm{BR2,CS_\T}}_{0}(v_{0}, w_{0}) = 
  &\sum_{\T \in \Tf}\int_\T \GRADhf v_{0} \GRADhf w_{0}, \label{eq:BR2_sipT} \\
  a^{\mathrm{BR2,CS_\sigma}}_{0}(v_{0}, w_{0}) = 
  & - \sum_{\F \in \Ff} \int_\F \left( \averg{\GRADhf v_{0}} \cdot \mathbf{n}_\F \; \jump{w_{0}} 
                                    + \jump{v_{0}}\; \averg{\GRADhf w_{0}} \cdot \mathbf{n}_\F \right) \label{eq:BR2_sipF}. 
\end{align}
Since $\Poly{}{k}(\Tf) \supset \Poly{}{k}(\Tl)$, we get 
\begin{align}
  a^{\mathrm{BR2,CS_\T}}_{0}(\It_{\ell}^0 v_{\ell}, \It_{\ell}^0 w_{\ell}) 
  &= \dissum{\T \in \Tf} \disint{\T} \GRADhf \Itvhlp \cdot \GRADhf \Itwhlp \nonumber \\
  &= \dissum{\T \in \Tl} \dissum{\T \in K_0^{\ell}} \disint{\T} \GRADhf \Itvhlp \cdot \GRADhf \Itwhlp \nonumber \\
  &= \dissum{\T \in \Tl} \disint{\T} \GRADhl v_{\ell} \cdot \GRADhl w_{\ell} \label{eq:br2cs_inh0}
  = a^{\mathrm{BR2,CS_\T}}_{\ell}(v_{\ell}, w_{\ell}). 
\end{align}
Since $\jump{\Itvhl}_{\F_0} = 0$ if $\sigma_0 \notin \Fl \cap \Ff$, we get 
$-a^{\mathrm{BR2,CS_\sigma}}_{0}(\It_{\ell}^0 v_{\ell}, \It_{\ell}^0 w_{\ell})= $  
\begin{align}
   = &\sum_{\F \in \Ff} \int_\F \averg{\GRADhf \Itvhlp} \cdot \mathbf{n}_\F \; \jump{\Itwhl} 
    &+&\sum_{\F \in \Ff} \int_\F \jump{\Itvhl}\; \averg{\GRADhf \Itwhlp} \cdot \mathbf{n}_\F  \nonumber \\
    = &\sum_{\F \in \Fl} \dissum{\F \in \Sigma_0^{\ell}} \int_\F \averg{\GRADhf \Itvhlp} \cdot \mathbf{n}_\F \jump{\Itwhl} 
     &+&\sum_{\F \in \Fl} \dissum{\F \in \Sigma_0^{\ell}} \int_\F \jump{\Itvhl} \averg{\GRADhf \Itwhlp} \cdot \mathbf{n}_\F   \nonumber \\
   = & \sum_{\F \in \Fl} \int_\F \averg{\GRADhl v_{\ell}} \cdot \mathbf{n}_\F \; \jump{w_{\ell}}
    &+& \sum_{\F \in \Fl} \int_\F \jump{v_{\ell}} \; \averg{\GRADhl w_{\ell}} \cdot \mathbf{n}_\F  \nonumber \\
   = & \int_\Omega \GRADh v_\ell \cdot \mathbf{R}(w_\ell) 
    &+& \int_\Omega \mathbf{R}(v_\ell) \cdot \GRADh w_\ell = -a^{\mathrm{BR2,CS_\sigma}}_{\ell}(v_{\ell}, w_{\ell})\label{eq:br2cs_inh1} 
\end{align}
The above result together with \eqref{eq:br2cs_inh0} prove \eqref{eq:ahcs_inh}.

Using the local lifting operator definitions \eqref{eq:rF} 
the stabilization term in \eqref{eq:stabBR2} can be rewritten as 
\begin{equation}
  s^{\mathrm{BR2}}_{0}(v_{0},w_{0}) = \sum_{\F \in \Ff} \eta_{\F}\int_{\F} \averg{\mathbf{r}_{\F}^k(\jump{v_{0}})}\SCAL \mathbf{n}_\F \; \jump{w_{0}}. \label{eq:BR2_sip_s}
\end{equation}
The inherited stabilization term reads 
\begin{align}
                   s^{\mathrm{BR2}}_{0}(\It_{\ell}^0 v_{\ell},\It_{\ell}^0 w_{\ell})   
                      & =\sum_{\F \in \Ff} \eta_{\F}\int_{\F} \averg{\mathbf{r}_{\F}^k(\jump{\It_{\ell}^0 v_{\ell}})}\SCAL \mathbf{n}_\F \jump{\It_{\ell}^0 w_{\ell}} \nonumber \\
                      & = \sum_{\F \in \Fl} \dissum{\F \in \Sigma_0^{\ell}} \eta_{\F}\int_{\F} 
		          \averg{\mathbf{r}_{\F}^k(\jump{\It_{\ell}^0 v_{\ell}})}\SCAL \mathbf{n}_\F \jump{\It_{\ell}^0 w_{\ell}} \nonumber\\
                      & = \sum_{\F \in \Fl} \dissum{\F \in \Sigma_0^{\ell}} \eta_{\F}\int_{\F} \averg{\mathbf{r}_{\F}^k(\jump{v_{\ell}})}\SCAL \mathbf{n}_\F \jump{w_{\ell}} \nonumber\\
                      & = \sum_{\F \in \Fl} \dissum{\F \in \Sigma_0^{\ell}} \eta_{\F}\int_{\Omega} \mathbf{r}_{\F}^k(\jump{v_{\ell}})\SCAL \mathbf{r}_{\F}^k(\jump{w_{\ell}}) \label{eq:br2_stabinh}
\end{align}
while its non-inherited counterpart is simply
\begin{equation}
s_{\ell}^{\mathrm{BR2}}(v_{\ell},w_{\ell}) = \sum_{\F \in \Fl} \eta_{\F} \int_{\Omega}\mathbf{r}_{\F_\ell}^k(\jump{v_\ell})\SCAL\mathbf{r}_{\F_\ell}^k(\jump{w_\ell}) \label{eq:stabBR2l}.
\end{equation}
The inherited stabilization term \eqref{eq:br2_stabinh} 
introduces an excessive amount of stabilization as compared to \eqref{eq:stabBR2l} 
having a detrimental effect on the spectral properties of inherited coarse grid operators, see Antonietti \ea \cite{AntoniettiSartiVerani}.

In order to recover the correct amount of stabilization we propose to rescale it introducing the scaling term 
\begin{equation}
\label{eq:rescFac}
\displaystyle \mathcal{H}_{\F_0}^{\F_{\ell}} \eqbydef \frac{\eta_{\F_{\ell}}}{\eta_{\F_0}} \frac{h_{\T_0,\Tpr_0}}{h_{\T_{\ell},\Tpr_{\ell}}},
\end{equation}
and defining the rescaled stabilization term
\begin{equation}
\widetilde{s}^{\mathrm{BR2}}_{0}(\Itvhl,\Itwhl) \eqbydef \dissum{\F \in \Fl} \dissum{\F \in \Sigma_0^{\ell}} 
               \mathcal{H}_{\F_0}^{\F_{\ell}} \; \eta_{\F} \int_{\Omega} \mathbf{r}_{\F_0}^k(\jump{\Itvhl}) \cdot \mathbf{r}_{\F_0}^k(\jump{\Itwhl}), \label{eq:rescStabs}
\end{equation}
such that 
\begin{align}
\widetilde{s}_{\ell}^{\mathrm{BR2}}(\Itvhl,\Itwhl) &\lesssim s_{\ell}^{\mathrm{BR2}}(v_{\ell},w_{\ell}), & \forall v_{\ell},w_{\ell} \in \Poly{}{k}(\Tl), \label{eq:rescStabCont}\\
\widetilde{s}_{\ell}^{\mathrm{BR2}}(\Itvhl,\Itvhl) &\gtrsim s_{\ell}^{\mathrm{BR2}}(v_{\ell},v_{\ell}), & \forall v_{\ell} \in \Poly{}{k}(\Tl).\label{eq:rescStabCoerc}
\end{align}
To prove \eqref{eq:rescStabCont} we recall the following bounds on the local lifting operator: let $\phi \in L^2(\F)$, for all $\F \in \Fl$ 
\begin{equation}
\label{eq:upliftbound}
C_{\mathrm{r}} h_{\T,\Tpr}^{-1/2} \| \jump{\phi} \|_{L^2(\F)} \leq \| \mathbf{r}_{\F}^k (\phi) \|_{[L^2 (\Omega)]^d} \leq C_\mathrm{tr} h_{\T,\Tpr}^{-1/2} \| \phi \|_{L^2(F)}, 
\end{equation}
where $h_{\T,\Tpr} = \min \left( h_\T, h_{\Tpr} \right)$,
see \eg \cite[Lemma 2]{Brezzi.Manzini.ea:2000}, \cite[Lemma 7.2]{Toselli03} or \cite[Lemma 4.33 and Lemma 5.18]{DiPiErn11} for a proof.
\alertblu{The constant $C_\mathrm{tr}$ depends on $d$, $k$ and the shape regularity of the elements sharing $\F$ and 
is inherited from the discrete trace inequality}: for all $\T \in \Tl$, $\F \in \FT$
\begin{equation}
\label{eq:traceIneq}
\| v_{\ell} \|_{L^2(\F)} \leq C_{\mathrm{tr}} h_{\T,\Tpr}^{-1/2} \| v_{\ell} \|_{L^2(\T)}
\end{equation}
While trace inequalities in the form of \eqref{eq:traceIneq} are commonly available in the context 
of simplicial and quadrilateral/hexahedral meshes we refer to \cite[Lemma 1.46]{DiPiErn11} for a version valid in the context 
of matching simplicial submeshes and to \cite{Cangiani14,Giani14}
for an optimal version derived in the context of polygonal/polyhedral element meshes.

Using \eqref{eq:upliftbound} we get the following bounds
\begin{align}
s_{\ell}^{\mathrm{BR2}}(v_{\ell},w_{\ell}) &= \sum_{\F \in \Fl} \eta_{\F_\ell} \int_{\Omega}\mathbf{r}_{\F_\ell}^k(\jump{v_\ell})\SCAL\mathbf{r}_{\F_\ell}^k(\jump{w_\ell}) \nonumber \\
		  &\leq  \sum_{\F \in \Fl} \eta_{\F_\ell} \| \mathbf{r}_{\F_\ell}^k (\jump{v_{\ell}}) \|_{ [ L^2(\Omega) ]^d} \; 
                                                                                  \| \mathbf{r}_{\F_\ell}^k (\jump{w_{\ell}}) \|_{ [ L^2(\Omega) ]^d} \nonumber \\
		  &\lesssim  \sum_{\F \in \Fl} \eta_{\F_\ell}  h_{\T_\ell,\Tpr_\ell}^{-1} \| \jump{v_{\ell}} \|_{ L^2(\sigma) } \; 
                                                                                                     \| \jump{w_{\ell}} \|_{ L^2(\sigma) } \nonumber 
\end{align}
\begin{align}
\widetilde{s}_{0}^{\mathrm{BR2}}(\Itvhl,\Itwhl) &=\sum_{\F \in \Fl} \dissum{\F \in \Sigma_0^{\ell}}  \mathcal{H}_{\F_0}^{\F_{\ell}} \;
                       \eta_{\F_0} \int_{\Omega}\mathbf{r}_{\F_0}^k(\jump{\Itvhlp})\SCAL\mathbf{r}_{\F_0}^k(\jump{\Itwhlp}) \nonumber \\
		  &\leq  \sum_{\F \in \Fl} \dissum{\F \in \Sigma_0^{\ell}} \frac{\eta_{\F_{\ell}}}{\eta_{\F_0}} \frac{h_{\T_0,\Tpr_0}}{h_{\T_{\ell},\Tpr_{\ell}}}
		            \eta_{\F_0} \| \mathbf{r}_{\F_0}^k (\jump{v_{\ell}}) \|_{ [ L^2(\Omega) ]^d} \; 
                                                                                  \| \mathbf{r}_{\F_0}^k (\jump{w_{\ell}}) \|_{ [ L^2(\Omega) ]^d} \nonumber \\
		  &\lesssim  \sum_{\F \in \Fl} \dissum{\F \in \Sigma_0^{\ell}}\eta_{\F_\ell}  \frac{h_{\T_0,\Tpr_0}}{h_{\T_{\ell},\Tpr_{\ell}}}
		                                                                    h_{\T_0,\Tpr_0}^{-1} \| \jump{v_{\ell}} \|_{ L^2(\sigma) } \; 
                                                                                                     \| \jump{w_{\ell}} \|_{ L^2(\sigma) } \nonumber \\
                  &\leq  \sum_{\F \in \Fl} \eta_{\F_\ell}  h_{\T_\ell,\Tpr_\ell}^{-1}
		                       \left(  \dissum{\F \in \Sigma_0^{\ell}}\| \jump{v_{\ell}} \|_{ L^2(\sigma) }^2 \right)^{\frac{1}{2}}\; 
                                       \left(  \dissum{\F \in \Sigma_0^{\ell}}\| \jump{w_{\ell}} \|_{ L^2(\sigma) }^2 \right)^{\frac{1}{2}} \nonumber \\
		  &= \sum_{\F \in \Fl}  \eta_{\F_\ell} h_{\T_\ell,\Tpr_{\ell}}^{-1}
		                                                           \|  \jump{v_{\ell}} \|_{ L^2(\sigma) } \; 
                                                                           \|  \jump{w_{\ell}} \|_{ L^2(\sigma) }  \nonumber 
\end{align}
which prove \eqref{eq:rescStabCont}.

In view of \eqref{eq:rescStabCoerc}, using \eqref{eq:upliftbound}, we now infer
\begin{align}    
\widetilde{s}_{0}^{\mathrm{BR2}}(\Itvhl,\Itvhl) |_{\F_{\ell}} &= \dissum{\F \in \Sigma_0^{\ell}} {\eta_{\F_{\ell}}} 
			     \frac{h_{\T_0,\Tpr_0}}{h_{\T_{\ell},\Tpr_{\ell}}} \| \mathbf{r}_{\F_0}^k(\jump{v_{\ell}}) \|^2_{[L^2(\Omega)]^d} \nonumber \\
			   &\geq \dissum{\F \in \Sigma_0^{\ell}} {\eta_{\F_{\ell}}} \frac{h_{\T_0,\Tpr_0}}{h_{\T_{\ell},\Tpr_{\ell}}} 
			     \frac{C_{\mathrm{r}_0} }{h_{\T_0,\Tpr_0}} \| \jump{v_{\ell}} \|^2_{L^2(\F_0)} \nonumber \\
			& = {\eta_{\F_{\ell}}} \frac{C_{\mathrm{r}_0}}{h_{\T_{\ell},\Tpr_{\ell}}} \| \jump{v_{\ell}} \|^2_{L^2(\F_{\ell})} \nonumber \\ 
			& \geq {\eta_{\F_{\ell}}} \frac{C_{\mathrm{r}_0}}{C_{\mathrm{tr}_\ell}} \| \mathbf{r}_{\F_{\ell}}^k(\jump{v_{\ell}}) \|^2_{[L^2(\Omega)]^d}.  \nonumber \\
			& = \frac{C_{\mathrm{r}_0}}{C_{\mathrm{tr}_\ell}} {s}_{\ell}^{\mathrm{BR2}}(v_{\ell},v_{\ell}) |_{\F_{\ell}} \label{eq:interlevelTrace} 
\end{align}
and summing over mesh faces on level $\ell$ we get the desired result.
\alertblu{As remarked by Antonietti \ea \cite{AntoniettiHoustonSartiVerani}, 
$C_{\mathrm{tr}_\ell}$ is influenced by the aspect ratio of the agglomerated element as well as by 
the ratio between the agglomerated element and the agglomerated face measure.
Interestingly enough MGridGen algorithms are designed to optimize the aspect ratio of agglomerates 
and minimize the number of graph neighbors, which should also limit the occurrence of 
small degenerate faces (note that according to the definitions given in Section \ref{sec:aggloTopo} 
the number of faces is equivalent to the number of element neighbors).}

Consider now the inherited coarse grid operators 
\begin{equation}
\Sys{\ell}^{\widetilde{\It},\mathrm{BR2}} \eqbydef \Sys{\ell}^{\It,\mathrm{BR2,CS}} + \Sys{\ell}^{\widetilde{\It},\mathrm{BR2,STB}} \label{def:modinhM}
\end{equation}
such that $\forall v_{\ell},w_{\ell} \in \Poly{}{k}(\Tl)$, $\ell=1,...,L$
\begin{equation}
\label{def:modinhM_AS}
\mbox{Mod-Inherited:} \; \; 
\begin{array}{ll}
  (\Sys{\ell}^{\It,\mathrm{BR2,CS}} v_{\ell}, w_{\ell})_{L^2(\Omega)} 
                       &\eqbydef a_0^{\mathrm{BR2,CS}} (\It_{\ell}^0 v_{\ell}, \It_{\ell}^0 w_{\ell}),  \\
  (\Sys{\ell}^{\widetilde{\It},\mathrm{BR2,STB}} v_{\ell}, w_{\ell})_{L^2(\Omega)} 
                       &\eqbydef \widetilde{s}_0^{\mathrm{BR2}} (\It_{\ell}^0 v_{\ell}, \It_{\ell}^0 w_{\ell}). 
\end{array}
\end{equation}

Since $a_{0}^{\mathrm{BR2,CS}}(\It_{\ell}^0 v_{\ell}, \It_{\ell}^0 w_{\ell}) + \widetilde{s}_{0}^{\mathrm{BR2}}(\Itvhl,\Itvhl) =$
\begin{align}
  &= \int_\Omega \left| \GRADh v_{\ell} - \mathbf{R}_{\ell}^k(v_{\ell})\right|^2 
                         + \sum_{\F \in \Fl} \dissum{\F \in \Sigma_0^{\ell}} 
			\mathcal{H}_{\F_0}^{\F_{\ell}} \eta_{\F_{0}}  
			  \int_{\Omega} \left| \mathbf{r}_{\F_0}^k(\jump{v_{\ell}}) \right|^2 
			 -\int_{\Omega} \left| \mathbf{R}_{\ell}^k(v_{\ell})        \right|^2 \nonumber \\
                         &\geq \| \GRADh v_{\ell} - \mathbf{R}_{\ell}^k(v_{\ell})\|^2_{[L^2(\Omega)]^d} \nonumber \\ 
			 &+  \sum_{\F \in \Fl} \left( {\eta_{\F_{\ell}}} 
			 \frac{C_{\mathrm{r}_0}}{C_{\mathrm{tr}_\ell}}    \| \mathbf{r}_{\F_\ell}^k(\jump{v_{\ell}}) \|^2_{[L^2(\Omega)]^d}
                         -  \max_{\T \in \TF}(\card(\FT)) \| \mathbf{r}_{\F_{\ell}}^k(\jump{v_{\ell}}) \|^2_{[L^2(\Omega)]^d} \right), \nonumber 
\end{align}
stability holds provided that $\eta_{\F_{\ell}} \frac{C_{\mathrm{r}_0}}{C_{\mathrm{tr}_l}} > \max_{\T \in \TFl}(\card(\FT))$.
In practice, motivated by the observation that the stabilization parameter choice 
suggested by theory is abundant, see \eg \cite{BassiFree14}, 
we deliberately neglected the dependence on $C_{\mathrm{tr}}$ and $C_{\mathrm{r}}$
in definition \eqref{eq:rescFac}. 
Note that a strategy for estimating $C_{\mathrm{tr}}$ over agglomerated element meshes has been proposed by
\cite{Cangiani14}.

As we already pointed out the main advantage of inherited multigrid 
is the possibility to build coarse grid operator by means of intergrid transfer operators, 
avoiding numerical integration over agglomerated elements.
The matrix restriction algorithm is described in \ref{sec:append}
and exploit the possibility to recursively inherit operators according to the following identities
\begin{align}
 (\Sys{\ell+1}^{{\It},\mathrm{BR2,CS}} v_{\ell+1}, w_{\ell+1})_{L^2(\Omega)} 
       & = (\It_{\ell}^{\ell+1} \Sys{\ell}^{\mathrm{BR2,CS}} \It_{\ell+1}^{\ell} v_{\ell+1}, w_{\ell+1})_{L^2(\Omega)} \label{eq:galproj}
\end{align}
\begin{align}
 \dissum{\F \in \Flpo} &(\Sys{\ell+1}^{\widetilde{\It},\mathrm{BR2,STB}} v_{\ell+1}, w_{\ell+1})_{L^2(\T_{\ell+1} \cup \Tpr_{\ell+1})} \nonumber \\
       & = \dissum{\F \in \Flpo} \; \dissum{\F \in \Sigma_{\ell}^{\ell+1}}( \mathcal{H}_{\F_\ell}^{\F_{\ell+1}}
                  \It_{\ell}^{\ell+1} \Sys{\ell}^{\mathrm{BR2,STB}} \It_{\ell+1}^{\ell} v_{\ell+1}, w_{\ell+1})_{L^2(\T_{\ell} \cup \Tpr_{\ell})} \label{eq:resgalproj}
\end{align}
where $\Itre$ and $\Itpr$ are the restriction and prolongation operators described in Section \ref{sec:intergrid}.

\subsubsection{Stokes dG discretization}
\label{sec:coarseStk}
Consider the Stokes operator $A^\mathrm{Stk}_{0}$ defined in \eqref{discr:stokesBlock}, the inherited coarse grid operators 
employed in this work read
\begin{equation}
\Amat_{\ell}^{\widetilde{\It},\mathrm{Stk}} =
\left[
\begin{array}{rc}
\Amat_{\ell}^{\widetilde{\It}},  & \Bmat_{\ell}^{\It,t} \\ 
\Bmat_{\ell}^{\It} & \Cmat_{\ell}^{\It}      
\end{array} \right].
\end{equation}

Consider the bilinear form $b_{0}(\vvec_{0},q_{0}):[\Poly{}{k}(\Tf)]^d \times \Poly{}{k}(\Tf) \rightarrow \mathbb{R}$ defined in \eqref{eq:Bdiv},
and the corresponding fine grid operator $B_{0} : [\Poly{}{k}(\Tf)]^d \rightarrow \Poly{}{k}(\Tf)$, see definition \eqref{discr:opB}.
The coarse grid operators $B_{\ell},B_{\ell}^{\It}: [\Poly{}{k}(\Tl)]^d \rightarrow \Poly{}{k}(\Tl), \; \ell=1,...,L$, read
\begin{align}
\mbox{Non-Inherited:} \; (B_{\ell} \vvec_{\ell}, q_{\ell})_{L^2(\Omega)} &\eqbydef b_{\ell}( \vvec_{\ell}, q_{\ell}), 
  & \forall (\vvec_{\ell}, q_{\ell}) \in [\Poly{}{k}(\Tl)]^d \times \Poly{}{k}(\Tl), \nonumber \\
\mbox{Inherited:} \; (B_{\ell}^{\It} \vvec_{\ell}, q_{\ell})_{L^2(\Omega)} &\eqbydef b_{0}(\It_{\ell}^0 \vvec_{\ell}, \It_{\ell}^0 q_{\ell}), 
  & \forall (\vvec_{\ell}, q_{\ell}) \in [\Poly{}{k}(\Tl)]^d \times \Poly{}{k}(\Tl),\nonumber
\end{align}
where the restriction of a vector function is performed componentwise  
$$\It_{\ell}^0 \vvec_{\ell} = \displaystyle \sum_{i=0}^d \It_{\ell}^0 v_{\ell,i}.$$
Proceeding as in Section \ref{sec:coarseBR2}, see in particular \eqref{eq:br2cs_inh1}, it is straightforward to show that $B_{\ell} = B_{\ell}^{\It}$.

According to definition \eqref{eq:BR2Stokes} the operator $A_{\ell}^{\widetilde{\It}}: [\Poly{}{k}(\Tl)]^d \rightarrow [\Poly{}{k}(\Tl)]^d$, read
\begin{align}
\mbox{Inherited:} && (\Sys{\ell}^{\widetilde{\It}} \vvec_{\ell}, \wvec_{\ell})_{L^2(\Omega)} &\eqbydef 
\sum_{i=1}^d (\Sys{\ell}^{\widetilde{\It},\mathrm{BR2}} v_{\ell,i}, w_{\ell,i})_{L^2(\Omega)}, &\forall \vvec_{\ell},\wvec_{\ell} \in [\Poly{}{k}(\Tl)]^d, \label{def:ninhMStk}
\end{align}
see \eqref{def:modinhM} for the definition of $\Sys{\ell}^{\widetilde{\It},\mathrm{BR2}}$.

To conclude, the coarse operators $C_{\ell}^{\It}: \Poly{}{k}(\Tl) \rightarrow \Poly{}{k}(\Tl)$, read
\begin{align}
\mbox{Inherited:} \; (C_{\ell}^{\It} q_{\ell}, r_{\ell})_{L^2(\Omega)} &\eqbydef c_{0}(\It_{\ell}^0 q_{\ell}, \It_{\ell}^0 r_{\ell}), 
  & \forall (q_{\ell}, r_{\ell}) \in \Poly{}{k}(\Tl) \times \Poly{}{k}(\Tl),\nonumber
\end{align}
see \eqref{eq:stabStokes} for the definition of the bilinear form $c_0(q_0, r_0):\Poly{}{k}(\Tf) \times \Poly{}{k}(\Tf) \rightarrow \mathbb{R}$.
Even if the inherited bilinear form introduces a different (read smaller) amount of stabilization 
as compared to its non-inherited counterpart the numerical test case corroborate the choice not 
to modify the scaling of $C_{\ell}^{\It}$.

Coarse grid operators are built by means of intergrid transfer operators,
exploiting the possibility to recursively inherit operators.
For example the operators $B_{\ell}^{\It}$ are such that, for all $(\vvec_{\ell}, q_{\ell}) \in [\Poly{}{k}(\Tl)]^d \times \Poly{}{k}(\Tl)$
\begin{align}
 (B_{\ell+1}^{{\It}} \vvec_{\ell+1}, q_{\ell+1})_{L^2(\Omega)} 
       & = (\It_{\ell}^{\ell+1} B_{\ell} \It_{\ell+1}^{\ell} \vvec_{\ell+1}, q_{\ell+1})_{L^2(\Omega)} \nonumber
\end{align}
where $\Itre: [\Poly{}{k}(\Tl)]^{d} \rightarrow [\Poly{}{k}(\Tlpo)]^{d}$ and $\Itpr: [\Poly{}{k}(\Tlpo)]^{d} \rightarrow [\Poly{}{k}(\Tl)]^{d}$ and $\ell = 0,...,L-1$.
Similarly to vector restriction and prolongation, matrix restriction can be performed matrix-free 
without requiring to assemble the matrices $\Itmatrixre$ and $\Itmatrixpr$. 
This practice yields large memory savings when the operators $\Itre,\Itpr$ act on vector functions.

\subsubsection{Navier-Stokes dG discretization}
Consider the Navier-Stokes operator $A^\mathrm{INS}_{0}$ defined in \eqref{discr:navierStokesBlock}, the inherited coarse grid operators 
employed in this work read
\begin{equation}
\Amat_{\ell}^{\widetilde{\It},\mathrm{INS}} =
\left[
\begin{array}{cr}
\Mmat_{\ell}^{\It} + \Jmat^{A,\widetilde{\It}}_{\ell} + \Jmat^{D,{\It}}_{\ell}  & \Jmat^{B^t,\It}_\ell \\ 
\Jmat_{\ell}^{B,\It}  & \Jmat_{\ell}^{C,\It}      
\end{array} \right].
\end{equation}
To inherit the viscous operators we follow the same path of the Stokes case. 
Accordingly we get $\Jmat^{A,\widetilde{\It}}_{\ell}= A^{\widetilde{\It}}_{\ell}$, $\ell=1,...L$, see Definition \eqref{def:ninhMStk}
and note that, according to Definition \eqref{eq:BR2Stokes}, $\Jmat^{A}_{0} = A_0$.

Regarding inviscid operators we consider the trilinear form 
$$j^{!\nu}_{0}(\wvec_0, \uvec_0, \vvec_0) = \displaystyle \sum_{i=1}^d \sum_{j=1}^d j^{!\nu}_{i,j}({\wvec}_0, \mathbf{\delta} w_{0,j}, k_{0,i}),$$ see Definition \eqref{eq:jacTrilF}, 
such that
$$(J^{D}_{0}(\wvec_0) \delta \uvec_0, \mathbf{v}_0)_{[{L^2}(\Omega)]^{d}} =j^{!\nu}_{0}(\wvec_0, \uvec_0, \vvec_0), 
  \;\;\; \forall \mathbf{\delta} \uvec_0,\mathbf{v}_0 \in [\Poly{}{k}(\mathcal{T}_0)]^{d}. $$
We remark that operators $J_0^B,J_0^{B^t}$ and $J_0^C$ can be restricted in a similar fashion.
The coarse grid operators $J^D_{\ell},J_{\ell}^{D,\It}: [\Poly{}{k}(\Tl)]^{d} \rightarrow [\Poly{}{k}(\Tl)]^{d}, \; \ell=1,...,L$, read
\begin{align}
\mbox{Non-Inherited:} && (J_{\ell}^D(\It^{\ell}_0 \wvec_0) \uvec_{\ell}, \vvec_{\ell})_{[L^2(\Omega)]^d} &\eqbydef 
                                                                j_{\ell}(\It^{\ell}_0 \wvec_{0}, \uvec_{\ell}, \vvec_{\ell}), 
   \label{def:inhMt}\\
\mbox{Inherited:} && (J_{\ell}^{D,\It}(\wvec_0) \uvec_{\ell}, \vvec_{\ell})_{[L^2(\Omega)]^d} &\eqbydef 
                                                                j_{0}(\wvec_0, \It_{\ell}^0 \uvec_{\ell}, \It_{\ell}^0 \vvec_{\ell}), 
   \label{def:ninhMt}
\end{align}
$\forall \, \uvec_{\ell},\vvec_{\ell} \in [\Poly{}{k}(\Tl)]^{d}, \wvec_0 \in [\Poly{}{k}(\Tf)]^{d+1}$.
The non-inherited version of coarse grid operators is not employed in this work but is included 
for the sake of comparison. 

In practice, given the fine grid operator $J^D_0(\wvec_0)$, 
the inherited coarse grid operators $J_{\ell}^{D,\It}(\wvec_0)$ 
are defined recursively by means of the Galerkin projection.
The operators $J_{\ell}^{D,\It}(\wvec_0)$ are such that, for all
$ \uvec_{\ell},\vvec_{\ell} \in [\Poly{}{k}(\Tl)]^{d}, \wvec_0 \in [\Poly{}{k}(\Tf)]^{d+1}$
\begin{align}
 (J_{\ell+1}^{D,\It}(\wvec_0) \uvec_{\ell+1}, \vvec_{\ell+1})_{[L^2(\Omega)]^d}
       & = (\It_{\ell}^{\ell+1} J_{\ell}^{D,\It}(\wvec_0) \It_{\ell+1}^{\ell} \uvec_{\ell+1}, \vvec_{\ell+1})_{[L^2(\Omega)]^d}. \nonumber
\end{align}
Accordingly
\begin{align}
\label{eq:matrestrJ}
\mbox{Galerkin projection:} \;\;  J_{\ell+1}^{\It}(\wvec_0)       &\eqbydef \Itre \;\, J_{\ell}^{\It}(\wvec_0) \;\, \Itpr, \;\; & \ell = 0,...,L-1,
\end{align}
where $\Itre$ and $\Itpr$ are the restriction and prolongation operators described in Section \ref{sec:intergrid}
and \eqref{eq:matrestrJ} is performed matrix-free. 

\section{Multigrid and Block preconditioners}
\label{sec:precINS}
In this work we consider multigrid preconditioners for the Navier-Stokes equations and 
block preconditioners for the dG discretization of the Stokes problem \eqref{discr:stokesBlock}.
Both the Stokes and the Navier-Stokes problem have a block structure that can be exploited 
to devise preconditioners based on Schur complement decompositions, nevertheless 
pressure Schur complement preconditioners are less trivial in the Navier-Stokes case than in the Stokes case \cite{Benzi05numericalsolution}.
Comparison between block and $h$-multigrid preconditioners will be performed on a Stokes model problem, while in the Navier-Stokes 
case we will focus on $h$-multigrid as a preconditioner of a FGMRES backward Euler iteration.

Incompressible flow problem dG discretizations in the form \eqref{discr:general} can be solved by preconditioned Krylov iterative methods, 
say $ksp(\Amat_0,\widehat{\Amat}_0)$, where the preconditioner $\widehat{\Amat}_0$ 
is a suitable approximation of $\Amat_0$ (such that the application of 
$\widehat{\Amat}_0^{-1}$ to a vector is cheap to compute).
For example an Incomplete Lower Upper (ILU) decomposition of the system matrix is a common preconditioner choice, 
 \ie $ksp\left(\Amat_0,\mathrm{ILU}(\Amat_0)\right)$.
Interestingly a Krylov iterative method, say $\widehat{ksp}(\Amat_0)$, can serve as a preconditioner 
by triggering convergence of the iteration on loose tolerances, \ie $ksp\left( \Amat_0,\widehat{ksp}({\Amat}_0) \right)$.
Note that in this case Flexible Generalized Minimal RESidual (FGMRES) is usually employed as a solver as the preconditioner varies at each outer
Krylov iteration.

Similarly, the multigrid V-cycle 
iteration of Section \ref{sec:multiVcicle} can be employed as preconditioner,
thus the solver strategy reads: $\mathrm{FGMRES}\left(\Amat_0, \mathrm{MG}_{\mathcal{V}}(\Amat_0)\right)$.
Building the coarse grid operators $\Amat_{\ell},~\ell=1,...,L$ as described in Section \ref{sec:coarseGridOp},
the multigrid V-cycle $\mathrm{MG}_{\mathcal{V}}(\Amat_0)$ can be applied as a preconditioner of the Stokes and Navier-Stokes operators $\Amat^{\mathrm{Stk}}_0$, $\Amat^{\mathrm{INS}}_0$.

Besides multigrid preconditioners, block preconditioners for the Stokes problem \eqref{discr:stokesBlock} are derived by noticing that $\Amat^{\mathrm{Stk}}$ admits the following LDU factorization
\begin{equation}
\Amat_0^{\mathrm{Stk}} =
\left[ \begin{array}{cc}
\Idmat & 0 \\
\Bmat_0 \Amat_0^{-1} & \Idmat
\end{array} \right]
\left[ \begin{array}{cc}
\Amat_0 & 0 \\
0 & \Smat_0 
\end{array} \right]
\left[ \begin{array}{cc}
\Idmat & \Amat_0^{-1} \Bmat_0^t \\
0 & \Idmat
\end{array} \right],
\end{equation}
where 
\begin{equation} 
\Smat_0 = \Cmat_0 - \Bmat_0 \Amat_0^{-1} \Bmat_0^t, \label{eq:schurS}
\end{equation}
is the \emph{pressure Schur complement} matrix.
Since 
\begin{equation}
\left(\Amat_0^{\mathrm{Stk}}\right)^{-1} =
\left[ \begin{array}{cc}
\Idmat & -\Amat_0^{-1} \Bmat_0^t \\
0 & \Idmat
\end{array} \right]
\left[ \begin{array}{cc}
\Amat_0^{-1} & 0 \\
0 & \Smat_0^{-1} 
\end{array} \right]
\left[ \begin{array}{cc}
\Idmat & 0 \\
-\Bmat_0 \Amat_0^{-1} & \Idmat
\end{array} \right],
\end{equation}
$\left(\widehat{\Amat}_0^{\mathrm{Stk}}\right)^{-1}$ can be obtained by replacing
$\Amat_0^{-1}$ and $\Smat_0^{-1}$ with preconditioned Krylov solvers, say
$ksp(\Amat_0,\widehat{\Amat}_0)$ and $ksp(\Smat_0,\widehat{\Smat}_0)$. 
Whereas computing $\Smat_0$ explicitly is not viable, 
an approximate solver $ksp(\widetilde{\Smat}_0,\widehat{\Smat}_0)$ can be employed 
with $$\widetilde{\Smat}_0 = \Cmat_0 - \Bmat_0 \; ksp(\Amat_0,\widehat{\Amat}_0) \; \Bmat_0^t.$$
Note that applying $\widetilde{\Smat}_0$ to a vector involves a nested Krylov iteration.
Clearly the performance of the outer solver $ksp(\Amat_0^{\mathrm{Stk}},\widehat{\Amat}_0^{\mathrm{Stk}})$ 
is strongly influenced by the availability of good preconditioners for the Laplace and the pressure 
Schur complement operators, read $\widehat{\Amat}_0$ and $\widehat{\Smat}_0$.

As suggested by Shahbazi \ea \cite{ethier07} , $\widehat{\Smat}_0$ can be constructed by
a dG discretization of the Laplace operator with homogeneous Neumann boundary conditions on Dirichlet boundaries.
Accordingly the operator $\widehat{S}_0$ is such that 
\begin{equation}
(\widehat{S}_{0} v_{0},w_{0})_{{L^2}(\Omega)} = -a^{\mathrm{BR2,hN}}_{0}(v_0,w_{0}), \; \forall v_{0},w_{0} \in \Poly{}{k}(\mathcal{T}_0) \label{eq:presSPrec}
\end{equation}
where
\begin{align}
  a^{\mathrm{BR2,hN}}_{0}(v_{0}, w_{0}) = & 
  \sum_{\T \in \Tf}\int_\T \GRADhf v_{0} \GRADhf w_{0} + \sum_{\F \in \Ffi} \eta_{\F}\int_{\F} \averg{\mathbf{r}_{\F}^k(\jump{v_{0}})}\SCAL \mathbf{n}_\F \; \jump{w_{0}} \nonumber \\
   & \; - \sum_{\F \in \Ffi} \int_\F \left( \averg{\GRADhf v_{0}} \cdot \mathbf{n}_\F \; \jump{w_{0}} 
                                    + \jump{v_{0}}\; \averg{\GRADhf w_{0}} \cdot \mathbf{n}_\F \right) \label{eq:BR2_neu}.
\end{align}
Note that \eqref{eq:BR2_neu} can be obtained from the BR2 bilinear form in \eqref{eq:ahBR2} 
using the local and global lifting operator definitions \eqref{eq:rF} and \eqref{eq:RF}
and dropping the boundary face terms.
As a preconditioner for ${\Amat}_0$ we employ the $h$-multigrid V-Cycle iteration described in Section \ref{sec:multiALL}
using the rescaled-inherited version of coarse grid operators defined in \eqref{def:ninhMStk}.

The solver and preconditioners options are summarized in what follows.
Richardson iteration serves as the outer loop, \ie $\mathrm{RCHRD}(\Amat_0^{\mathrm{Stk}},\widehat{\Amat}_0^{\mathrm{Stk}})$.
The application of the block preconditioner reads: $\left(\widehat{\Amat}_0^{Stk}\right)^{-1}=$ 
\begin{equation}
\left[ \begin{array}{cc}
ksp(\Amat_0,\widehat{\Amat}_0) & 0  \\
0 & \Idmat
\end{array} \right]
\left[ \begin{array}{cc}
\Idmat & - \Bmat_0^t \\
0 & \Idmat
\end{array} \right]
\left[ \begin{array}{cc}
\Idmat & 0 \\
0 & ksp(\widetilde{\Smat}_0,\widehat{\Smat}_0)
\end{array} \right]
\left[ \begin{array}{cc}
\Idmat & 0 \\
-\Bmat_0 ksp(\Amat_0,\widehat{\Amat}_0) & \Idmat
\end{array} \right],
\end{equation}
see PETSc User manual \cite{petsc-user-ref}, where
\begin{align}
ksp(\Amat_0,\widehat{\Amat}_0) &= \mathrm{FGMRES}\left(\Amat_0,\mathrm{MG}_{\mathcal{V}}(\Amat_0)\right), \nonumber\\
ksp(\widetilde{\Smat}_0,\widehat{\Smat}_0) &= \mathrm{GMRES}\left( \widetilde{\Smat}_0
								   ,\mathrm{ILU}(\widehat{\Smat}_0)\right),  \nonumber 
\end{align}
and $\widetilde{\Smat}_0 = \Cmat_0 - \Bmat_0 \; ksp(\Amat_0,\widehat{\Amat}_0) \; \Bmat_0^t$.

\section{Numerical results}
\label{sec:numres}
\subsection{BR2 dG discretization}
\label{sec:numresBR2}
In this section we apply the $h$-multigrid V-cycle iteration of Algorithm \ref{eq:vcic} for solving a Poisson problem discretized by means of the BR2 dG formulation.
For the sake of comparison we consider the three strategies for defining coarse grid operators introduced in Section \ref{sec:coarseBR2}, that is:
\begin{inparaenum}[i)]
\item
non-inherited operators defined assembling bilinear forms on each mesh level
\item 
inherited operators defined by means of a Galerkin projection
\item
the newly introduced inherited operators with stability rescaling.
\end{inparaenum}
We compare these approaches on the basis of convergence rate and computation time
and we assess the benefits of using $h$-multigrid as compared to state-of-the-art single grid solvers like the preconditioned Conjugate-Gradient (CG) method
and the preconditioned Generalized Minimal RESidual (GMRES) method.

We consider the Poisson problem in \eqref{eq:lap} on the bi-unit square and cube, $\Omega = [-1,1]^d$ with $d=2$ and $d=3$, respectively, 
where the forcing term is imposed according to the following smooth analytical solution
\begin{equation}
  \label{eq:laptest2}
   u = \prod_{i=1}^d \mathrm{sin}(\pi x_i),
\end{equation}
and homogeneous boundary conditions are imposed on $\partial \Omega$.

\begin{figure}[H]
\begin{tabular}{lr}
\includegraphics[width=0.47 \textwidth]{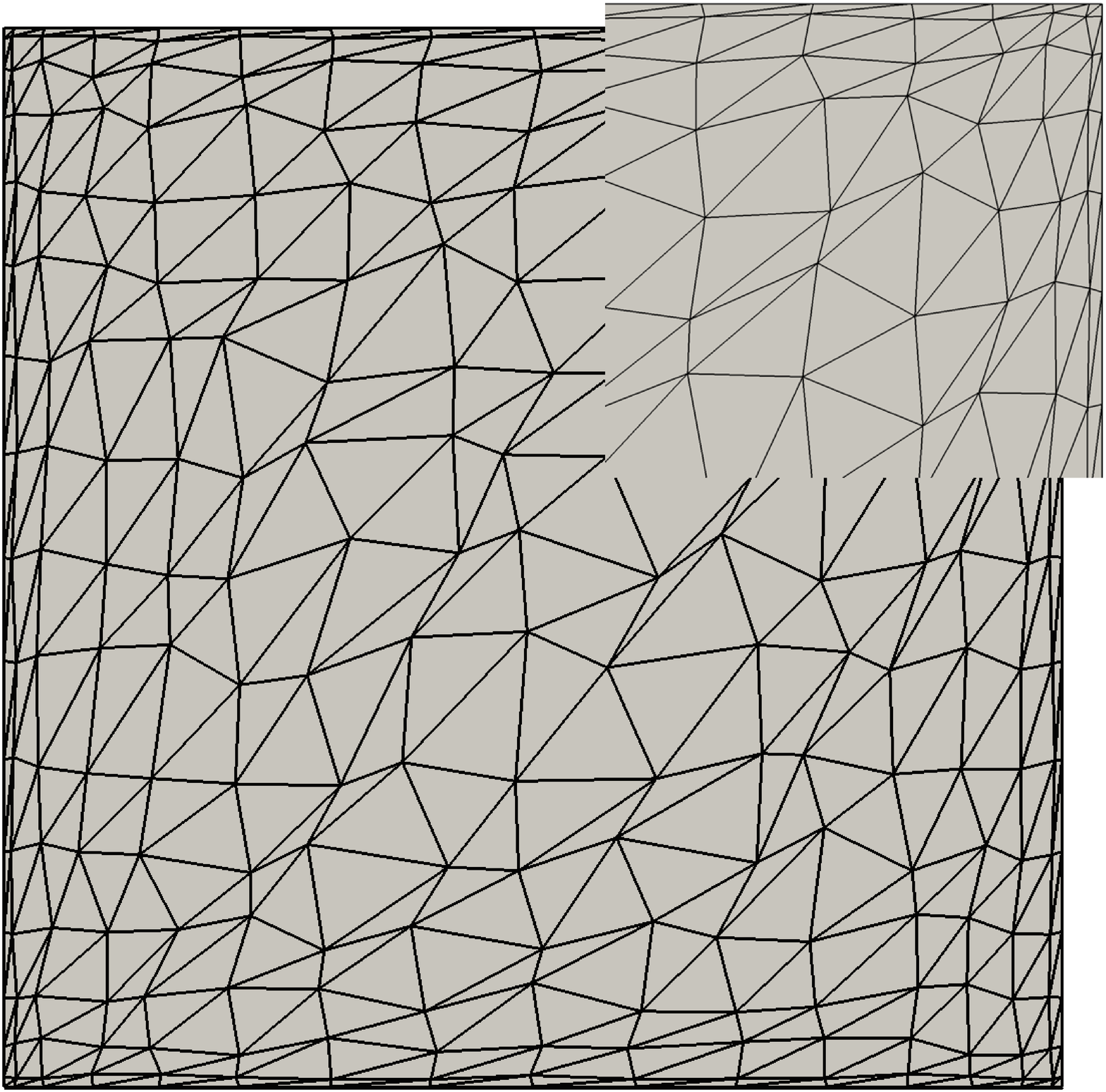} &
\includegraphics[width=0.47 \textwidth]{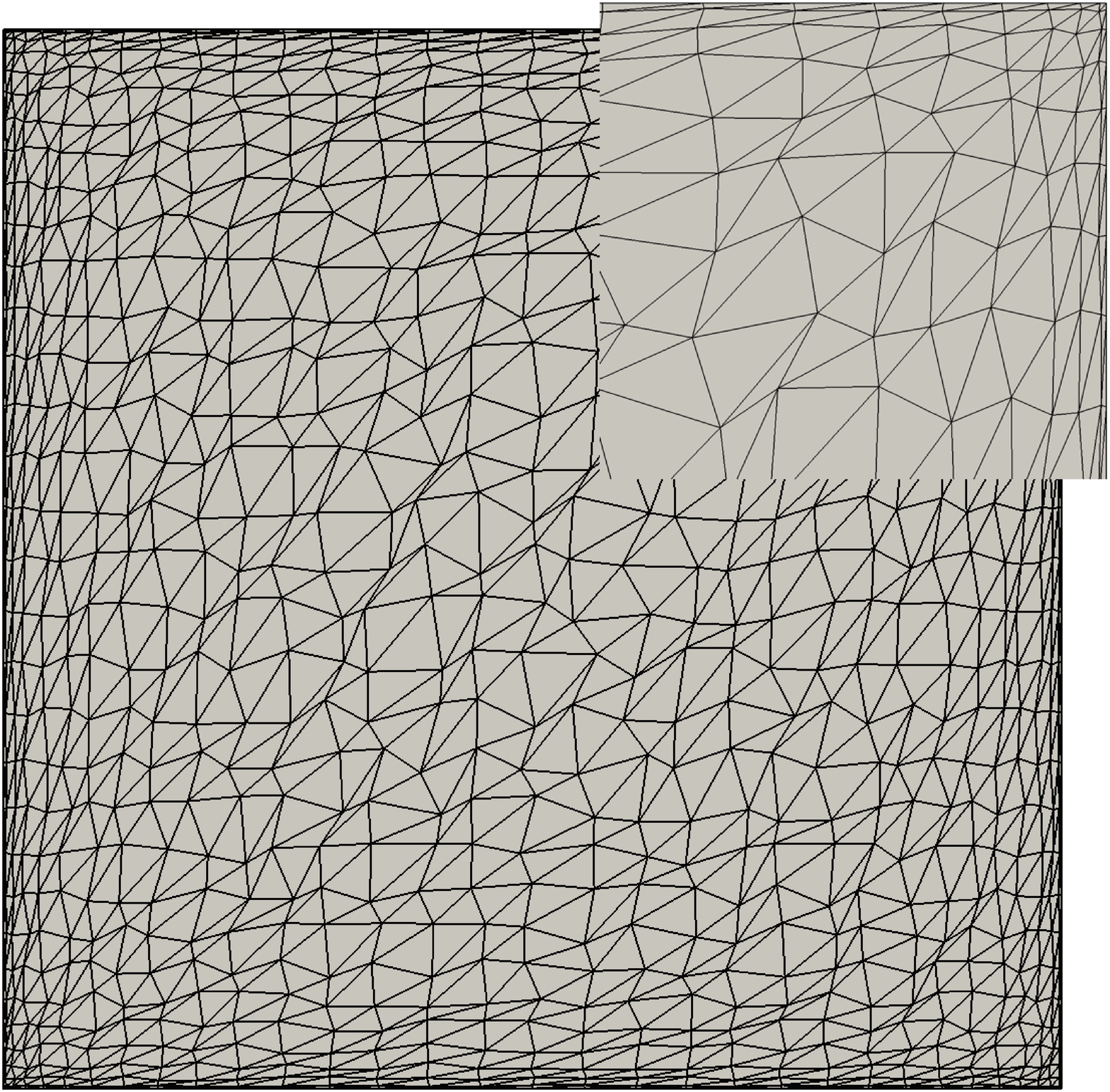} \\
\end{tabular}
\caption{Two grids of the distorted and graded triangular mesh sequence. \emph{Left}: $2\,(32^2)$ grid and \emph{right}: $2\,(64^2)$ grid. The square corner detail allows to appreciate 
         that the aspect ratio of triangular elements increases moving towards the domain boundaries.  
	 \label{fig:grid_tri}}
\end{figure}
\begin{figure}[H]
\centering
\includegraphics[width=0.8 \textwidth]{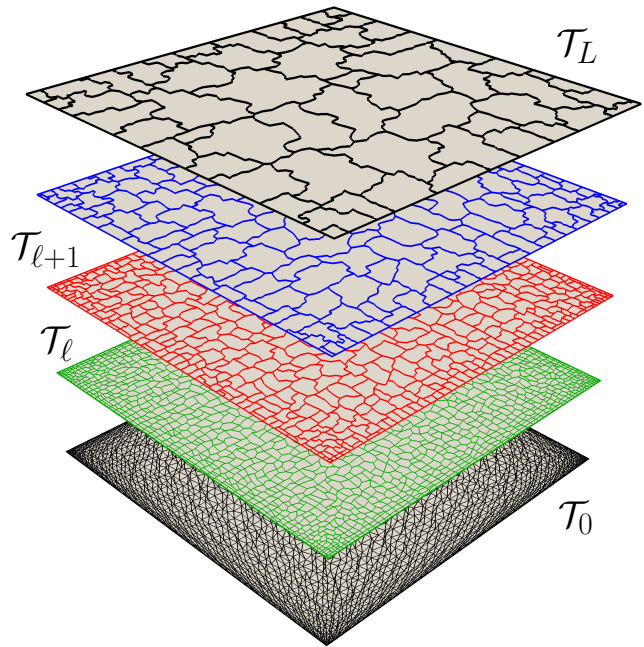} \\
\caption{Example of a five levels ($L=4$) $h$-coarsened mesh sequence generated on top the $2(64^2)$ triangular elements grid of Figure \ref{fig:grid_tri}. 
		       \label{fig:grid_seq_two}}
\end{figure}
In order to investigate the growth of computational costs while increasing the mesh size,
2D solutions are computed on three uniform quadrilateral elements meshes of size $(128 \cdot 2^n)^2, n=\{0,1,2\}$
and three distorted and graded triangular meshes of size $2(64 \cdot 2^n)^2, n=\{0,1,2\}$, see Figure \ref{fig:grid_tri}.
As for 3D solutions we consider a $128^3$ grid, counting of more than two million hexahedral elements, and we
investigate parallel performance of the multigrid algorithm running on up to 128 processes.
In both two and three space dimensions we check the influence of raising the polynomial degree 
on the convergence rate and the computational expense considering $k=\{1,2,3\}$, that is first, second and third
polynomial degree dG discretizations.
The $L^2$ error norm is on the order of $10^{-11}$ for the fourth-order accurate dG discretization on the $512^2$ quadrilateral grid.
We do not consider a further raise of the polynomial degree since for higher-order discretizations 
$p$-multigrid or $hp$-multigrid solution strategies might be best suited. 
Indeed $h$-multigrid is to be applied in the context of large scale computations where 
the mesh size is constrained by the need to accurately discretize a complex computational domain,
note that the design of coarse high-order meshes suited for higher-order discretizations is an open field of research, see \eg \cite{Toulorge20138}.
We remark that the agglomeration strategy does not take advantage of the triviality of the geometry here considered.
In Section \ref{sec:hemo} the multigrid strategy will be applied without any modification to unstructured, 
possibly hybrid, meshes of complex computational domains.

\begin{table}[!htbp]
\centering
    \begin{tabular}{@{}c|ccccc@{}}
        \multicolumn{6}{c}{$h$-coarsened quadrilateral mesh sequences} \\\hline
        $\card({\Tf})$          & \multicolumn{5}{c}{$\card(\Tl)$} \\ 
                                & $\ell=1$ & $\ell=2$ & $\ell=3$ & $\ell=4$ & $\ell=5$ \\ \hline
        $64^2$                  &1208      &365       &109      &34     &11     \\ 
        $128^2$                 &4824      &1447      &437      &136    &41     \\ 
        $256^2$                 &19218     &5791      &1754     &535    &161    \\ 
        $512^2$                 &76880     &23087     &6976     &2116   &643    \\ 
        \multicolumn{6}{c}{$h$-coarsened triangular mesh sequences} \\\hline
        $\card({\Tf})$          & \multicolumn{5}{c}{$\card(\Tl)$} \\ 
                                & $\ell=1$ & $\ell=2$ & $\ell=3$ & $\ell=4$ & $\ell=5$ \\ \hline
        $2(32^2)$               &541       &157       &46       &13     &4     \\ 
        $2(64^2)$               &2290      &660       &194      &57     &17     \\ 
        $2(128^2)$              &9287      &2683      &780      &231    &66    \\ 
        $2(256^2)$              &37551     &10956     &3214     &935    &274    \\ 
    \end{tabular}
\caption{\label{tab:gridAgglo2d} Six levels $h$-coarsened agglomerated elements mesh sequences of the bi-unit square.
                                 Number of agglomerated elements at each mesh level $\ell=0,...,5$.}
\end{table} 

\begin{table}[!htbp]
\centering
    \begin{tabular}{@{}c|c|ccccc@{}} 
        \multicolumn{7}{c}{3D $h$-coarsened mesh sequences, grid partition size} \\\hline
        $\card(\Tf)$  & processes ($np$) 
	              & \multicolumn{5}{c}{$\displaystyle \genfrac{}{}{0pt}{}{\max_{i=1,...,np}}{\min_{i=1,...,np}} \; \card(\Tl^i) $} \\ 
		      &  & $\ell=1$ & $\ell=2$ & $\ell=3$ & $\ell=4$ & $\ell=5$ \\ \hline
 \multirow{2}{*}{$128^3$}         & \multirow{2}{*}{16}   &131078&19378&2895&434&64\\ 
                                  &                       &131066&19258&2846&416&60\\ \hline
 \multirow{2}{*}{$128^3$}         & \multirow{2}{*}{32}   &65541 &9763 &1468&219&34\\ 
                                  &                       &65531 &9666 &1426&207&28\\ \hline
 \multirow{2}{*}{$128^3$}         & \multirow{2}{*}{64}   &32770 &4878 &734 &111&17\\ 
                                  &                       &32766 &4809 &708 &104&14\\ \hline
 \multirow{2}{*}{$128^3$}         & \multirow{2}{*}{128}  &16392 &2454 &372 &57 &9 \\ 
                                  &                       &16380 &2407 &353 &50 &7 \\ 
    \end{tabular}
\caption{\label{tab:gridAgglo3d} Five levels ($\ell=1,...,5$) $h$-coarsened mesh sequences agglomerated on top of a $128^3$ hexahedral elements grids of the bi-unit cube. 
                                 Maximum and minimum number of elements among the distributed grid partitions $\Tl^i,\; i=1,...,np$,
                                 obtained running in parallel with $np$ processes.}
\end{table} 
To investigate the influence of the number of coarse levels on the convergence rate we consider $L=\{2,3,4,5\}$ and $L=\{2,3,4\}$ for $d=2$
and $d=3$, respectively, that is we consider a stack of 3 to 6 grids in two space dimensions and a stack of 3 to 5 grids in three space dimensions, 
see Figure \ref{fig:grid_seq} and Figure \ref{fig:grid_seq_two}. 
The number of mesh elements at each level $\ell$ and 
the maximum and the minimum number of elements among the distributed grid 
partitions at level $\ell$ is reported in Table \eqref{tab:gridAgglo2d} and Table \eqref{tab:gridAgglo3d}, respectively. 
It is interesting to remark that in three (resp. two) space dimensions a 6.7 (resp. 3.3) fold decrease of the number of elements is obtained 
at each agglomeration step, whereas an 8 (resp. 4) fold decrease would be required in order 
to halve the mesh step size in uniform hexahedral (resp. quadrilateral) elements mesh sequences. 
Nevertheless, since agglomerated elements have very general shapes, the element size $h_{\kappa}$ is not uniform and, as demonstrated in \cite{BassiFlexy12}, 
the maximum and average mesh step size are usually bigger as compared to quadrilateral elements meshes of the same cardinality.
The (user provided) upper bound for the number of sub-elements to be clustered together by means of the MGridGen library, see Section \ref{sec:aggloTopo} for details,
should guarantee that $\frac{h_{\kappa_{\ell+1}}}{h_{\kappa_{\ell}}} \lessapprox 2$ for each $\kappa_\ell \in \Tl$.

We complete the definition of the V-cycle preconditioned FGMRES iteration, 
\ie $\mathrm{FGMRES}\left( \Amat^{\mathrm{BR2}}_0, \mathrm{MG}_\mathcal{V}(\Amat^{\mathrm{BR2}}_0) \right)$ 
specifying the relevant solver and preconditioner options.
We employ a single iteration of multigrid V-cycle as a preconditioner for a Flexible GMRES solver (restarted FGMRES with 60 Krylov spaces) \cite{SaadFlexy93}.
High-order modes of the error are smoothed with a single iteration of a right preconditioned GMRES solver. 
In 2D serial runs we employ an Incomplete Lower-Upper (ILU) preconditioner 
while in 3D parallel runs we opt for an Additive Schwarz domain decomposition Method (ASM) with one level of overlap between sub-domains and 
an ILU decomposition for each sub-domain matrix.
On the coarsest level $L$ linear systems are solved with a direct solver in 2D. For parallel 3D runs we 
rely on the solver employed in smoothing steps but, instead of a single iteration, 
we impose a four order of magnitude decrease of the relative residual norm, that is $\frac{\| f_L - A^{\mathrm{BR2}}_L \bar{u}_L \|}{\| f_L \|} \leq 10^{-4}$.
The FGMRES solver is forced to reach tight relative residual tolerance, in particular
the linear system solution converges in $N_\mathrm{it}$ iterations
if at the $i$-th iterate $\| \hat{r}_0^i \| = \frac{\| f_0 - A_0^{\mathrm{BR2}} \bar{u}_0^i \|}{\| f_0 \|} \leq 10^{-10}$.

The numerical results reported in the next section have been computed by exploiting the PCMG multigrid 
preconditioner framework available in the \texttt{PETSc} library~\cite{petsc-web-page,petsc-user-ref,petsc-efficient}.
The \texttt{MOAB}~\cite{MOAB} library is employed for storing distributed mesh data at all mesh levels
and \texttt{METIS}~\cite{Metis} library is employed to partition the mesh in case of parallel computations.
\subsubsection{2D Poisson problem}
\label{sec:poisson2D}
\begin{table}[!htbp]
\small
\centering
    \begin{tabular}{@{}l|ccc|ccc|ccc@{}} 
                                \multicolumn{10}{c}{Linear solver iterations, 2D Poisson problem, quadrilateral mesh sequences} \\ \hline
                   operators           & \multicolumn{3}{c|}{inherited} & \multicolumn{3}{c|}{non-inherited} & \multicolumn{3}{c}{rescaled-inherited}\\
                   grid ($\card{(\Tf)}=(\cdot)^2$)               & $128$ &  $256$   & $512$ & $128$ &  $256$   & $512$ & $128$ &  $256$   & $512$ \\\hline
                         $\mathbf{k=1}$ & \multicolumn{3}{c|}{}& \multicolumn{3}{c|}{}& \multicolumn{3}{c}{}\\
                   FGMRES $\mathrm{MG}_{\mathcal{V}}$ $\nlev$ = 2    & 12      &  12        & 13      & 10       &  10         & 10       & 10      &  10        & 10      \\
                   FGMRES $\mathrm{MG}_{\mathcal{V}}$ $\nlev$ = 3    & 15      &  15        & 15      & 10       &  10         & 10       & 10      &  11        & 11      \\
                   FGMRES $\mathrm{MG}_{\mathcal{V}}$ $\nlev$ = 4    & 19      &  19        & 20      & 10       &  10         & 10       & 10      &  11        & 11      \\
                   FGMRES $\mathrm{MG}_{\mathcal{V}}$ $\nlev$ = 5    & 24      &  24        & 25      & 10       &  10         & 10       & 11      &  11        & 11      \\
		   \cline{2-10}
                   CG ILU(0)                                    & \multicolumn{6}{c}{}    &  206     &  398       & 743     \\
                   GMRES(120) ILU(0)                            & \multicolumn{6}{c}{}    &  229     &  566       & 1255    \\
		   \hline
                                   $\mathbf{k=2}$ & \multicolumn{3}{c|}{}& \multicolumn{3}{c|}{}& \multicolumn{3}{c}{}\\
                   FGMRES $\mathrm{MG}_{\mathcal{V}}$ $\nlev$ = 2    & 9       &  9         & 9       & 8        &  8          & 8        & 8       &  8         & 8       \\
                   FGMRES $\mathrm{MG}_{\mathcal{V}}$ $\nlev$ = 3    & 16      &  15        & 15      & 8        &  8          & 8        & 8       &  8         & 8       \\
                   FGMRES $\mathrm{MG}_{\mathcal{V}}$ $\nlev$ = 4    & 27      &  27        & 26      & 9        &  8          & 8        & 8       &  8         & 8       \\
                   FGMRES $\mathrm{MG}_{\mathcal{V}}$ $\nlev$ = 5    & 60      &  45        & 49      & 9        &  9          & 8        & 9       &  9         & 8       \\
		   \cline{2-10}
                   CG ILU(0)                                    & \multicolumn{6}{c}{}   & 230     &  446       & 827     \\
                   GMRES(120) ILU(0)                            & \multicolumn{6}{c}{}   & 312     &  811       & 1925    \\
		   \hline
                                   $\mathbf{k=3}$ & \multicolumn{3}{c|}{}& \multicolumn{3}{c|}{}& \multicolumn{3}{c}{}\\
                   FGMRES $\mathrm{MG}_{\mathcal{V}}$ $\nlev$ = 2    & 8       &  7         & 7       & 7        &  7          & 6        & 7       &  6         & 6       \\
                   FGMRES $\mathrm{MG}_{\mathcal{V}}$ $\nlev$ = 3    & 12      &  12        & 13      & 7        &  7          & 6        & 7       &  7         & 6       \\
                   FGMRES $\mathrm{MG}_{\mathcal{V}}$ $\nlev$ = 4    & 20      &  21        & 23      & 8        &  7          & 7        & 8       &  7         & 7       \\
                   FGMRES $\mathrm{MG}_{\mathcal{V}}$ $\nlev$ = 5    & 37      &  39        & 42      & 8        &  8          & 7        & 8       &  8         & 7       \\
		   \cline{2-10}
                   CG ILU(0)                                    & \multicolumn{6}{c}{}     & 250     &  480       & 899     \\
                   GMRES(120) ILU(0)                            & \multicolumn{6}{c}{}     & 310     &  1015      & 2000$^*$\\
		   \hline
    \end{tabular}
    \caption{Number of iterations required to solve the 2D model Poisson problem \eqref{eq:lap}-\eqref{eq:laptest2} 
             with an FGMRES solver preconditioned with $h$-multigrid. Quadrilateral mesh sequence.
	     Linear system relative residual tolerance is $10^{-10}$, see text for details. \label{tab:it_multasprec}}
\end{table} 
\begin{table}[!htbp]
\small
\centering
    \begin{tabular}{@{}l|ccc|ccc|ccc@{}} 
                                \multicolumn{10}{c}{Linear solver iterations, 2D Poisson problem, triangular mesh sequences} \\ \hline
                   operators           & \multicolumn{3}{c|}{inherited} & \multicolumn{3}{c|}{non-inherited} & \multicolumn{3}{c}{rescaled-inherited}\\
                   grid ($\card{(\Tf)}=2(\cdot)^2$)   & $64$ &  $128$   & $256$ &  $64$ &  $128$   & $256$ &  $64$ &  $128$   & $256$\\\hline
                         $\mathbf{k=1}$ & \multicolumn{3}{c|}{}& \multicolumn{3}{c|}{}& \multicolumn{3}{c}{}\\
                   FGMRES $\mathrm{MG}_{\mathcal{V}}$ $\nlev$ = 2    & 19      &  22        & 28      & 24       &  29       & 37       & 19      &  23        & 30     \\
                   FGMRES $\mathrm{MG}_{\mathcal{V}}$ $\nlev$ = 3    & 20      &  23        & 28      & 24       &  29       & 38       & 19      &  24        & 30     \\
                   FGMRES $\mathrm{MG}_{\mathcal{V}}$ $\nlev$ = 4    & 22      &  24        & 29      & 24       &  29       & 38       & 19      &  23        & 30     \\
                   FGMRES $\mathrm{MG}_{\mathcal{V}}$ $\nlev$ = 5    & 26      &  29        & 32      & 24       &  29       & 38       & 19      &  23        & 30     \\
		   \cline{2-10}
                   CG ILU(0)                                         & \multicolumn{6}{c}{}  & 250     &  509       & 1011     \\
                   GMRES(120) ILU(0)                                 & \multicolumn{6}{c}{}  & 298     &  677       & 1520    \\
		   \hline
                                   $\mathbf{k=2}$ & \multicolumn{3}{c|}{}& \multicolumn{3}{c|}{}& \multicolumn{3}{c}{}\\
                   FGMRES $\mathrm{MG}_{\mathcal{V}}$ $\nlev$ = 2    & 17      &  19        & 24      &  21       & 26       & 37       & 17       &  20       & 27        \\
                   FGMRES $\mathrm{MG}_{\mathcal{V}}$ $\nlev$ = 3    & 20      &  22        & 26      &  21       & 26       & 37       & 17       &  21       & 27        \\
                   FGMRES $\mathrm{MG}_{\mathcal{V}}$ $\nlev$ = 4    & 26      &  32        & 37      &  21       & 26       & 38       & 17       &  21       & 27        \\
                   FGMRES $\mathrm{MG}_{\mathcal{V}}$ $\nlev$ = 5    & 44      &  61        & 77      &  21       & 27       & 38       & 17       &  21       & 27        \\
		   \cline{2-10}
                   CG ILU(0)                                         & \multicolumn{6}{c}{}  & 297     &  599       & 1197     \\
                   GMRES(120) ILU(0)                                 & \multicolumn{6}{c}{}  & 330     &  866       & 2000$^*$  \\
		   \hline
                                   $\mathbf{k=3}$ & \multicolumn{3}{c|}{}& \multicolumn{3}{c|}{}& \multicolumn{3}{c}{}\\
                   FGMRES $\mathrm{MG}_{\mathcal{V}}$ $\nlev$ = 2    & 16      &  17        & 20      & 17        &  20      & 28       & 15       &  18       & 21   \\
                   FGMRES $\mathrm{MG}_{\mathcal{V}}$ $\nlev$ = 3    & 18      &  19        & 23      & 17        &  21      & 29       & 16       &  18       & 21   \\
                   FGMRES $\mathrm{MG}_{\mathcal{V}}$ $\nlev$ = 4    & 25      &  25        & 31      & 18        &  21      & 29       & 16       &  18       & 22   \\
                   FGMRES $\mathrm{MG}_{\mathcal{V}}$ $\nlev$ = 5    & 71      &  43        & 67      & 18        &  22      & 29       & 17       &  19       & 22   \\
		   \cline{2-10}
                   CG ILU(0)                                         & \multicolumn{6}{c}{}  & 334     &  664       & 1324    \\
                   GMRES(120) ILU(0)                                 & \multicolumn{6}{c}{}  & 396     &  911       & 2000$^*$\\
		   \hline
    \end{tabular}
    \caption{Number of iterations required to solve the 2D model Poisson problem \eqref{eq:lap}-\eqref{eq:laptest2} 
             with an FGMRES solver preconditioned with $h$-multigrid. Distorted triangular mesh sequence. 
	     Linear system relative residual tolerance is $10^{-10}$, see text for details. \label{tab:it_multasprec_unstr}}
\end{table} 

Tables \ref{tab:it_multasprec} and \ref{tab:it_multasprec_unstr} report the number of iterations 
required to solve the Poisson problem \eqref{eq:lap}-\eqref{eq:laptest2} discretized with the BR2 method.
Single grid solver options mimic multigrid ones: we impose a relative residual decrease of $10^{-10}$ and employ 
ILU (right) preconditioned Conjugate Gradient (CG) and GMRES solvers setting the number of Krylov spaces to 120 for GMRES.

As expected only non-inherited and inherited multigrid with stabilization term rescaling (\emph{rescaled-inherited}) are 
able to provide uniform convergence with respect to the number of levels, note that this is the case even on bad quality triangular meshes.
Comparison of the number of iterations on quadrilateral elements meshes
highlights that the performance of single grid solvers worsen on finer meshes 
while the rescaled-inherited multigrid iteration is almost grid independent.
Moving towards finer distorted triangular meshes also the number of multigrid iterations increases,
but far less dramatically than with single grid solvers.
Interestingly the number of iterations of single grid solvers is also affected by raising the polynomial degree
while the convergence rates of $h$-multigrid improve increasing the polynomial degree
when non-inherited and rescaled-inherited multigrid are employed.
Note that for $k=2$ and $k=3$ the average residual decrease exceeds one order of magnitude at each V-cycle iteration on quadrilateral mesh sequences.

\begin{table}[!htbp]
\small
\centering
    \begin{tabular}{@{}l|ccc|ccc|ccc@{}} 
                                \multicolumn{10}{c}{Total CPU time (s), 2D Poisson problem, quadrilateral mesh sequences} \\ \hline
 operators             & \multicolumn{3}{c|}{inherited}        & \multicolumn{3}{c|}{non-inherited}        & \multicolumn{3}{c}{rescaled-inherited}    \\
 grid ($\card{(\Tf)}=(\cdot)^2$) & $128$ &$256$& $512$             & $128$ &$256$&$512$                 & $128$&$256$&$512$             \\\hline
 $\mathbf{k=1}$        & \multicolumn{3}{c|}{}                  & \multicolumn{3}{c|}{}                     & \multicolumn{3}{c}{}                \\ 
 FGMRES $\mathrm{MG}_{\mathcal{V}}$ $\nlev$ = 2 & 1.58       &  \bf{5.95}  & \bf{27.4}  & 2.03       &  \bf{7.23}  & \bf{31.7}     & 1.49      &  \bf{5.70}  & \bf{24.5} \\
 FGMRES $\mathrm{MG}_{\mathcal{V}}$ $\nlev$ = 3 & 1.72       &  6.70       & 28.1       & 2.10       &  7.89       & 32.3          & 1.69      &  6.11       & 25.9      \\
 FGMRES $\mathrm{MG}_{\mathcal{V}}$ $\nlev$ = 4 & 1.92       &  7.78       & 31.2       & 2.27       &  8.61       & 34.6          & 1.60      &  6.29       & 25.1      \\
 FGMRES $\mathrm{MG}_{\mathcal{V}}$ $\nlev$ = 5 & 2.14       &  8.32       & 36.0       & 2.67       &  8.95       & 36.9          & 1.57      &  6.69       & 26.3      \\
		   \cline{2-10}
 CG ILU(0)             & \multicolumn{6}{c}{}      & \bf{1.45} &  8.01       & 48.7      \\
 GMRES(120) ILU(0)     & \multicolumn{6}{c}{}      & 3.01      &  28.1       & 230       \\
		   \hline
 $\mathbf{k=2}$        & \multicolumn{3}{c|}{}                  & \multicolumn{3}{c|}{}                     & \multicolumn{3}{c}{}                \\
 FGMRES $\mathrm{MG}_{\mathcal{V}}$ $\nlev$ = 2 & \bf{2.75}  &  \bf{11.3}  & 54.4       & \bf{3.30}  &  \bf{13.6}  & 64.4          & \bf{2.43} &  10.3       & 51.1      \\
 FGMRES $\mathrm{MG}_{\mathcal{V}}$ $\nlev$ = 3 & 3.49       &  12.0       & \bf{54.2}  & 3.34       &  14.1       & 58.3          & \bf{2.43} &  9.4        & 42.1      \\
 FGMRES $\mathrm{MG}_{\mathcal{V}}$ $\nlev$ = 4 & 4.12       &  16.7       & 73.9       & 3.36       &  13.8       & \bf{57.9}     & \bf{2.43} &  \bf{9.3}   & \bf{41.4} \\
 FGMRES $\mathrm{MG}_{\mathcal{V}}$ $\nlev$ = 5 & 7.18       &  24.3       & 115        & 3.74       &  15.6       & 60.3          & \bf{2.43} &  9.7        & 41.7      \\
		   \cline{2-10}
 CG ILU(0)             & \multicolumn{6}{c}{}       & 3.64      &  22.1       & 143       \\
 GMRES(120) ILU(0)     & \multicolumn{6}{c}{}       & 7.90      &  86.7       & 811       \\
		   \hline
 $\mathbf{k=3}$        & \multicolumn{3}{c|}{}                  & \multicolumn{3}{c|}{}                     & \multicolumn{3}{c}{}                \\
 FGMRES $\mathrm{MG}_{\mathcal{V}}$ $\nlev$ = 2 & 6.62       &  30.3       & 195        & 8.40       &  37.5       & 226           & 6.65      &  29.4       & 191       \\
 FGMRES $\mathrm{MG}_{\mathcal{V}}$ $\nlev$ = 3 & \bf{6.03}  &  \bf{26.1}  & \bf{126}   & \bf{7.46}  &  \bf{30.0}  & 133           & \bf{5.07} &  22.0       & 96.2        \\
 FGMRES $\mathrm{MG}_{\mathcal{V}}$ $\nlev$ = 4 & 7.91       &  34.6       & 160        & 7.59       &  32.1       & \bf{129}      & 5.30      &  \bf{20.9}  & 88.9        \\
 FGMRES $\mathrm{MG}_{\mathcal{V}}$ $\nlev$ = 5 & 11.9       &  52.5       & 242        & 8.17       &  33.4       & 134           & 5.38      &  21.8       & \bf{87.7}  \\
		   \cline{2-10}
 CG ILU(0)             & \multicolumn{6}{c}{}      & 11.0      &  68.4       & 475        \\
 GMRES(120) ILU(0)     & \multicolumn{6}{c}{}      & 19.3      &  240        & 1851$^*$   \\
		   \hline
    \end{tabular}                                                               
    \caption{BR2 dG discretization of the 2D model Poisson problem \eqref{eq:lap}-\eqref{eq:laptest2}.
             Total CPU times (assembly plus solution times) required to reach a relative residual tolerance of $10^{-10}$,
             see text for details. Bold text highlights the best result column-wise.\label{tab:cpu_tot}} 
\end{table} 
Total CPU times (total means the sum of solution and assembly CPU times) 
reported in Table \ref{tab:cpu_tot} demonstrate that the newly introduced rescaled-inherited multigrid strategy is the best performing.
From the solution time viewpoint inherited and rescaled-inherited multigrid are almost indistinguishable 
while inherited multigrid is largely affected by the performance degradation increasing the number of grid levels. 
From the assembly time viewpoint inherited and rescaled-inherited multigrid avoids the burden of 
numerically integrating bilinear forms over agglomerated elements meshes.
Non-inherited multigrid assembly times strongly increase with the number of levels 
(note that assembly can be twice as expensive than solution) negatively impacting total CPU times.
We remark that, if quadrature formulas are defined over sub-elements as described in \eqref{eq:intRefAgglo}, 
the number of quadrature points is the same at each mesh level, irrespectively of the mesh density.

Rescaled-inherited total CPU times are almost independent of the number of levels provided that the coarsest grid on level $L$ is coarse enough.
This is a very important result in view of applying multigrid in real-world computations since it basically removes the burden of choosing of the number of grid levels. 
Most importantly both non-inherited and rescaled-inherited multigrid are close to the optimal multigrid efficiency. 
They lead to a four-fold increase of the total computation time with a four-fold increase of the mesh size 
at all the polynomials degrees, provided that $L$ is chosen large enough.

\begin{table}
\small
\centering
    \begin{tabular}{@{}l|ccc|ccc|ccc@{}} 
      $\mathrm{MG}_{\mathcal{V}}(k)$ vs CG$(k)$ & \multicolumn{9}{c}{Total CPU time speedup}\\ \cline{1-10}
      degree               & \multicolumn{3}{c|}{$k=1$} & \multicolumn{3}{c|}{$k=2$} & \multicolumn{3}{c}{$k=3$} \\
      quad grid & $128$ &$256$& $512$  & $128$ &$256$&$512$ & $128$&$256$&$512$             \\
      CG/MG step time    & 0.92    &  1.2  & 1.8    &  1.5  & 2.3   & 3.4       & 2.0  & 3.1 & 5.4 \\ \hline
      tri grid & $64$ &$128$& $256$   & $64$ &$128$&$256$  & $64$&$128$&$256$             \\
      CG/MG step time  & 0.95    &  1.6  & 2.4    &  1.4  & 2.3   & 3.5       & 1.7  & 3.1 & 5.0 \\ \hline
    \end{tabular}
    \caption{BR2 dG discretization of the 2D model Poisson problem \eqref{eq:lap}-\eqref{eq:laptest2}.
             Comparison of total CPU times (assembly plus solution times) required to reach a relative residual tolerance of $10^{-10}$ with Conjugate Gradient and $h$-multigrid ($L=5$),
             see text for details. \label{tab:cpu_comp_BR2}} 
\end{table}
\begin{table}
\centering
\small
  \begin{tabular}{@{}l|cc|cc|cc@{}} 
    CG($k_{\mathrm{CG}}$) vs $\mathrm{MG}_{\mathcal{V}}(k_{\mathrm{MG}})$ &\multicolumn{6}{c}{Total CPU time ratio}  \\ \cline{1-7}   
    solver(degree)          & \multicolumn{2}{c|}{CG(1)/$\mathrm{MG}_{\mathcal{V}}$(2)} & \multicolumn{2}{c|}{CG(2)/$\mathrm{MG}_{\mathcal{V}}$(3)} & \multicolumn{2}{c}{CG(1)/$\mathrm{MG}_{\mathcal{V}}$(3)}   \\
    finest grid &     quad & tri & quads & tri & quads & tri \\
    CG/MG step time     & 0.95    &  1.0   &  1.6  & 1.3  &  0.55 & 0.37 \\ \hline
 \end{tabular}
    \caption{BR2 dG discretization of the 2D model Poisson problem \eqref{eq:lap}-\eqref{eq:laptest2}.
             Comparison of total CPU times (assembly plus solution times) required to reach a relative residual tolerance of $10^{-10}$ with Conjugate Gradient and $h$-multigrid ($L=5$).
	     The multigrid solution strategy is applied to an higher polynomial degree dG discretization ($k_{\mathrm{MG}}>k_{\mathrm{CG}}$).
             \label{tab:cpu_comp_kBR2}} 
\end{table}  
Since CPU times for the 2D Poisson problem are measured running on a 2010 laptop, we recommend not to consider absolute values but rather relative gains.
In this regard, the gains with respect to the best performing single grid solver (an ILU preconditioned Conjugate Gradient iteration) are significant, 
especially at the highest polynomial degrees on fine meshes, see Table \ref{tab:cpu_comp_BR2}.
As a result, if we consider the finest quadrilateral and triangular meshes, 
solving a second degree polynomial degree BR2 dG discretization with $h$-multigrid 
is comparable to solving a first degree dG discretization with CG, see Table \ref{tab:cpu_comp_kBR2}.
Interestingly the time required for solving a third polynomial degree BR2 dG discretization 
with $h$-multigrid is twice the time required for solving 
a first degree dG discretization with Conjugate Gradient, 
which is quite impressive considering the accuracy gap.

\begin{table}[!htbp]
\small
    \begin{tabular}{@{}c|ccc|ccc|ccc|ccc@{}} 
                                               \multicolumn{13}{c}{Preprocessing CPU time (s), quadrilateral elements mesh sequence} \\ \cline{1-13}
                         task                  &\multicolumn{3}{c|}{grid topology} & \multicolumn{9}{c}{orthogonalization and intergrid operators}\\ \cline{2-13}
                                               & \multicolumn{3}{c|}{}      & \multicolumn{3}{c|}{k=1} & \multicolumn{3}{c|}{k=2} & \multicolumn{3}{c}{k=3} \\
                         grid                  & $128$&$256$&$512$   & $128$&$256$&$512$  & $128$&$256$&$512$  & $128$&$256$& $512$    \\\hline
                         $\nlev$ = 2           & 0.72 & 3.28 & 22.1        & 0.17 & 0.64 & 2.79       & 0.34 & 1.37 & 5.36       & 0.72 & 2.63 & 10.5 \\
                         $\nlev$ = 3           & 0.75 & 3.84 & 25.1        & 0.21 & 0.80 & 4.01       & 0.45 & 1.68 & 6.82       & 0.97 & 3.72 & 15.1 \\
                         $\nlev$ = 4           & 0.82 & 4.22 & 26.8        & 0.25 & 1.00 & 4.68       & 0.58 & 2.11 & 8.24       & 1.26 & 4.76 & 19.3 \\
                         $\nlev$ = 5           & 0.88 & 4.65 & 28.1        & 0.29 & 1.19 & 5.46       & 0.67 & 2.53 & 10.4       & 1.57 & 5.95 & 24.2 \\
                         $\nlev$ = 0          & 0.09 & 0.29 & 1.13        & 0.03 & 0.10 & 0.38       & 0.07 & 0.22 & 0.83       & 0.13 & 0.42 & 1.77 \\
    \end{tabular}
    \caption{Preprocessing phases required for $h$-multigrid computations as compared to single grid computations ($\nlev = 0$). 
             CPU times for generation of $h$-coarsened mesh sequences (grid topology computation task),   
             orthogonalization of shape functions and computation of intergrid operators, see text for details. \label{tab:cpu_pre}} 
\end{table} 
To conclude we also report preprocessing CPU times including generation of $h$-coarsened mesh sequence
and orthogonalization of shape functions together with computation and storage of intergrid transfer operators.
In Table \ref{tab:cpu_pre} it is possible to appreciate that both these operations are time consuming as compared 
to setup times of single grid computations. Nevertheless for $k>1$ dG discretizations, even considering preprocessing times, multigrid outperforms single grid solvers.

\subsubsection{3D Poisson problem}
\label{sec:poisson3D}
The Poisson problem in three space dimensions is here considered to assess the performance
of the $h$-multigrid solution strategy in parallel computations.
The $128^3$ hexahedral mesh is first partitioned and distributed across the processes, thus
each process build an $h$-coarsened mesh sequence of its own partition.
This practice is optimal from the distribution of computational load viewpoint
but requires ad hoc strategies to deal with agglomerated elements whose faces are shared between partitions.
While in single grid dG solvers the replication of a single layer of cells (the so called \emph{ghost cells})
across partition boundaries ensures that the \emph{stencil} of the discretization is fully accessible,
the definition of ghost agglomerated cells is more tricky. 
Depending on the number of grid levels and the shape of agglomerated elements, 
many layers of cells of the fine grid might be involved in the process.

Nevertheless, it should be remarked that only non-inherited multigrid requires
to numerically integrate bilinear forms over internal faces located on partition boundaries
and, contextually, access the discretization stencil at all the mesh levels.
As opposite, whenever coarse grid operators are defined restricting the fine grid operator, 
like in inherited and rescaled-inherited multigrid, only intergrid operators associated to ghost cells are required.
In this regard the design decision of storing intergrid transfer operators in preprocessing is very handy, see Section \ref{sec:intergrid}.
Indeed, intergrid transfer operators associated to ghost cells
can be communicated across partitions without needing to actually build ghost cells, only adjacencies informations are required.
As a consequence the implementation of inherited and rescaled-inherited multigrid is simpler in parallel.

\begin{table}[!htbp]
\small
\centering
    \begin{tabular}{@{}l|ccc|ccc|ccc@{}} 
                                \multicolumn{10}{c}{Linear solver iterations, 3D Poisson problem in parallel} \\ \hline
 k                            & \multicolumn{3}{c|}{1}         & \multicolumn{3}{c|}{2        } & \multicolumn{3}{c}{3} \\
 processes                    & 8       &  16        & 32      & 16      &  32        & 64      & 32       &  64         & 128   \\\hline
 FGMRES $\mathrm{MG}_{\mathcal{V}}$  $\nlev$ =  2      & 11      &  11        & 11      & 10      &  10        & 10      & 9        &  9          & 10    \\
 FGMRES $\mathrm{MG}_{\mathcal{V}}$  $\nlev$ =  3      & 12      &  12        & 12      & 11      &  11        & 11      & 10       &  10         & 11    \\
 FGMRES $\mathrm{MG}_{\mathcal{V}}$  $\nlev$ =  4      & 12      &  12        & 12      & 11      &  11        & 11      & 10       &  11         & 11    \\
 CG BJACOBI(ILU)              & 488     &  509       & 530     & 680     &  708       & 687     & 567      &  856        & 898   \\
 GMRES ASM(1,ILU)             & 432     &  467       & 455     & 675     &  699       & 614     & 868      &  794        & 745   \\
 \hline
    \end{tabular}
    \caption{BR2 dG discretization of the 3D model Poisson problem \eqref{eq:lap}-\eqref{eq:laptest2}.
             Number of iterations required by a $h$-multigrid preconditioned FMGMRES solver and block-ILU preconditioned single grid solvers, see text for details.
	     Linear system relative residual tolerance is $10^{-10}$. \label{tab:it_multvssingle_3d}} 
\end{table} 
We consider the rescaled-inherited strategy for defining coarse grid operators which has demonstrated 
to provide uniform convergence with respect to the number of levels and affordable assembly times, see Section \ref{sec:poisson2D}.
Beside the baseline computations performed with 8, 16 and 32 
processes at first, second and third polynomial degree, respectively, 
we double the number of processes two times for a total of three runs at each polynomial degree.
Thanks to the use of an ASM preconditioner for the GMRES smoother the number of FGMRES iterations is
independent from the number of processes, see Table \ref{tab:it_multvssingle_3d}.
The single grid GMRES solver uses the same kind of ASM preconditioner employed by the GMRES smoothers,
that is an ASM preconditioner with one level of overlap between the sub-domains and an ILU decomposition 
in each sub-domain matrix. The single grid CG solver employs a block-Jacobi preconditioner 
with an ILU decomposition in each sub-domain matrix (but no overlap between sub-domains).

\begin{table}[!htbp]
\small
\centering
    \begin{tabular}{@{}l|ccc|ccc|ccc@{}} \hline
                         CPU time (s)         & \multicolumn{3}{c|}{solution} & \multicolumn{3}{c|}{assembly} & \multicolumn{3}{c}{total}\\
                         $\mathbf{k=1}$ & \multicolumn{3}{c|}{}& \multicolumn{3}{c|}{}& \multicolumn{3}{c}{}\\
                         processes             & 8        & 16    & 32        & 8       &   16   & 32      & 8     &   16  & 32  \\\hline
                         FGMRES $\mathrm{MG}_{\mathcal{V}}$ $\nlev$ = 2 & 12.2    &  5.94  & 5.11      & 28.3    &  13.2  & 6.26    & 40.5  &  19.2 & 11.4    \\
                         FGMRES $\mathrm{MG}_{\mathcal{V}}$ $\nlev$ = 3 & 11.6    &  5.85  & 3.12      & 28.8    &  12.8  & 6.72    & 40.4  &  18.7 & 9.84    \\
                         FGMRES $\mathrm{MG}_{\mathcal{V}}$ $\nlev$ = 4 & 11.6    &  5.82  & 3.26      & 28.8    &  12.9  & 6.27    & 37.4  &  18.7 & 9.53    \\
                         CG BJACOBI(ILU)       & 55.7    &  29.3  & 15.1      & 25.6    &  12.8  & 6.71    & 88.4  &  42.2 & 21.8    \\
			 GMRES(120) ASM(1,ILU) & 144     &  80.8  & 41.3      & 25.9    &  12.9  & 6.74    & 170   &  93.8 & 48.0    \\
                         $\mathbf{k=2}$ & \multicolumn{3}{c|}{}& \multicolumn{3}{c|}{}& \multicolumn{3}{c}{}\\
                         processes             & 16    & 32   & 64     &   16   & 32    & 64        &   16  & 32   & 64        \\\hline
                         FGMRES $\mathrm{MG}_{\mathcal{V}}$ $\nlev$ = 2 &  32.7 & 26.1 & 17.9   &  45.6  & 23.3  & 11.8      &  78.3 & 49.3 & 29.8      \\
                         FGMRES $\mathrm{MG}_{\mathcal{V}}$ $\nlev$ = 3 &  26.4 & 14.3 & 7.95   &  47.2  & 24.1  & 11.8      &  73.6 & 38.4 & 19.7       \\
                         FGMRES $\mathrm{MG}_{\mathcal{V}}$ $\nlev$ = 4 &  25.8 & 14.2 & 7.92   &  48.2  & 24.4  & 12.5      &  74.0 & 38.6 & 20.4       \\
                         CG BJACOBI(ILU)       &  224  & 109  & 54.6   &  45.4  & 22.9  & 11.4      &  269  & 132  & 66.1       \\
			 GMRES(120) ASM(1,ILU) &  411  & 237  & 100    &  45.4  & 23.2  & 11.4      &  457  & 261  & 111       \\
                         $\mathbf{k=3}$ & \multicolumn{3}{c|}{}& \multicolumn{3}{c|}{}& \multicolumn{3}{c}{}\\
                         processes             & 32     & 64     & 128      & 32    &  64    &  128        & 32     &   64   &  128   \\\hline
                         FGMRES $\mathrm{MG}_{\mathcal{V}}$ $\nlev$ = 2 & 111    & 54.2   & 36.9     & 98.4  &  49.6  &  25.1       & 209    &   104  &  62.0  \\
                         FGMRES $\mathrm{MG}_{\mathcal{V}}$ $\nlev$ = 3 & 60.9   & 24.1   & 15.1     & 101   &  50.2  &  26.1       & 162    &   74.2 &  41.2  \\
                         FGMRES $\mathrm{MG}_{\mathcal{V}}$ $\nlev$ = 4 & 53.8   & 30.2   & 14.7     & 103   &  51.9  &  26.4       & 157    &   81.4 &  41.1  \\
                         CG BJACOBI(ILU)       & 472    & 305    & 172      & 94.5  &  46.5  &  23.5       & 567    &   352  &  195   \\
			 GMRES(120) ASM(1,ILU) & 729    & 386    & 236      & 94.6  &  47.3  &  24.1       & 824    &   433  &  262   \\
    \end{tabular}
    \caption{BR2 dG discretization of the 3D model Poisson problem \eqref{eq:lap}-\eqref{eq:laptest2}.
             Solution, assembly and total CPU times (total means assembly plus solution times) required to reach a relative residual tolerance of $10^{-10}$,
             see text for details.\label{tab:cpu_tot_3d}} 
\end{table} 
The CPU times reported in Table \ref{tab:cpu_tot_3d} demonstrate that the $h$-multigrid efficiency 
does not deteriorate increasing the number of processes. The gains with respect to single grid solvers
are comparable to those observed in serial computations in two space dimensions, see Section \ref{sec:poisson2D}. 
Similarly, a sufficient number of grid levels must be employed to ensure that 
the grid on level $L$ is coarse enough, note the poor performance for $L=2$. 
Even if a scalability analysis would require to further increase the number of processes 
we observe a strong linear scaling, that is the computation times halves doubling the number of processes.

\begin{table}[!htbp]
\centering
    \begin{tabular}{@{}l|ccccc@{}} \hline
                         CPU time (s)          &\multicolumn{5}{c}{grid topology computation} \\ \cline{1-6}
                                               & \multicolumn{5}{c}{}             \\
                         processes        &  8   &  16  &  32  &  64  &  128    \\\hline
                         $\nlev$ = 2      & 31.3 & 17.1 & 7.15 & 2.88 & 1.77     \\
                         $\nlev$ = 3      & 35.2 & 19.4 & 8.94 & 3.65 & 2.13     \\
                         $\nlev$ = 4      & 41.7 & 22.5 & 10.2 & 4.90 & 2.52     \\
                         $\nlev$ = 0      & 6.05 & 4.91 & 1.71 & 0.75 & 0.45     \\
    \end{tabular}
    \caption{Preprocessing phases required for $h$-multigrid computations as compared to single grid computations ($\nlev = 0$). 
             CPU times for generation of $h$-coarsened mesh sequences (grid topology computation task), see text for details. \label{tab:cpu_pre_3d}} 
\end{table} 
\begin{table}[!htbp]
\centering
    \begin{tabular}{@{}l|ccc|ccc|ccc@{}} \hline
                         CPU time (s)     & \multicolumn{9}{c}{orthonormalization and intergrid operators}\\ \cline{1-10}
                                          & \multicolumn{3}{c|}{k=1} & \multicolumn{3}{c|}{k=2} & \multicolumn{3}{c}{k=3} \\
                         processes        & 8    &  16  & 32         &  16  & 32   & 64         & 32   & 64   & 128 \\\hline
                         $\nlev$ = 2      & 5.58 & 2.78 & 1.51       & 9.65 & 4.95 & 2.88       & 24.4 & 12.3 & 6.69 \\
                         $\nlev$ = 3      & 7.75 & 3.88 & 2.02       & 13.9 & 7.73 & 3.65       & 37.6 & 19.7 & 10.7 \\
                         $\nlev$ = 4      & 11.5 & 5.81 & 2.95       & 22.8 & 9.85 & 5.58       & 61.9 & 31.1 & 18.4 \\
                         $\nlev$ = 0      & 1.03 & 0.52 & 0.26       & 1.88 & 0.96 & 0.48       & 4.06 & 2.04 & 1.11 \\
    \end{tabular}
    \caption{Preprocessing phases required for $h$-multigrid computations as compared to single grid computations ($\nlev = 0$). 
             CPU times for orthogonalization of shape functions and computation of intergrid operators, see text for details. \label{tab:cpu_pre_3d_2}} 
\end{table} 
To demonstrate that also the preprocessing phase is scalable, in Table \ref{tab:cpu_pre_3d} and Table \ref{tab:cpu_pre_3d_2} 
we report CPU times for the generation of $h$-coarsened mesh sequence
and orthogonalization of shape functions together with computation and storage of intergrid transfer operators, respectively.
Remarkably, since the problem size has increased as compared to 2D computations, 
multigrid outperforms single grid solvers in terms of overall computation time (that is considering assembly, solution and preprocessing CPU times)
at all polynomials degrees.

\begin{table}
\small
\centering
    \begin{tabular}{@{}l|ccc|ccc|ccc@{}} 
      $\mathrm{MG}_{\mathcal{V}}(k)$ vs CG$(k)$ & \multicolumn{9}{c}{Total CPU time speedup, 2M hex elems grid }\\ \cline{1-10}
      degree               & \multicolumn{3}{c|}{$k=1$} & \multicolumn{3}{c|}{$k=2$} & \multicolumn{3}{c}{$k=3$} \\
      processes            & 8    &  16   & 32         &  16  & 32   & 64         & 32   & 64   & 128 \\\hline
      CG/MG step time      & 2.4  &  2.3  & 2.3        &  3.6  & 3.4   & 3.2       & 3.6  & 4.3 &4.7 \\
    \end{tabular}
    \caption{BR2 dG discretization of the 3D model Poisson problem \eqref{eq:lap}-\eqref{eq:laptest2}.
             Comparison of total CPU times (assembly plus solution times) required to reach a relative residual tolerance of $10^{-10}$ with Conjugate Gradient and $h$-multigrid ($L=4$),
             see text for details. \label{tab:cpu_comp_BR23D}} 
\end{table}
\begin{table}
\small
\centering
  \begin{tabular}{@{}l|c|c|c@{}} 
    CG($k_{\mathrm{CG}}$) vs $\mathrm{MG}_{\mathcal{V}}(k_{\mathrm{MG}})$ &\multicolumn{3}{c}{Total CPU time ratio}  \\ \cline{1-4}   
    solver(degree)          & \multicolumn{1}{c|}{CG(1)/$\mathrm{MG}_{\mathcal{V}}$(2)} & \multicolumn{1}{c|}{CG(2)/$\mathrm{MG}_{\mathcal{V}}$(3)} & \multicolumn{1}{c}{CG(1)/$\mathrm{MG}_{\mathcal{V}}$(3)}   \\
    CG/MG step time     & 1.1    &   1.6   & 0.53\\ \hline
 \end{tabular}
    \caption{BR2 dG discretization of the 3D model Poisson problem \eqref{eq:lap}-\eqref{eq:laptest2}.
             Comparison of total CPU times (assembly plus solution times) required to reach a relative residual tolerance of $10^{-10}$ with Conjugate Gradient and $h$-multigrid.
	     The multigrid solution strategy is applied to an higher polynomial degree dG discretization ($k_{\mathrm{MG}}>k_{\mathrm{CG}}$).
             \label{tab:cpu_comp_kBR23D}} 
\end{table}  
The gains with respect to the best performing single grid solver (the preconditioned Conjugate Gradient iteration)
are on pair with those observed in 2D computations and do not deteriorate increasing the number of processes, see Table \ref{tab:cpu_comp_BR23D}.
Similarly to the 2D case solving a second degree polynomial degree BR2 dG discretization with $h$-multigrid 
is comparable to solving a first degree dG discretization with CG
and the time required for solving a third polynomial degree BR2 dG discretization 
with $h$-multigrid is twice the time required for solving 
a first degree dG discretization with Conjugate Gradient, see Table \ref{tab:cpu_comp_kBR2}.

\subsection{Stokes dG discretization}
\label{sec:numresStokes}
In this section we tackle the solution of
a model Stokes problem discretized by means of the dG formulation in \eqref{discr:stokes}.
The computational domain is the bi-unit square, $\Omega = [-1,1]^2$,
and we impose Dirichlet boundary conditions on $\partial \Omega$ according to the following smooth analytical solution
\begin{align}
  \label{eq:laptest2}
   \mathbf{u} &= \left[- e^x \, (y \, \cos(y) + \sin(y)) \, \mathbf{i}, e^x \, (y \, \sin(y)) \, \mathbf{j}\right],\nonumber\\
   p &= 2 \, e^x \, \sin(y) , \nonumber
\end{align}
In order to investigate the growth of computational costs while increasing the mesh size,
solutions are computed on four uniform quadrilateral elements meshes of size $(64 \cdot 2^n)^2, n=\{0,1,2,3\}$
and four distorted and graded triangular meshes of size $2(32 \cdot 2^n)^2, n=\{0,1,2,3\}$, see Figure \ref{fig:grid_tri}.
We check the influence of raising the polynomial degree and the number of coarse levels
considering $k=\{1,2,3\}$ and $L=\{2,3,4,5\}$.  
The number of mesh elements at each level 
$\ell$ of the stack of grids is reported in Table \eqref{tab:gridAgglo2d}.

For the sake of comparison we consider the $h$-multigrid V-cycle preconditioned FGMRES iteration 
and the Pressure Schur Complement preconditioned Richardson iteration 
of Section \ref{sec:precINS}. 
The relevant solvers options are summarized in what follows.

\paragraph{$\mathrm{MG}_\mathcal{V}$ preconditioned FGMRES solver}
One iteration of $\mathrm{MG}_\mathcal{V}$ cycle is used as a preconditioner for the FMGRES(60) iteration.  
On quadrilateral meshes high-order modes of the error are smoothed with a single iteration of a right ILU preconditioned GMRES solver 
while on distorted and graded triangular meshes we consider one and two smoothing iterations.
On the coarsest level $L$ we employ the same solver but, instead of fixing the iteration number, 
we impose a four order of magnitude decrease of the relative residual norm, that is $\frac{\| f_L - A^{\mathrm{Stk}}_L \bar{w}_L \|}{\| f_L \|} \leq 10^{-4}$.
The FGMRES iteration is forced to reach tight relative residual tolerance, in particular
the linear system solution converges in $N_\mathrm{it}$ iterations
if at the $i$-th iterate $\| \hat{r}_0^i \| = \frac{\| f_0 - A_0^{\mathrm{Stk}} \bar{w}_0^i \|}{\| f_0 \|} \leq 10^{-12}$.
\paragraph{Pressure Schur Complement $\mathrm{MG}_\mathcal{V}$ preconditioned Richardson solver}
To approximatively invert the discrete vector Laplace operator $\Amat_0$ 
and the pressure Schur complement $\widetilde{\Smat}_0$ appearing in the block factorization of $\Amat_0^{\mathrm{Stk}}$, see Section \ref{sec:precINS}, 
we employ a FGMRES(60) and a GMRES(60) solver, respectively.
We set a two order of magnitude decrease of the relative residual norm and limit the maximum number of iterations to 2 and 40, respectively.
The pressure Schur Complement GMRES solver is preconditioned with 
an ILU decomposition of the operator $\widehat{\Smat}_0$ in \eqref{eq:presSPrec}. 
The FMGRES solver acting on the discrete Laplace operator is preconditioned with one and two iteration of multigrid V-cycle 
on quadrilateral and triangular grids, respectively. Smoothing options are the same of Section \ref{sec:numresBR2}. 
The Richardson iteration is forced to reach a relative residual tolerance of $10^{-12}$.

We compare the solvers on the basis of convergence rate and computation time
and we compare execution times with a Lower Upper (LU) decomposition direct solver and the ILU preconditioned GMRES(200) solver.
All the numerical results have been computed by exploiting the PCMG multigrid 
preconditioner and the PCFIELDSPLIT block preconditioner frameworks available in the \texttt{PETSc} library~\cite{petsc-web-page,petsc-user-ref}.
Provided that the iterative solver's convergence criterion is satisfied and the LU factorisation is computed without running out of memory, 
the same error with respect to the exact solution is measured. 
For the $k=3$ dG discretization on the $128^2$ quadrilateral grid
the $L^2$ error norm is on the order of $10^{-10}$ and $10^{-8}$ for velocity and pressure, respectively.

The number of iteration reported in Table \ref{tab:it_multasprec_stokes_quad} and Table \ref{tab:it_multasprec_stokes_tri} confirm
uniform converge with respect to the number of levels for multigrid preconditioned solvers, both on quadrilateral and triangular mesh sequences.
Nevertheless, while on uniform quadrilateral elements grids the convergence is grid independent,
on distorted and graded triangular meshes the number of iterations increases on finer grids. 
Moreover, only on quadrilateral elements meshes increasing the polynomial degree entails less iterations.
This can be better appreciated by inspecting the average residual decrease or \emph{convergence factor} 
$$ \rho = exp\left(\frac{1}{N_{\mathrm{it}}} ln \frac{\| r_0 \|}{\| r_{N_\mathrm{it}}\|}\right)$$
reported in Table \ref{tab:conv_multasprec_stokes}.
Convergence failure after 2000 ILU preconditioned GMRES iterations reads $2000^*$ in Tables \ref{tab:it_multasprec_stokes_quad}-\ref{tab:it_multasprec_stokes_tri}.

Looking at the wall clock times (solution times plus assembly times) reported in Table \ref{tab:cpu_multasprec_stokes_quad} 
it is clear that both multigrid preconditioned iterative solvers
yield significant execution times gains with respect to direct solver on the quadrilateral mesh sequence.
Since we get a four-to-five-fold increase 
of the total computation time with a four-fold increase of the mesh size 
at all the polynomials degrees, optimal multigrid efficiency is approached. 
Note that in case of the $h$-multigrid preconditioned FGMRES solver the number of levels 
$L$ must be chosen large enough because of the poor performance of the coarse grid GMRES solver, even on relatively coarse meshes.
In this regard uniform convergence with respect to the number of levels is highly beneficial.
 
\begin{table}[!htbp]
\small
\centering
    \begin{tabular}{@{}l|cccc|cccc|c@{}} 
                                \multicolumn{10}{c}{Linear solver iterations, 2D Stokes problem, quadrilateral mesh sequence} \\ \hline
                   solver           & \multicolumn{4}{c|}{FGMRES $\mathrm{MG}_{\mathcal{V}}$} & \multicolumn{4}{c|}{SchurCompl $\mathrm{MG}_{\mathcal{V}}$} & \multicolumn{1}{c}{GMRES ILU}\\
                   $\nlev$          &  2 &   3   &   4 &    5  &  2 &   3   &   4 &    5  & 0 \\
		   \cline{1-10}                                                                                              
                    grid &   \multicolumn{9}{c}{$\mathbf{k=1}$} \\
		   \cline{2-10}
                    $64$       & 25      &  26        & 27      & 27       &  11       & 11       & 11      &  11   &  1013 \\
                    $128$      & 25      &  27        & 27      & 27       &  11       & 12       & 11      &  11   &  2000$^*$\\
                    $256$      & 26      &  27        & 28      & 28       &  11       & 12       & 12      &  12   &  2000$^*$\\
                    $512$      & 27      &  27        & 28      & 28       &  11       & 12       & 12      &  12   &  2000$^*$\\
		   \cline{2-10}                                                                                              
                     &   \multicolumn{9}{c}{$\mathbf{k=2}$} \\
		   \cline{2-10}
                    $64$       & 16      &  16        & 17      &  17       & 10       & 10       & 10       &  10   & 531         \\
                    $128$      & 16      &  16        & 17      &  17       & 10       & 10       & 10       &  10   & 2000$^*$ \\
                    $256$      & 16      &  16        & 17      &  17       & 10       & 10       & 10       &  10   & 2000$^*$ \\
                    $512$      & 16      &  16        & 17      &  17       & 10       & 10       & 10       &  10   & 2000$^*$ \\
		   \cline{2-10}                                                                                              
                     &   \multicolumn{9}{c}{$\mathbf{k=3}$} \\
		   \cline{2-10}
                    $64$    & 13      &  14        & 14      & 14        &  10      & 10       & 10       &  10   & 597 \\
                    $128$   & 13      &  14        & 14      & 14        &  10      & 10       & 10       &  10   & 2000$^*$ \\
                    $256$   & 13      &  14        & 14      & 14        &  10      & 10       & 10       &  10   & 2000$^*$ \\
                    $512$   & 18      &  14        & 14      & 14        &  10      & 10       & 10       &  10   & 2000$^*$ \\
		   \cline{2-10}
		   \hline
    \end{tabular}
    \caption{Comparison of the number of iterations required to solve a 2D model Stokes problem, see text for details.
	     Linear system relative residual tolerance is $10^{-12}$. \label{tab:it_multasprec_stokes_quad}}
\end{table} 

\begin{table}[!htbp]
\small
\centering
    \begin{tabular}{@{}l|cccc|cccc|cccc|c@{}} 
                                \multicolumn{14}{c}{Linear solver iterations, 2D Stokes problem, triangular mesh sequence} \\ \hline
                   solver           & \multicolumn{8}{c|}{FGMRES $\mathrm{MG}_{\mathcal{V}}$} & 
		                      \multicolumn{4}{c|}{SchurCompl $\mathrm{MG}_{\mathcal{V}}$} & 
				      \multicolumn{1}{c}{GMRES}\\
		   \cline{2-14}                                                                                              
                   $SM$ it     & \multicolumn{4}{c|}{1} & \multicolumn{4}{c|}{2} &\multicolumn{4}{c|}{2} & \\
                   $\nlev$          &  2 &   3   &   4 &    5  &  2 &   3   &   4 &    5  &  2 &   3   &   4 &    5  & 0 \\
		   \cline{1-14}                                                                                              
                    grid &   \multicolumn{13}{c}{$\mathbf{k=1}$} \\
		   \cline{2-14}
                   $32$        & 35      &  35        & 35      & 35       &  20       & 20       & 20      &  20   & 28 & 29 & 29 & 30    & 464   \\
                   $64$        & 39      &  39        & 39      & 39       &  23       & 23       & 23      &  23   & 44 & 45 & 45 & 46    & 1433   \\
                   $128$       & 46      &  47        & 47      & 47       &  26       & 26       & 26      &  26   & 70 & 70 & 70 & 71    & 2000$^*$   \\
                   $256$       & 51      &  51        & 51      & 51       &  29       & 29       & 29      &  29   & 86 & 86 & 86 & 86    & 2000$^*$   \\
		   \cline{2-14}                                                                                               
                     &   \multicolumn{13}{c}{$\mathbf{k=2}$} \\
		   \cline{2-14}
                   $32$       & 29      &  29        & 29      &  29       & 16       & 16       & 16       &  16   & 29 & 29 & 29 & 29   & 374    \\
                   $64$       & 36      &  36        & 36      &  36       & 22       & 22       & 22       &  22   & 45 & 45 & 45 & 45   & 1526    \\
                   $128$      & 55      &  55        & 55      &  55       & 28       & 28       & 28       &  28   & 81 & 80 & 81 & 81   & 2000$^*$     \\
                   $256$      & 73      &  70        & 75      &  76       & 41       & 41       & 41       &  41   & 64 & 63 & 66 & 65   & 2000$^*$     \\
		   \cline{2-14}                                                                                                
                     &   \multicolumn{13}{c}{$\mathbf{k=3}$} \\
		   \cline{2-14}
                  $32$      & 27      &  27        & 27      & 27        &  14      & 14       & 14       &  14   &  39  & 39 & 39 & 39    & 437    \\
                  $64$      & 36      &  36        & 36      & 36        &  19      & 19       & 19       &  19   &  57  & 57 & 57 & 57    & 1863    \\
                  $128$     & 44      &  44        & 44      & 44        &  26      & 26       & 26       &  26   &  72  & 72 & 72 & 72    & 2000$^*$     \\
                  $256$     & 63      &  63        & 63      & 63        &  40      & 40       & 40       &  40   &  119 & 119 & 119 & 120 & 2000$^*$     \\
		   \cline{2-14}
		   \hline
    \end{tabular}
    \caption{Comparison of the number of iterations required to solve a 2D model Stokes problem, see text for details.
	     One and two smoothing iterations ($SM$ it) are considered for triangular meshes to improve the performance of multigrid preconditioners.
	     Linear system relative residual tolerance is $10^{-12}$. \label{tab:it_multasprec_stokes_tri}}
\end{table} 
\begin{table}[!htbp]
\small
\centering
    \begin{tabular}{@{}l|cc|cc|cc|cc@{}} 
                                \multicolumn{9}{c}{Convergence factor $\rho$, 2D Stokes problem} \\ \hline
                   grid      & \multicolumn{4}{c|}{quadrilateral meshes} & \multicolumn{4}{c}{(triangular meshes)} \\
		   \cline{2-9}                                                                                   
                   solver      & \multicolumn{2}{c|}{FGMRES $\mathrm{MG}_{\mathcal{V}}$} & \multicolumn{2}{c|}{SchurCompl $\mathrm{MG}_{\mathcal{V}}$} 
                               & \multicolumn{2}{c|}{FGMRES $\mathrm{MG}_{\mathcal{V}}$} & \multicolumn{2}{c}{SchurCompl $\mathrm{MG}_{\mathcal{V}}$}\\
                   $SM$ it     & \multicolumn{2}{c|}{1} & \multicolumn{2}{c|}{1} &\multicolumn{2}{c|}{1} & \multicolumn{2}{c}{2}\\
                   $\nlev$             &  2    &    5      &  2         &    5      &   2       &    5      &  2         &    5 \\
		   \cline{1-9}                                                                                   
                                                                                                                              grid size &   \multicolumn{8}{c}{$\mathbf{k=1}$} \\
		   \cline{2-9}                                                                                   
                    $64$  $\;\,(32)$     & .324  & .349       &  .070     &  .075     & .456      & .456       &  .372     &  .387\\
                    $128$ $(64)$       & .328  & .354       &  .078     &  .079     & .497      & .498       &  .536     &  .544\\
                    $256$ $(128)$      & .337  & .364       &  .078     &  .089     & .554      & .556       &  .676     &  .679\\
                    $512$ $(256)$      & .348  & .370       &  .079     &  .087     & .583      & .584       &  .728     &  .727\\
		   \cline{2-9}                                                                                   
                                                                                                                                  &   \multicolumn{8}{c}{$\mathbf{k=2}$} \\
		   \cline{2-9}                                                                                          
                    $64$  $\;\,(32 )$    & .162  &  .186      &  .053     &  .055     & .382      &  .383      &  .383     &  .383 \\
                    $128$ $(64 )$      & .165  &  .193      &  .052     &  .053     & .470      &  .471      &  .544     &  .543 \\
                    $256$ $(128)$      & .165  &  .196      &  .055     &  .056     & .608      &  .606      &  .711     &  .712 \\
                    $512$ $(256)$      & .167  &  .196      &  .051     &  .054     & .687      &  .695      &  .651     &  .658 \\
		   \cline{2-9}                                                                                   
                                                                                                                                  &   \multicolumn{8}{c}{$\mathbf{k=3}$} \\
		   \cline{2-9}                                                                                          
                    $64$  $\;\,(32 )$    & .116  & .124       &  .060     &  .061     & .349      & .349       &  .495     &  .494   \\
                    $128$ $(64 )$      & .117  & .127       &  .051     &  .051     & .464      & .465       &  .616     &  .617   \\
                    $256$ $(128)$      & .118  & .129       &  .053     &  .051     & .539      & .540       &  .682     &  .682   \\
                    $512$ $(256)$      & .205  & .130       &  .051     &  .051     & .648      & .648       &  .792     &  .794   \\
		   \cline{2-9}
		   \hline
    \end{tabular}
    \caption{2D model Stokes problem. Convergence factors of a FGMRES solver preconditioned with $h$-multigrid and
             a Richardson solver with a Pressure Schur Complement Block preconditioner, see text for details. 
	     Quadrilateral and distorted triangular mesh sequences.
	     One and two smoothing iterations ($SM$ it) are considered for triangular meshes to improve the performance of multigrid preconditioners.
	     Linear system relative residual tolerance is $10^{-12}$. \label{tab:conv_multasprec_stokes}}
\end{table} 

The wall clock times measured on the triangular mesh sequence and reported in Table \ref{tab:cpu_multasprec_stokes_tri}
demonstrate that only the preconditioned FGMRES solver allows to bit direct solvers.
The performance of the Schur Complement block preconditioner is hit by the poor performance of the Schur complement subsolver
which fails to lower the residual by two orders of magnitude in 40 iterations
(we verified that increasing the maximum number of iteration beyond 40 is not beneficial in terms of execution times).
As opposite the multigrid preconditioned FGMRES solver employed for the Laplace 
operator performs fairly well on distorted triangular meshes, see Table \ref{tab:it_multasprec_unstr}.
Also the performance of $h$-multigrid FGMRES degrade as compare to quadrilateral elements meshes: 
we observe a six-fold increase of the total computation time with a four-fold increase of the mesh size.
Worsening of convergence factors is awaited given that the number of stretched triangles increases 
and their aspect ratios worsen when the mesh is refined, see Figure \ref{fig:grid_tri}.

\begin{table}[!htbp]
\small
\centering
    \begin{tabular}{@{}l|cccc|cccc|c|c@{}} 
                                \multicolumn{11}{c}{Total CPU time (s), 2D Stokes problem, quadrilateral mesh sequence} \\ \hline
                   solver           & \multicolumn{4}{c|}{FGMRES $\mathrm{MG}_{\mathcal{V}}$} & \multicolumn{4}{c|}{SchurCompl $\mathrm{MG}_{\mathcal{V}}$} & GMRES & LU\\
                   $\nlev$          &  2 &   3   &   4 &    5  &  2 &   3   &   4 &    5 & 0 & 0 \\
		   \cline{1-11}                                                                                              
                    grid &   \multicolumn{10}{c}{$\mathbf{k=1}$} \\
		   \cline{2-11}
                    $64$    & 1.03      & 0.91        & 0.91      & 0.89      &  1.61   & 2.0    & 1.86   & 1.92 &  11.1    & 1.33  \\
                    $128$   & 5.07      & 3.87        & 3.83      & 3.89      &  7.57   & 8.42   & 7.62   & 7.58 &  90.4*   & 10.7  \\
                    $256$   & 42.2      & 17.6        & 16.8      & 16.8      &  31.0   & 36.3   & 35.4   & 37.0 &  349*    & 84.0  \\
                    $512$   & 480       & 99.4        & 75.9      & 75.7      &  137    & 151    & 153    & 149  &  1278*   & 691  \\
		   \cline{2-11}                                                                                              
                    &   \multicolumn{10}{c}{$\mathbf{k=2}$} \\
		   \cline{2-11}
                   $64$    & 2.81    & 1.84    & 1.83   & 1.89    & 5.86 & 5.97 & 6.1   & 6.22    & 11.2 & 6.06  \\    
                   $128$   &  13.76   &  7.86   &  7.78  &  7.83   & 25.7 & 27.1 & 28.6  & 28.0   & 168* & 49.9  \\     
                   $256$   & 113     & 39.9    & 33.1   & 32.7   & 114  & 110  & 109   & 112    & 713* & 409  \\
                   $512$   &  977   &  288    &  172   &  159    &  497 &  494 &  485  &  547  & 2690* &   \\
   
		   \cline{2-11}                                                                                              
                    &   \multicolumn{10}{c}{$\mathbf{k=3}$} \\
		   \cline{2-11}
                   $64$    & 6.19   & 4.47    & 4.44   & 4.51  &  18.0  &  18.1 &  18.5 &  18.6   & 30.2 & 17.2   \\
                   $128$   & 37.2  &  20.8    & 18.6   & 18.2  & 79.8   & 82.4   & 85.6   & 81.8   &  365* & 146\\
                   $256$   & 301    & 103     & 77.8   & 74.8   & 326    & 325   & 325   & 324   & 1533* &   \\
                   $512$   & 3012   & 734     & 381    & 347   & 1419   & 1360  & 1355  & 1350  & 5973* &   \\
		   \cline{1-11}
    \end{tabular}
    \caption{Comparison of wall clock time (solution plus assembly times) required to solve a 2D model Stokes problem, see text for details. 
	     Linear system relative residual tolerance is $10^{-12}$. \label{tab:cpu_multasprec_stokes_quad}}
\end{table} 

\begin{table}[!htbp]
\small
\centering
    \begin{tabular}{@{}l|cccc|cccc|c|c@{}} 
                                \multicolumn{11}{c}{Total CPU time (s), 2D Stokes problem, triangular mesh sequence} \\ \hline
                   solver           & \multicolumn{4}{c|}{FGMRES $\mathrm{MG}_{\mathcal{V}}$} & \multicolumn{4}{c|}{SchurCompl $\mathrm{MG}_{\mathcal{V}}$} & GMRES & LU\\
                   $\nlev$          &  2 &   3   &   4 &    5  &  2 &   3   &   4 &    5 & 0 & 0 \\
		   \cline{1-11}                                                                                              
                    grid &   \multicolumn{10}{c}{$\mathbf{k=1}$} \\
		   \cline{2-11}
                   $32$   & 0.42      & 0.42        & 0.42      & 0.42        &  7.63   & 8.33   & 8.75   & 9.33       &  1.70    & 0.26  \\
                   $64$   & 2.08      & 1.93        & 1.90      & 1.94        &  51.0   & 55.1   & 56.1   & 57.7       &  21.8   & 2.22  \\
                   $128$  & 11.1      & 9.04        & 9.10      & 8.74        &  329    & 346    & 356    & 357        &  121*    & 18.9  \\
                   $256$  & 80.2      & 44.5        & 40.3      & 40.6        &  1798   & 1870   & 1895   & 1879       &  522*   & 158  \\
		   \cline{2-11}                                                                                                  
                    &   \multicolumn{10}{c}{$\mathbf{k=2}$} \\
		   \cline{2-11}
                   $32$   & 1.04      & 0.99        & 0.99      & 1.10        & 19.5 & 20.6 & 21.6  & 22.1             & 3.30  & 1.65  \\    
                   $64$   & 6.49      & 5.53        & 5.57      & 5.57        & 136  & 144  & 138   & 137              & 58.6  & 14.9  \\     
                   $128$  & 39.8      & 29.5        & 28.0      & 26.5        & 1000 & 1082 & 1123  & 1098             & 315*  & 129  \\
                   $256$  & 290       & 181         & 164       & 156         & 3433 & 3315 & 3458  & 3397             & 1305* & 1087  \\
   
		   \cline{2-11}                                                                                              
                    &   \multicolumn{10}{c}{$\mathbf{k=3}$} \\
		   \cline{2-11}
                   $32$    & 2.73   & 2.74    & 2.57   & 2.64                 & 70.6   & 72.7   & 73.8  & 74.4            & 9.45  & 4.16   \\
                   $64$    & 17.6   & 14.4    & 14.1   & 13.9                 & 457    & 472    & 480   & 510             & 160   & 39.5   \\
                   $128$   & 116.8  & 77.5    & 72.1   & 71.0                 & 2473   & 2444   & 2459  & 2472            & 689*  & 343    \\
                   $256$   & 973    & 484     & 419    & 434                  & 12112  & 12360  & 12355 & 12350            & 2782* &   \\
		   \cline{1-11}
    \end{tabular}
    \caption{Comparison of wall clock time (solution plus assembly times) required to solve a 2D model Stokes problem, see text for details. 
	     Linear system relative residual tolerance is $10^{-12}$. \label{tab:cpu_multasprec_stokes_tri}}
\end{table} 

The comparison between direct solver and $h$-multigrid FGMRES CPU times proposed in Table \ref{tab:cpu_comp_stk}
confirms that strong gains can be attained by means of multilevel preconditioners, even on  
unstructured meshes composed of stretched and skewed elements.
On fine enough uniform quadrilateral mesh solving a first degree dG discretization with an LU solver is
comparable to solving a third degree dG discretization with a multigrid preconditioned FGMRES solver.
Similarly, on a fine enough distorted triangular grids, $k+1$ and $k$ degree dG 
discretizations are comparable in terms of exacution time if solved with multigrid preconditioned FGMRES and 
direct LU solver, respectively.

Since the multigrid V-cycle preconditioner is the best performing and
is reliable on low quality grids, in the next section the strategy will be applied for solving 
non-linear incompressible flow problems.

\begin{table}
\small
    \begin{tabular}{@{}l|cccc|cccc|ccc@{}} 
      $\mathrm{MG}_{\mathcal{V}}(k)$ vs LU$(k)$ & \multicolumn{9}{c}{Total CPU time speedup}\\ \cline{1-12}
      degree               & \multicolumn{4}{c|}{$k=1$} & \multicolumn{4}{c|}{$k=2$} & \multicolumn{3}{c}{$k=3$} \\
      quad grid & $64$ & $128$ &$256$& $512$ & $64$  & $128$ &$256$&$512$ & $64$ & $128$&$256$             \\
      LU/MG step time &    1.5    &  2.7  & 5  & 9.1 & 3.2 &  6.4  & 12.5   &   &  3.8      & 8  &  \\ \hline
      tri grid & $32$ & $64$ &$128$& $256$   & $32$ & $64$ &$128$&$256$  & $32$ & $64$&$128$             \\
      LU/MG step time &    0.6    &  1.1  & 2.2  & 3.9 & 1.5 &  2.7  & 4.9   & 7       & 1.6  & 2.8 & 4.8 \\ \hline
    \end{tabular}
    \caption{2D model Stokes problem. 
             Comparison of total CPU times (assembly plus solution times) required to solve with a direct solver and  
	     with $h$-multigrid preconditioned FGMRES ($L=5$), see text for details. 
	     Linear system relative residual tolerance is $10^{-12}$ in case of FGMRES. \label{tab:cpu_comp_stk}} 
\end{table}

\subsection{Navier-Stokes dG discretization}
\label{sec:numresNS}
In this section we assess the performance of the multigrid preconditioned FGMRES solver applied to 
repeatedly solve the linearized system of equation in \eqref{eq:nsbe}, as required for advancing in time 
the dG discretization of the incompressible Navier-Sokes equation in \eqref{discr:ns}
by means of the Backward Euler method.

We consider the 2D Kovasznay and 2D-3D Lid-driven cavity problems, admitting a steady state solution at
the Reynolds numbers considered in this work. Thus we tackle a real-life transient hemodynamic application:
we consider the possibility to simulate the blood flow behavior all along the cardiac cycle 
in a 3D cerebral aneurysm geometry reconstructed from medical images.

We remark that the Backward Euler method can be modified to implement a pseudo-transient continuation strategy, $\Psi_{tc}$ see \eg \cite{KK_PsiTC_98},
that can be employed to seek steady state solutions of the incompressible Navier-Stokes equations.
Roughly speaking it is sufficient to omit the while loop in Algorithm \ref{algo:be} 
(which is done for efficiency purposes since accuracy of the time integration is unnecessary)
and introduce a time step adaptation strategy, \eg the Successive Evolution Relaxation Strategy \cite{mulder1985}, 
which allows to progressively enlarge the pseudo time step (starting from a sufficiently small initial guess) 
when the steady state solution is approached. 
$\Psi_{tc}$ is a globalization of Newton method that guarantees convergence 
even when the tentative solution is far from the sought steady state solution. 
Moreover the favourable convergence rates of the Newton method can be exploited
when the pseudo time step is large enough.

\subsubsection{Kovasznay test case}
To assess convergence with respect to the number of levels we consider the 2D Kovasznay problem \cite{Kovasznay48} at Reynolds 40.
Dirichlet boundary conditions are imposed according to the exact solution and we seek for the steady state solution 
starting from fluid at rest. 
Since the flow regime is diffusion dominated we set the initial pseudo-time step of the continuation strategy to a very large 
value ($10^{13}$) and fall back to pure Newton for the steady Navier-Stokes equations.
We impose a four order of magnitude decrease of the relative residual norm at each Newton iteration:
\ie at the $n$-th Newton iterate the $i$-th iterate of the multigrid preconditioned FGMRES solver has congerged if
$\| \hat{r}_0^{n,i} \| = \frac{\| f_0(w_0^{n}) - A_0^{\mathrm{INS}}(w_0^{n}) \overline{\delta w}_0^{n,i} \|}{\| f_0(w_0^{n}) \|} \leq 10^{-4}$.
Convergence is achieved is six Newton iterations, the steady state solution $w_0^{6}$ is such that $ | f_0(w_0^{6}) | \leq 10^{-12} $.

The smoothing and solver option for the multigrid preconditioner are the same that in the Stokes case.
One iteration of $\mathrm{MG}_\mathcal{V}$ cycle is used as a preconditioner for the FMGRES(60) iteration.
Smoothing is performed with a single iteration of a right ILU preconditioned GMRES solver while
for the ILU preconditioned GMRES solver on level $L$ we impose a four order of magnitude residual decrease.
Besides the multilevel V-cycle iteration, for the Kovasznay test case we include the results 
obtained with the W-cycle iteration, see \eg \cite{TrottenbergMultigridBook:2001}.
The V- and W-cycle iterations differ in terms of the coarse grid correction of Algorithm \ref{eq:vcic},
as outlined below
\begin{equation} 
\nonumber
\begin{split}
&\text{\underline{\emph{Coarse grid correction (V-cycle)}}}      \\  
&r_{\ell} = f_{\ell} - A_{\ell} \overline{w}_{\ell}           \\  
&r_{\ell+1} = \mathcal{I}_{\ell}^{\ell+1} r_{\ell}             \\
&{e}_{\ell+1} = \mathrm{MG}_{\mathcal{V}}(\ell+1,r_{\ell+1},0) \\
&             \\
&\widehat{w}_{\ell} = \overline{w}_{\ell} + \mathcal{I}_{\ell+1}^{\ell} e_{\ell+1} 
\end{split}
\quad \quad \quad
\begin{split}
&\text{\underline{\emph{Coarse grid correction (W-cycle)}}}  \\
&r_{\ell} = f_{\ell} - A_{\ell} \overline{w}_{\ell} \\
&r_{\ell+1} = \mathcal{I}_{\ell}^{\ell+1} r_{\ell}             \\
&{\widehat{e}}_{\ell+1} = \mathrm{MG}_{\mathcal{V}}(\ell+1,r_{\ell+1},0) \\
&{e}_{\ell+1} = \mathrm{MG}_{\mathcal{V}}(\ell+1,r_{\ell+1},\widehat{e}_{\ell+1}) \\
&\widehat{w}_{\ell} = \overline{w}_{\ell} + \mathcal{I}_{\ell+1}^{\ell} e_{\ell+1} 
\end{split}
\end{equation}

In order to investigate the growth of computational costs while increasing the mesh size,
2D solutions are computed on three uniform quadrilateral elements meshes of size $(128 \cdot 2^n)^2, n=\{1,2,3\}$
of the bi-unit square domain $[-0.5,1.5]\times[0,2]$.
We check the influence of raising the polynomial degree 
on the convergence rate and the computational expense considering $k=\{1,2,3\}$.
To investigate the influence of the number of coarse levels on the convergence rate we consider $L=\{2,3,4,5\}$.

Since we are considering the performance of a linear multigrid iteration applied to each of six Newton method steps required to reach the 
steady state solution all the numerical results presented in what follows are averaged over the six steps.
The number of linear iterations reported in Table \ref{tab:it_multasprec_ns} and the convergence factors reported in Table \ref{tab:conv_multasprec_ns}
show that only the W-cycle iteration yields uniform convergence with respect to the number of levels. 
The influence of the number of levels on the V-cycle iteration is not dramatic
but clearly noticeable.
The number of iterations is not grid independent but moving from a $128^2$ to a $512^2$ quadrilateral elements mesh 
(a sixteen-fold increase of the number of elements) the iterations increase is less than two-fold.
Also the polynomial degree dependence is mild: similarly to the Stokes case a slight worsening of the convergence rates is observed for $k=2$.

\begin{table}[!htbp]
\small
\centering
    \begin{tabular}{@{}l|ccc|ccc|ccc@{}} \hline
                                \multicolumn{10}{c}{Average linear solver iterations} \\ \hline
       k                         & \multicolumn{3}{c|}{1}         & \multicolumn{3}{c|}{2}         & \multicolumn{3}{c}{3}          \\
       grid                                               & $128$   &  $256$     & $512$   & $128$  &  $256$     & $512$   & $128$   &  $256$     & $512$ \\\hline 
       FGMRES $\mathrm{MG}_{\mathcal{V}}$ $\nlev$ = 2     & 7       &  8         & 9       & 8      &  11        & 16      & 8       &  9         & 12    \\                
       FGMRES $\mathrm{MG}_{\mathcal{V}}$ $\nlev$ = 3     & 8       &  9         & 10      & 10     &  12        & 17      & 9       &  11        & 13    \\      
       FGMRES $\mathrm{MG}_{\mathcal{V}}$ $\nlev$ = 4     & 9       &  10        & 12      & 11     &  14        & 18      & 11      &  13        & 16    \\      
       FGMRES $\mathrm{MG}_{\mathcal{V}}$ $\nlev$ = 5     & 10      &  11        & 13      & 12     &  15        & 20      & 12      &  14        & 18    \\      
       FGMRES $\mathrm{MG}_{\mathcal{W}}$ $\nlev$ = 2     & 5       &  6         & 8       & 7      &  9         & 13      & 6       &  8         & 11    \\                
       FGMRES $\mathrm{MG}_{\mathcal{W}}$ $\nlev$ = 3     & 5       &  6         & 8       & 7      &  9         & 13      & 6       &  8         & 11    \\      
       FGMRES $\mathrm{MG}_{\mathcal{W}}$ $\nlev$ = 4     & 5       &  6         & 8       & 7      &  9         & 13      & 6       &  8         & 11    \\      
       FGMRES $\mathrm{MG}_{\mathcal{W}}$ $\nlev$ = 5     & 5       &  6         & 8       & 7      &  9         & 13      & 6       &  8         & 11    \\      
       GMRES(200) ILU(0)                                  & 464     &  1589$^*$  &         & 449    &  1518      &         & 579     &  1669$^*$  &         \\      
    \end{tabular}
    \caption{2D Kovasznay problem. Number of iterations of a FGMRES solver preconditioned with a V-cycle and a W-cycle $h$-multigrid iteration (one iteration), see text for details. 
	     Linear system relative residual tolerance is $10^{-4}$. Average linear iterations over the six Newton steps required to find the steady state solution. 
	     \label{tab:it_multasprec_ns}}
\end{table} 
\begin{table}[!htbp]
\small
\centering
    \begin{tabular}{@{}l|ccc|ccc|ccc@{}} \hline
                                \multicolumn{10}{c}{Average convergence factor, $ \rho$} \\ \hline
       k                         & \multicolumn{3}{c|}{1}         & \multicolumn{3}{c|}{2}         & \multicolumn{3}{c}{3}          \\
       grid                                               & $128$  &  $256$      & $512$   & $128$   &  $256$     & $512$     & $128$     &  $256$   & $512$ \\\hline 
       FGMRES $\mathrm{MG}_{\mathcal{V}}$ $\nlev$ = 2     & .23   &  .26       & .34    & .31    &  .40      & .53      & .28      &  .34   & .43      \\    
       FGMRES $\mathrm{MG}_{\mathcal{V}}$ $\nlev$ = 3     & .29   &  .33       & .39    & .36    &  .44      & .55      & .35      &  .41   & .47      \\    
       FGMRES $\mathrm{MG}_{\mathcal{V}}$ $\nlev$ = 4     & .33   &  .37       & .43    & .40    &  .49      & .58      & .40      &  .46   & .52      \\    
       FGMRES $\mathrm{MG}_{\mathcal{V}}$ $\nlev$ = 5     & .36   &  .41       & .47    & .43    &  .52      & .61      & .42      &  .49   & .57      \\    
       FGMRES $\mathrm{MG}_{\mathcal{W}}$ $\nlev$ = 2     & .14   &  .20       & .28    & .22    &  .33      & .45      & .20      &  .28   & .40           \\    
       FGMRES $\mathrm{MG}_{\mathcal{W}}$ $\nlev$ = 3     & .14   &  .20       & .28    & .23    &  .33      & .45      & .20      &  .28   & .40           \\    
       FGMRES $\mathrm{MG}_{\mathcal{W}}$ $\nlev$ = 4     & .14   &  .20       & .28    & .23    &  .33      & .45      & .21      &  .28   & .40           \\    
       FGMRES $\mathrm{MG}_{\mathcal{W}}$ $\nlev$ = 5     & .14   &  .20       & .28    & .23    &  .33      & .45      & .21      &  .28   & .40           \\    
    \end{tabular}
    \caption{2D Kovasznay problem. Convergence factors of a FGMRES solver preconditioned with a V-cycle and a W-cycle $h$-multigrid iteration (one iteration), see text for details. 
	     Linear system relative residual tolerance is $10^{-4}$. Average convergence factors over the six Newton steps required to find the steady state solution. 
	       \label{tab:conv_multasprec_ns} }
\end{table} 

The wall clock times comparison of Table \ref{tab:cpu_multasprec_ns_quad} confirms that 
strong gains can be obtained as compared to the ILU preconditioned GMRES(200) iteration.
Interestingly, even if the W-cycle iteration is the best performing in terms of convergence 
rates, the increased computational cost as compared to the V-cycle penalizes execution times. 
Similarly to the Stokes case we get a four-to-five fold increase of the computational cost 
with a four fold increase of the number of levels.
\begin{table}[!htbp]
\small
\centering
    \begin{tabular}{@{}l|cccc|cccc|c@{}} 
                                \multicolumn{10}{c}{Total CPU time (s), 2D Kovasznay problem, quadrilateral mesh sequence} \\ \hline
                   solver           & \multicolumn{4}{c|}{FGMRES $\mathrm{MG}_{\mathcal{V}}$} & \multicolumn{4}{c|}{FGMRES $\mathrm{MG}_{\mathcal{W}}$} & GMRES \\
                   $\nlev$          &  2 &   3   &   4 &    5  &  2 &   3   &   4 &    5 & 0  \\
		   \cline{1-10}                                                                                              
                    grid &   \multicolumn{9}{c}{$\mathbf{k=1}$} \\
		   \cline{2-10}
                    $128$   & 3.18      & 2.90        & 3.08      & 3.24      &  2.86   & 2.47   & 2.48   & 2.56 &  16.5    \\
                    $256$   & 25.6      & 13.4        & 13.3      & 14.0      &  26.2   & 12.4   & 11.0   & 11.1 &  229     \\
                    $512$   & 229       & 79.1        & 59.8      & 62.3      &  294    & 108    & 58.9   & 51.8 &          \\
		   \cline{2-10}                                                                                              
                    &   \multicolumn{9}{c}{$\mathbf{k=2}$} \\
		   \cline{2-10}
                   $128$   &  11.1   &  8.14   &  8.34  &  8.58   & 13.4 & 8.46 & 8.15  & 8.27   & 46.0  \\     
                   $256$   & 109     & 42.9    & 37.9   & 39.2   & 138  & 58.1  & 42.5   & 40.8  & 586   \\
                   $512$   &  1090   &  313    &  176   &  179    &  1589 &  627 &  269  & 215   &    \\
   
		   \cline{2-10}                                                                                              
                    &   \multicolumn{9}{c}{$\mathbf{k=3}$} \\
		   \cline{2-10}
                   $128$   & 29.4   & 19.6    & 19.8   & 20.8  & 35.1   & 22.6   & 19.8  & 19.8   & 107    \\
                   $256$   & 283    & 108     & 94.1   & 97.6   & 361    & 152   & 106   & 94.8   & 1640   \\
                   $512$   & 2218   & 829     & 481    & 471    & 3399  &  1563  & 754  & 550     &        \\
		   \cline{1-10}
    \end{tabular}
    \caption{2D Kovasznay problem. Comparison of wall clock time (solution plus assembly times) 
             required to solve linearized systems of Newton method, see text for details. 
	     Linear system relative residual tolerance is $10^{-4}$. \label{tab:cpu_multasprec_ns_quad}.
	     Average CPU times over the six Newton steps required to find the steady state solution}. 
\end{table} 

As observed for the Stokes problem, the assembly and solution wall clock times of Table \ref{tab:cpuall_multasprec_ns} confirms that 
it is important to choose a sufficiently high number of levels not to get penalized by the poor performance of the 
ILU preconditioned GMRES coarse grid solver.
\begin{table}[!htbp]
\small
\centering
    \begin{tabular}{@{}l|ccc|ccc|ccc@{}} \hline
                         CPU time (s)         & \multicolumn{3}{c|}{solution} & \multicolumn{3}{c|}{assembly} & \multicolumn{3}{c}{total}\\
                         grid                 & $128$ &$256$& $512$   & $128$ &$256$&$512$   & $128$&$256$&$512$  \\\hline
                         $\mathbf{k=1}$ & \multicolumn{3}{c|}{}& \multicolumn{3}{c|}{}& \multicolumn{3}{c}{}\\
                         FGMRES MG $\nlev$ = 2 & 1.64    &  19.3 & 204       & 1.54    &  6.25  & 24.8    & 3.19  &  25.6 & 229    \\
                         FGMRES MG $\nlev$ = 3 & 1.14    &  6.34 & 50.8      & 1.75    &  7.01  & 28.2    & 2.90  &  13.4 & 79.1    \\
                         FGMRES MG $\nlev$ = 4 & 1.18    &  5.68 & 29.3      & 1.89    &  7.59  & 30.5    & 3.08  &  13.3 & 59.8    \\
                         FGMRES MG $\nlev$ = 5 & 1.25    &  5.97 & 30.1      & 1.99    &  8.01  & 32.1    & 3.24  &  14.0 & 62.3    \\
			 GMRES(200) ILU(0)     & 15.8    &  226  &           & 0.70    &  2.78  &         & 16.5  &  229  &         \\\hline
                         $\mathbf{k=2}$ & \multicolumn{3}{c|}{}& \multicolumn{3}{c|}{}& \multicolumn{3}{c}{}\\
                         FGMRES MG $\nlev$ = 2 & 8.18    &  97.4 & 1048      & 2.90    &  11.7  & 42.5    & 11.1  &  109  & 1090     \\
                         FGMRES MG $\nlev$ = 3 & 4.89    &  30.0 & 271       & 3.24    &  12.9  & 41.4    & 8.14  &  42.9 & 313     \\
                         FGMRES MG $\nlev$ = 4 & 4.88    &  24.5 & 131       & 3.07    &  13.5  & 44.3    & 8.34  &  37.9 & 176     \\
                         FGMRES MG $\nlev$ = 5 & 4.96    &  24.9 & 131       & 3.62    &  14.2  & 47.0    & 8.58  &  39.2 & 179     \\
			 GMRES(200) ILU(0)     & 44.4    &  580  &           & 1.58    &  5.90  &         & 46.0  &  586  &          \\\hline
                         $\mathbf{k=3}$ & \multicolumn{3}{c|}{}& \multicolumn{3}{c|}{}& \multicolumn{3}{c}{}\\
                         FGMRES MG $\nlev$ = 2 & 23.6    &  260    & 2121    & 5.81    &  23.5  & 96.7    & 29.5  &  283  & 2218     \\
                         FGMRES MG $\nlev$ = 3 & 13.2    &  81.8   & 720     & 6.40    &  25.9  & 109     & 19.6  &  108  & 829     \\
                         FGMRES MG $\nlev$ = 4 & 12.9    &  66.3   & 364     & 6.89    &  27.7  & 115     & 19.8  &  94.1 & 480     \\
                         FGMRES MG $\nlev$ = 5 & 13.5    &  68.1   & 349     & 7.27    &  29.6  & 121     & 20.8  &  97.6 & 471    \\
			 GMRES(200) ILU(0)     & 104     &  1626   &         & 3.01    &  13.4  &         & 107   &  1640 &         \\\hline
    \end{tabular}
    \caption{2D Kovasznay problem. Solution, assembly and total (solution plus assembly) times for solving linearized systems of Newton method with with 
             a FGMRES solver preconditioned with a V-cycle $h$-multigrid iteration, see text for details. 
	     Linear system relative residual tolerance is $10^{-4}$. 
	     Average CPU times over the six Newton steps required to find the steady state solution. 
	     \label{tab:cpuall_multasprec_ns}}
\end{table}

\subsubsection{Lid-Driven cavity test case}
To investigate the influence of the Reynolds number on the convergence rates and the performance in three space dimensions we consider 
the lid-driven cavity problem.
We rely on a uniform $100^2$ quadrilateral and a uniform $80^3$ hexahedral grid of the unit square and the unit cube, respectively.
We check the influence of raising the polynomial degree 
on the convergence rate considering $k=\{1,2,3,4\}$ in 2D but omitting $k=4$ in 3D.
We consider two Reynolds numbers, $ Re=1000$ and $Re=5000$, in 2D and $Re=1000$ in 3D.

\begin{table}[!htbp]
\small
\centering
    \begin{tabular}{@{}l|cc|cc@{}} \hline
                               & \multicolumn{2}{c|}{$\Psi_{tc}$ iterations}          & \multicolumn{2}{c}{convergence factor $\rho$ max(avg)} \\ \hline
       Re                      & 1000                 & 5000                         & 1000                            & 5000              \\
       $\mathbf{k=1}$          &                      &                              &                                 &                   \\
       FGMRES MG $\nlev$ = 3   & \multirow{2}{*}{24}  & \multirow{2}{*}{45}          & .556 (.395)                   & .815 (.537)     \\ 
       FGMRES MG $\nlev$ = 4   &                      &                              & .646 (.451)                   & .883 (.567)     \\
       $\mathbf{k=2}$          &                      &                              &                               &                 \\
       FGMRES MG $\nlev$ = 3   & \multirow{2}{*}{23}  & \multirow{2}{*}{53}          & .321 (.228)                   & .638 (.349)     \\ 
       FGMRES MG $\nlev$ = 4   &                      &                              & .463 (.306)                   & .753 (.394)     \\
       $\mathbf{k=3}$          &                      &                              &                               &                 \\
       FGMRES MG $\nlev$ = 3   & \multirow{2}{*}{23}  & \multirow{2}{*}{79}          & .171 (.126)                   & .483 (.228)     \\ 
       FGMRES MG $\nlev$ = 4   &                      &                              & .313 (.199)                   & .682 (.284)     \\
       $\mathbf{k=4}$          &                      &                              &                               &                 \\
       FGMRES MG $\nlev$ = 3   & \multirow{2}{*}{23}  & \multirow{2}{*}{67}          & .126 (.090)                   & .469 (.225)     \\ 
       FGMRES MG $\nlev$ = 4   &                      &                              & .201 (.132)                   & .654 (.289)     \\
    \end{tabular}
    \caption{2D lid-driven cavity problem. Number of pseudo-transient continuation iterations and maximum/average convergence factors ($\rho$)
             measured over the $\Psi_{tc}$ iterations. Linearized systems of $\Psi_{tc}$ method are solved with 
             a FGMRES solver preconditioned with a V-cycle $h$-multigrid iteration, see text for details. 
	     \label{tab:conv_multasprec_lid}}
\end{table} 
Since the Reynolds number is higher than in the Kovasznay case and we approach convection dominated flow regimes, 
we seek for a steady state solution starting from fluid at rest by means of 
of the pseudo-transient continuation strategy with SER time stepping.
Besides adapting the time step, it is convenient to adapt the \emph{forcing terms}, that is the relative relative tolerance triggering 
convergence of the linear system at each continuation step,
we adopt the strategy proposed in \cite{botti2015713}.
The goal is to avoid \emph{oversolving} of the linear system when the linearization 
of the residual $f(w^{n+1}) = f(w^n) + J(w^n) \overline{\delta w}^n$ 
is not sufficiently accurate to pay off in terms of convergence towards the steady state. 

\begin{table}[!htbp]
\small
\centering
    \begin{tabular}{@{}l|c|c|c|c@{}} \hline
       Re 1000                & $np$             & avg $\card(\Tl^i)$   & $\Psi_{tc}$ it       & $\rho$ max(avg)  \\ \hline
       $\mathbf{k=1}$         &                       &                    &                       &                     \\
       FGMRES MG $\nlev$ = 2  & \multirow{2}{*}{32}   &16000/2350/350      & \multirow{2}{*}{33}   & .806 (.509)       \\ 
       FGMRES MG $\nlev$ = 3  &                       &16000/2350/350/53   &                       & .942 (.589)       \\
       $\mathbf{k=2}$         &                       &                    &                       &                   \\
       FGMRES MG $\nlev$ = 2  & \multirow{2}{*}{64}   &8000/1175/170       & \multirow{2}{*}{33}   & .535 (.356)       \\ 
       FGMRES MG $\nlev$ = 3  &                       &8000/1175/170/25    &                       & .822 (.487)       \\
       $\mathbf{k=3}$         &                       &                    &                       &                   \\
       FGMRES MG $\nlev$ = 2  & \multirow{2}{*}{128}  &4000/585/90         & \multirow{2}{*}{34}   & .392 (.259)       \\ 
       FGMRES MG $\nlev$ = 3  &                       &4000/585/90/13      &                       & .630 (.390)       \\
    \end{tabular}
    \caption{3D lid-driven cavity problem. Number of processes ($np$), average grid partition cardinality on each level of the $h$-coarsened mesh sequence, 
             number of pseudo-transient continuation iterations and maximum/average convergence factors ($\rho$) 
             measured over the $\Psi_{tc}$ iterations. Linearized systems of $\Psi_{tc}$ method are solved with 
             a FGMRES solver preconditioned with a V-cycle $h$-multigrid iteration, see text for details. 
	     \label{tab:conv_multasprec_lid3D}}
\end{table} 
The smoothing and solver option for the multigrid preconditioner are the same that in the Stokes and Kovasznay case.
One iteration of $\mathrm{MG}_\mathcal{V}$ cycle is used as a preconditioner for the FMGRES(60) iteration.
High-order modes of the error are smoothed with a single iteration of a right ILU preconditioned GMRES solver while
we require a four order of magnitude residual decrease for the ILU preconditioned GMRES solver on level $L$.
In parallel computations ILU preconditioners are replaced with ASM preconditioners with one level of overlap, 
as we did for solving elliptic problems in parallel in Section \ref{sec:numresBR2}.

The average and maximum convergence factors measured over the $\Psi_{tc}$ iterations are reported in Table \ref{tab:conv_multasprec_lid}.
While the maximum convergence factors, usually observed in the terminal phase of the convergence (that is when the time step is large),
are significantly affected by raising the Reynolds number, the average convergence factors are satisfactorily small at Reynold 5000.
Interestingly increasing the polynomial degree is beneficial from the convergence rates viewpoint, very good performances are observed for $k=4$.

Parallel 3D computations demonstrate that the convergence rates do not degrade, even if 
the number of mesh elements in each grid partition is remarkably small on the coarsest level, see Table \ref{tab:conv_multasprec_lid3D}. 
The trend observed in 2D is confirmed, raising the polynomial degree is advantageous from the convergence rate viewpoint.
We remark that the third polynomial degree dG discretization 
on the $80^3$ hexahedral elements grid tops at approximatively 10M unknowns. 

\subsubsection{Cerebral aneurysm hemodynamics}
\label{sec:hemo}
\begin{figure}[H]
\begin{tabular}{lr}
\includegraphics[width=0.47 \textwidth]{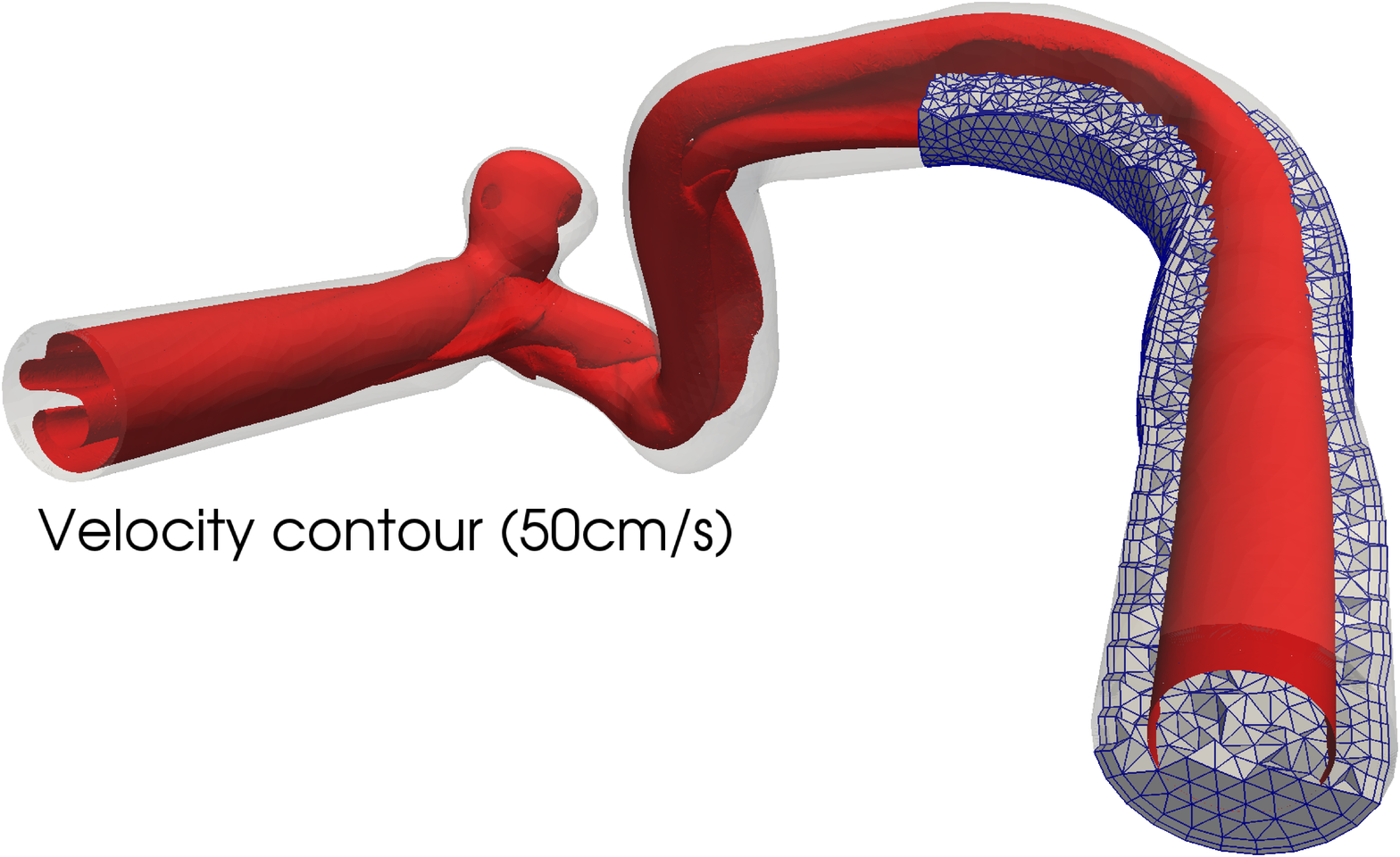} &
\includegraphics[width=0.47 \textwidth]{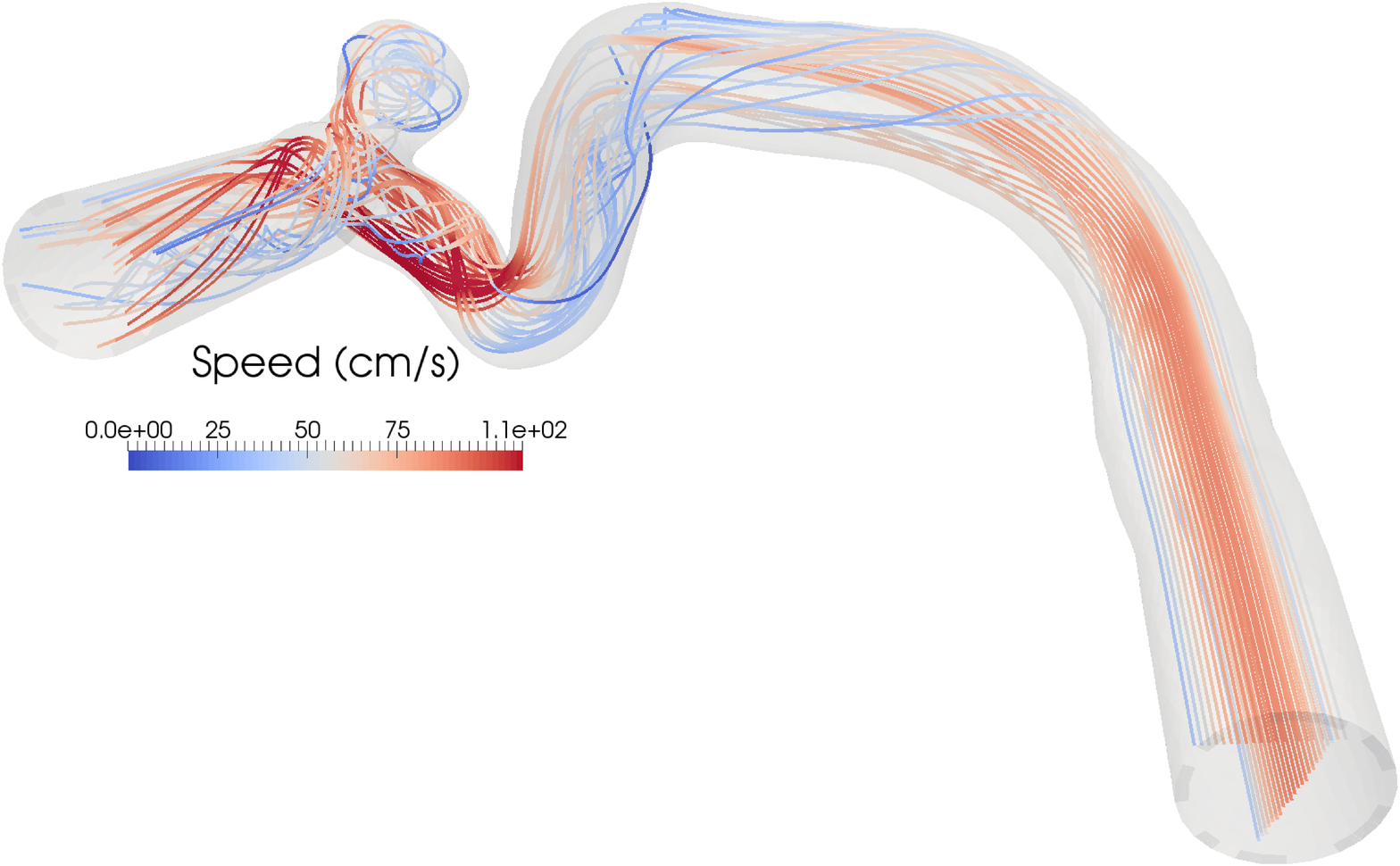} \\
\end{tabular}
\caption{ Hemodynamics of a cerebral aneurysm reconstructed from medical images.
         \emph{Left}, hybrid (tetrahedral and prismatic) 270k elements grid and velocity contour. 
	 \emph{Right}, streamlines computed at the systolic peak. 
	 \label{fig:grid_aneu}}
\end{figure}

In this section we apply the Backward Euler time integration strategy of Algorithm \ref{algo:be}
to approximate the blood flow field in a pathological Internal Carotid Artery (ICA) 
reconstructed from medical images, see Figure \ref{fig:grid_aneu}. 
In order to take into account the pulsatile flow behaviour Dirichlet boundary conditions are 
imposed at the circular inflow section relying on the Womersley analytical solution \cite{Womersley55} and considering a physiological 
flow rate all along the cardiac cycle \cite{Cezeaux97}. The average Reynolds number $Re_{avg} = 500$.
Stress-free boundary conditions are imposed at the outflow section and no-slip boundary conditions are imposed at the vessel walls.
We apply $k=\{1,2,3\}$ polynomial degree dG discretizations over the 270K hybrid grid 
generated with the open-source Vascular Modeling Toolkit (VMTK) \cite{Antiga2008}.
Simulations are performed running in parallel on 16, 32 and 64 processes for 
first, second and third degree dG discretizations, respectively.
The fixed time steps is chosen such that 150 numerical solution are computed 
in each cardiac cycle.
The time integration strategy is initialized with fluid at rest and conducted for three cardiac cycles.

\begin{table}[!htbp]
\small
\centering
    \begin{tabular}{@{}l|c|c|c|c@{}} \hline
       $Re_\mathrm{avg} = 500$  & $np$             &  avg $\card(\Tl^i)$     & time steps         & $\rho$ max(avg)\\ \hline
       $\mathbf{k=1}$         &      &                       &      &                     \\
       FGMRES MG $\nlev$ = 2  & 16   &10600/1560/230         & 150   & .644 (.547)       \\ 
       $\mathbf{k=2}$         &      &                       &      &                   \\
       FGMRES MG $\nlev$ = 2  & 32   &5300/770/114           & 150   & .678 (.540)       \\ 
       $\mathbf{k=3}$         &      &                       &      &                   \\
       FGMRES MG $\nlev$ = 2  & 64   &2650/390/57            & 150   & .631 (.477)       \\ 
    \end{tabular}
    \caption{Cerebral aneurysm hemodynamics. Number of processes ($np$), average grid partition cardinality on each level of the $h$-coarsened mesh sequence, 
             number of times steps per cardiac cycle and maximum/average convergence factors ($\rho$) 
             measured over the Backward Euler time integration strategy. Linearized systems of $\Psi_{tc}$ method are solved with 
             a FGMRES solver preconditioned with a V-cycle $h$-multigrid iteration, see text for details. 
	     \label{tab:conv_multasprec_aneu}}
\end{table} 
For solving the linearized systems of the BE method \eqref{eq:nsbe} 
one iteration of $\mathrm{MG}_\mathcal{V}$ cycle is used as a preconditioner for the FMGRES(60) solver.
High-order modes of the error are smoothed with a single iteration of an ASM preconditioned GMRES solver while
we require a four order of magnitude residual decrease for the ASM preconditioned GMRES solver on level $L$.
ASM preconditioners employ one level of overlap between sub-domains and an ILU decompositions for each sub-domain matrix. 

Table \ref{tab:conv_multasprec_aneu} reports the maximum and average convergence rates measured 
over the third cardiac cycle. 
Since the time step is fixed, the gap between maximum and average converge factors is narrower 
than in the pseudo-transient continuation strategy, \cf Table \ref{tab:conv_multasprec_lid3D},
and reflects the influence of varying the Reynolds number.
In particular the maximum convergence factor is recorded during systole where 
convection is more pronounced as compared to diastole.

Even if the average convergence rates reported in Table \ref{tab:conv_multasprec_aneu}
are less satisfactory that in the lid-driven cavity case, the fact that the linear system residual halves 
at each FGMRES iteration is a significant achievement. 
Hemodynamic computations are considered very challenging from the numerical solution viewpoint,
to the point that even segregated Pressure Corrections strategies might require ad-hoc preconditioners \cite{Cebral10}. 

\section{Conclusions}
This work demonstrates the feasibility and
effectiveness of $h$-multigrid preconditioners applied to high-order accurate dG discretizations
of incompressible flow problems.
In view of efficiency agglomeration based $h$-multigrid strategies with inherited coarse grid operators are attractive
because the expensive process of numerically integrating over agglomerated elements 
can be avoided in all but the preprocessing phase.
Indeed, intergrid transfer operators can be computed once prior 
to the non-linear iteration, and stored for later use.
In this work we introduced an effective strategy for improving performance of inherited coarse grid operators 
which exploits a rescaled Galerkin projection of the BR2 dG discretization stabilization term.
Using a single iteration of preconditioned GMRES as smoothing strategy,
the multigrid convergence is uniform with respect to the number of levels
and the \emph{typical multigrid efficiency} is closely approached on model problems.
The ability to beat direct solvers on arbitrarily unstructured low quality grids
and the appealing performance obtained on parallel real-life computations might revert the 
common belief that discontinuous Galerkin discretizations are more expensive to solve as compared 
to standard finite element and finite volume formulations.

\appendix
\section{Implementation details: restriction of BR2 operators}
\label{sec:append}
We provide implementations details about the matrix-free implementation of the restriction of coarse grid operators.
For the sake of brevity we consider inheritance of the BR2 bilinear forms, the Stokes and Navier-Stokes coarse grid operators 
can be obtained in a similar fashion. 
Note that BR2 coarse grid operators involve Galerkin projections for consistency terms, see Equation \eqref{eq:galproj} 
and the rescaled Galerkin projection for the stabilization term, see Equation \eqref{eq:resgalproj}.

The matrices counterparts $\SysM{\ell}^{\mathrm{BR2}}$ of the operators $\Sys{\ell}^{\mathrm{BR2}}$ 
are sparse block matrices of size $(\card(\Tl) \; N^{\T}_{\mathrm{dof}})^2$ (the block size is $(N^{\T}_{\mathrm{dof}})^2$)
composed of diagonal blocks $\SysM{\T_{\ell},\T_{\ell}}$ and off-diagonal blocks $\SysM{\T_{\ell},\Tprl}$.
Off-diagonal blocks are responsible of the coupling
between neighboring elements $\T_\ell,\Tpr_\ell$ sharing a face $\F_\ell$.
The coarse operators $\Sys{\ell}^{\widetilde{\It},\mathrm{BR2}}$ 
are obtained matrix-free as described in Algorithms \ref{algo:f2ctwo_m} and \ref{algo:f2ctwo_m_two}.
\begin{algorithm}
\caption{Inherited BR2 (mesh elements and boundary faces)}
 \begin{algorithmic}[0]
  \label{algo:f2ctwo_m}
  \FOR {$\T_{0} \in \Tf$}
    \STATE {assemble $(\SysM{\T_{0}, \T_{0}}^{\mathrm{CS_{\kappa}}})_{i,j} = a^{\mathrm{CS}_\kappa}_{0}(\varphi_{i}^{\kappa_0}, \varphi_{j}^{\kappa_0})$} 
    \FOR {$\ell = 0,...,L-1$}
       \STATE {find $\T_{\ell+1} \in \Tlpo$ such that $\T_{\ell} \in K_{\ell}^{\ell+1}$}
       \STATE {$ \SysM{\T_{\ell+1},\T_{\ell+1}} \mathrel{+}= \M_{\T_{\ell+1},\T_{\ell}} \; (\SysM{\T_{\ell}, \T_{\ell}}^{\mathrm{CS}_\kappa}) \;  \M_{ \T_{\ell+1},\T_{\ell}}^T$}
   \ENDFOR
  \ENDFOR
  \FOR {$\F_{0} \in \Ffb$}
     \STATE {find $\T_{0}$ such that $ \F_{0} = \partial \T_{0} \cap \partial \Omega$}
   \STATE {assemble $(\SysM{\T_{0}, \T_{0}}^{\mathrm{CS}_{\sigma}})_{i,j} = a^{\mathrm{CS}_\sigma}_{\ell}(\varphi_{i}^{\kappa_0}, \varphi_{j}^{\kappa_0})$}
   \STATE {assemble $(\SysM{\T_{0}, \T_{0}}^{\mathrm{STB}})_{i,j} = s_{\ell}(\varphi_{i}^{\kappa_0}, \varphi_{j}^{\kappa_0})$}
   \FOR {$\ell = 0,...,L-1$}
     \STATE {find $\T_{\ell+1}$ such that $\T_{\ell} \in K_{\ell}^{\ell+1}$}
     \STATE {find $\F_{\ell+1}$ such that $\F_{\ell} \in \Sigma_{\ell}^{\ell+1}$}
        \STATE{
         $ \SysM{\T_{\ell+1},\T_{\ell+1}}^{\mathrm{CS}_\sigma}  = \M_{\T_{\ell+1},\T_{\ell}} \; (\SysM{\T_{\ell}, \T_{\ell}} ^{\mathrm{CS}_\sigma}) \; \M_{\T_{\ell+1},\T_{\ell}}^T$\\
         $ \SysM{\T_{\ell+1},\T_{\ell+1}}^{\mathrm{STB}}        = \mathcal{H}_{\F_\ell}^{\F_{\ell+1}}
	                                                        \left( \M_{\T_{\ell+1},\T_{\ell}} \; (\SysM{\T_{\ell}, \T_{\ell}} ^{\mathrm{STB}}) \; \M_{\T_{\ell+1}, \T_{\ell}}^T  \right)$\\
	 $\SysM{\T_{\ell+1},\T_{\ell+1}} \mathrel{+}= \left( \SysM{\T_{\ell+1},\T_{\ell+1}}^{\mathrm{CS}_\sigma} + \SysM{\T_{\ell+1},\T_{\ell+1}}^{\mathrm{STB}} \right) $\\
         }
   \ENDFOR
 \ENDFOR
 \end{algorithmic}
\end{algorithm}

\begin{algorithm}
\caption{Inherited BR2 (internal faces)}
 \begin{algorithmic}[0]
  \label{algo:f2ctwo_m_two}
  \FOR {$\F_{0} \in \Ffi$}
     \STATE {find $\T_{0}, \Tpr_{0} $ such that $\T_{0} \neq \Tpr_{0}\; \mathrm{and}\; \F_{0} = \partial \T_{0} \cap \partial \Tpr_0$}
   \STATE {assemble $\SysM{\T_{0}, \T_{0}}^    {\mathrm{CS}_{\sigma}},
                     \SysM{\Tpr_{0}, \Tpr_{0}}^{\mathrm{CS}_{\sigma}},
                     \SysM{\T_{0}, \Tpr_{0}}^  {\mathrm{CS}_{\sigma}}$ and 
	             $(\SysM{\Tpr_{0}, \T_{0}}^  {\mathrm{CS}_{\sigma}})_{i,j} = a^{\mathrm{CS}_\sigma}_{0}(\varphi_{i}^{\Tpr_0}, \varphi_{j}^{\kappa_0})$} 
   \STATE {assemble $\SysM{\T_{0}, \T_{0}}^    {\mathrm{STB}},
                     \SysM{\Tpr_{0}, \Tpr_{0}}^{\mathrm{STB}},
		     \SysM{\T_{0}, \Tpr_{0}}^  {\mathrm{STB}}$ and
		    $(\SysM{\Tpr_{0}, \T_{0}}^  {\mathrm{STB}})_{i,j} = s_{0}(\varphi_{i}^{\Tpr_0}, \varphi_{j}^{\kappa_0})$} 
   \FOR {$\ell = 0,...,L-1$}
     \STATE {find $\T_{\ell+1}, \Tprlpo $ such that $\T_{\ell} \in K_{\ell}^{\ell+1}, \; \Tprl \in K{'}_{\ell}^{\ell+1} $}
     \STATE {find $\F_{\ell+1}$ such that $\F_{\ell} \in \Sigma_{\ell}^{\ell+1}$}
        \IF {$\T_{\ell+1} = \Tprlpo$}
	  \STATE {\textbf{break }\COMMENT {ignore contributions of internal facets $\sigma_l \not\in \partial K_l^{l+1} \cap \partial K{'}_{\ell}^{\ell+1}$}}
	\ELSE
        \STATE{
         $ \SysM{\T_{\ell+1},\Tprlpo} ^{\mathrm{CS}_\sigma}         = \M_{\T_{\ell+1},\T_{\ell}} \; (\SysM{\T_{\ell}, \Tprl}^{\mathrm{CS}_\sigma})      \; \M_{\Tprlpo, \Tprl}^T$\\
         $ \SysM{\Tprlpo,\T_{\ell+1}} ^{\mathrm{CS}_\sigma}         = \M_{ \Tprlpo, \Tprl}       \; (\SysM{\Tprl, \T_{\ell}}^{\mathrm{CS}_\sigma})      \; \M_{\T_{\ell+1},\T_{\ell}}^T$\\
         $ \SysM{\T_{\ell+1},\T_{\ell+1}}^{\mathrm{CS}_\sigma}      = \M_{\T_{\ell+1},\T_{\ell}} \; (\SysM{\T_{\ell}, \T_{\ell}} ^{\mathrm{CS}_\sigma}) \; \M_{\T_{\ell+1},\T_{\ell}}^T$\\
         $ \SysM{\Tprlpo,\Tprlpo}  ^{\mathrm{CS}_\sigma}            = \M_{ \Tprlpo, \Tprl}       \; (\SysM{\Tprl, \Tprl} ^{\mathrm{CS}_\sigma})         \; \M_{\Tprlpo, \Tprl}^T$\\
         $ \SysM{\T_{\ell+1},\Tprlpo} ^{\mathrm{STB}}    = \mathcal{H}_{\F_\ell}^{\F_{\ell+1}} \left( \M_{\T_{\ell+1},\T_{\ell}} \; 
	                                                                                      (\SysM{\T_{\ell}, \Tprl}^{\mathrm{STB}})            
							     				 \; \M_{\Tprlpo, \Tprl}^T  \right)$\\
         $ \SysM{\Tprlpo,\T_{\ell+1}} ^{\mathrm{STB}}    = \mathcal{H}_{\F_\ell}^{\F_{\ell+1}} \left( \M_{ \Tprlpo, \Tprl}      
	                                                                                    \; (\SysM{\Tprl, \T_{\ell}}^{\mathrm{STB}})            
							     			       \; \M_{\T_{\ell+1},\T_{\ell}}^T \right)$\\
         $ \SysM{\T_{\ell+1},\T_{\ell+1}}^{\mathrm{STB}} = \mathcal{H}_{\F_\ell}^{\F_{\ell+1}} \left( \M_{\T_{\ell+1},\T_{\ell}} 
	                                                                                    \; (\SysM{\T_{\ell}, \T_{\ell}} ^{\mathrm{STB}})       
							     			       \; \M_{\T_{\ell+1}, \T_{\ell}}^T  \right)$\\
         $ \SysM{\Tprlpo,\Tprlpo}  ^{\mathrm{STB}}       = \mathcal{H}_{\F_\ell}^{\F_{\ell+1}} \left( \M_{ \Tprlpo, \Tprl}       
	                                                                                               \; (\SysM{\Tprl, \Tprl} ^{\mathrm{STB}})               
												       \; \M_{\Tprlpo,\Tprl}^T   \right)$ \\
	 $\SysM{\T_{\ell+1},\Tprlpo}     \mathrel{+}= \left( \SysM{\T_{\ell+1},\Tprlpo}^{\mathrm{CS}_\sigma} + \SysM{\T_{\ell+1},\Tprlpo}^{\mathrm{STB}} \right) $\\
	 $\SysM{\Tprlpo,\T_{\ell+1}}     \mathrel{+}= \left( \SysM{\Tprlpo,\T_{\ell+1}}^{\mathrm{CS}_\sigma} + \SysM{\Tprlpo,\T_{\ell+1}}^{\mathrm{STB}} \right) $\\
	 $\SysM{\T_{\ell+1},\T_{\ell+1}} \mathrel{+}= \left( \SysM{\T_{\ell+1},\T_{\ell+1}}^{\mathrm{CS}_\sigma} + \SysM{\T_{\ell+1},\T_{\ell+1}}^{\mathrm{STB}} \right) $\\
	 $\SysM{\Tprlpo,\Tprlpo}         \mathrel{+}= \left( \SysM{\Tprlpo,\Tprlpo}^{\mathrm{CS}_\sigma} + \SysM{\Tprlpo,\Tprlpo}^{\mathrm{STB}} \right) $\\
         }
        \ENDIF
 \ENDFOR
\ENDFOR
 \end{algorithmic}
\end{algorithm}
Matrix restriction is performed contextually to fine matrix assembly 
so that stability term contributions $\SysM{}^{\mathrm{STB}}$, 
and consistency-symmetry terms contributions $\SysM{}^{\mathrm{CS}_\kappa}$ 
and $\SysM{}^{\mathrm{CS}_\sigma}$, see Section \ref{sec:coarseBR2}, 
are restricted separately, before being collected into diagonal and off-diagonal blocks of the fine matrix.

It is interesting to remark that only a subset of the internal faces contributions on level $\ell$ is restricted on level $\ell+1$. 
In particular we remark that all diagonal and off-diagonal contributions $\SysM{}^{\mathrm{STB}},\SysM{}^{\mathrm{CS}_\sigma}$ associated to facets $\sigma \in \Fli$ that do not belong 
to the boundary of agglomerated elements on level $\ell+1$ are ignored in Algorithm \ref{algo:f2ctwo_m_two}.
This optimization is permitted thanks to the local conservation properties of dG formulations.

\section*{Acknowledgements}
We acknowledge the CINECA HPC facility for the availability of high performance computing resources and support
within the agreement ``Convenzione di Ateneo Universit\`a degli Studi di Bergamo''.


\section*{References}

\begin{thebibliography}{10}
\expandafter\ifx\csname url\endcsname\relax
  \def\url#1{\texttt{#1}}\fi
\expandafter\ifx\csname urlprefix\endcsname\relax\def\urlprefix{URL }\fi
\expandafter\ifx\csname href\endcsname\relax
  \def\href#1#2{#2} \def\path#1{#1}\fi

\bibitem{Benzi05numericalsolution}
M.~Benzi, G.~H. Golub, J.~Liesen, Numerical solution of saddle point problems,
  ACTA NUMERICA 14 (2005) 1--137.

\bibitem{Fidkowski05}
K.~J. Fidkowski, T.~A. Oliver, J.~Lu, D.~L. Darmofal, p-multigrid solution of
  high-order discontinuous {G}alerkin discretizations of the compressible
  {N}avier-{S}tokes equations, J. Comput. Phys. 207~(1) (2005) 92--113.
\newblock \href {http://dx.doi.org/10.1016/j.jcp.2005.01.005}
  {\path{doi:10.1016/j.jcp.2005.01.005}}.

\bibitem{Nastase06}
C.~R. Nastase, D.~J. Mavriplis, High-order discontinuous {G}alerkin methods
  using an hp-multigrid approach, Journal of Computational Physics 213~(1)
  (2006) 330 -- 357.
\newblock \href {http://dx.doi.org/10.1016/j.jcp.2005.08.022}
  {\path{doi:10.1016/j.jcp.2005.08.022}}.

\bibitem{BassiGhidoni09}
F.~Bassi, A.~Ghidoni, S.~Rebay, P.~Tesini, High-order accurate p-multigrid
  discontinuous {G}alerkin solution of the {E}uler equations, International
  Journal for Numerical Methods in Fluids 60~(8) (2009) 847--865.
\newblock \href {http://dx.doi.org/10.1002/fld.1917}
  {\path{doi:10.1002/fld.1917}}.

\bibitem{ShahbaziMavriplis-MGAlgorithms:2009}
K.~Shahbazi, D.~J. Mavriplis, N.~K. Burgess, Multigrid algorithms for
  high-order discontinuous {G}alerkin discretizations of the compressible
  {N}avier–{S}tokes equations, Journal of Computational Physics 228~(21)
  (2009) 7917--7940.
\newblock \href {http://dx.doi.org/10.1016/j.jcp.2009.07.013}
  {\path{doi:10.1016/j.jcp.2009.07.013}}.

\bibitem{KanschatMultiAdvDiff}
J.~Gopalakrishnan, G.~Kanschat, A multilevel discontinuous {G}alerkin method,
  Numerische Mathematik 95~(3) (2003) 527--550.
\newblock \href {http://dx.doi.org/10.1007/s002110200392}
  {\path{doi:10.1007/s002110200392}}.

\bibitem{Brenner11}
S.~Brenner, J.~Cui, T.~Gudi, L.-Y. Sung, Multigrid algorithms for symmetric
  discontinuous {G}alerkin methods on graded meshes, Numerische Mathematik
  119~(1) (2011) 21--47.
\newblock \href {http://dx.doi.org/10.1007/s00211-011-0379-y}
  {\path{doi:10.1007/s00211-011-0379-y}}.

\bibitem{AntoniettiSartiVerani}
P.~F. Antonietti, M.~Sarti, M.~Verani, Multigrid algorithms for
  hp-discontinuous {G}alerkin discretizations of elliptic problems, SIAM
  Journal on Numerical Analysis 53~(1) (2015) 598--618.
\newblock \href {http://dx.doi.org/10.1137/130947015}
  {\path{doi:10.1137/130947015}}.

\bibitem{PrillHartmann}
F.~Prill, M.~Luk\'{a}\v{c}ov\'{a}-Medvidov\'{a}, R.~Hartmann, Smoothed
  aggregation multigrid for the {D}iscontinuous {G}alerkin method, SIAM Journal
  on Scientific Computing 31~(5) (2009) 3503--3528.
\newblock \href {http://dx.doi.org/10.1137/080728457}
  {\path{doi:10.1137/080728457}}.

\bibitem{AntoniettiGianiHouston2013}
P.~F. Antonietti, S.~Giani, P.~Houston, \$hp\$-version composite discontinuous
  {G}alerkin methods for elliptic problems on complicated domains, SIAM Journal
  on Scientific Computing 35~(3) (2013) A1417--A1439.
\newblock \href {http://dx.doi.org/10.1137/120877246}
  {\path{doi:10.1137/120877246}}.

\bibitem{Antonietti2014}
P.~F. Antonietti, S.~Giani, P.~Houston, Domain decomposition preconditioners
  for discontinuous {G}alerkin methods for elliptic problems on complicated
  domains, Journal of Scientific Computing 60~(1) (2014) 203--227.
\newblock \href {http://dx.doi.org/10.1007/s10915-013-9792-y}
  {\path{doi:10.1007/s10915-013-9792-y}}.

\bibitem{AntoniettiHoustonSmears2016}
P.~F. Antonietti, P.~Houston, I.~Smears, A note on optimal spectral bounds for
  nonoverlapping domain decomposition preconditioners for hp-version
  discontinuous {G}alerkin methods, International Journal of Numerical Analysis
  and Modeling 13~(4) (2016) 513--524.

\bibitem{AntoniettiHoustonSartiVerani}
P.~F. Antonietti, P.~Houston, X.~Hu, M.~Sarti, M.~Verani, Multigrid algorithms
  for hp-version {I}nterior {P}enalty {D}iscontinuous {G}alerkin methods on
  polygonal and polyhedral meshes, eprint {\tt arXiv:1412.0913}, submitted for
  publication.

\bibitem{WallraffLeicht14}
M.~Wallraff, T.~Leicht, Higher order multigrid algorithms for a discontinuous
  {G}alerkin {RANS} solver, in: 52nd Aerospace Sciences Meeting, no. 936 in
  AIAA SciTech, American Institute of Aeronautics and Astronautics, 2014, pp.
  1055--1072.
\newblock \href {http://dx.doi.org/10.2514/6.2014-0936}
  {\path{doi:10.2514/6.2014-0936}}.

\bibitem{WallraffHartmannLeicht15}
M.~Wallraff, R.~Hartmann, T.~Leicht, Multigrid solver algorithms for {DG}
  methods and applications to aerodynamic flows, in: N.~Kroll, C.~Hirsch,
  F.~Bassi, C.~Johnston, K.~Hillewaert (Eds.), IDIHOM: Industrialization of
  High-Order Methods - A Top-Down Approach, Vol. 128 of Notes on Numerical
  Fluid Mechanics and Multidisciplinary Design, Springer International
  Publishing, 2015, pp. 153--178.
\newblock \href {http://dx.doi.org/10.1007/978-3-319-12886-3_9}
  {\path{doi:10.1007/978-3-319-12886-3_9}}.

\bibitem{BassiFlexy12}
F.~Bassi, L.~Botti, A.~Colombo, D.~A. {D}i Pietro, P.~Tesini, On the
  flexibility of agglomeration based physical space discontinuous {G}alerkin
  discretizations, Journal of Computational Physics 231~(1) (2012) 45 -- 65.
\newblock \href {http://dx.doi.org/10.1016/j.jcp.2011.08.018}
  {\path{doi:10.1016/j.jcp.2011.08.018}}.

\bibitem{BassiFree14}
F.~Bassi, L.~Botti, A.~Colombo, Agglomeration based physical frame {dG}
  discretizations: An attempt to be mesh free, Mathematical Models and Methods
  in Applied Sciences 24~(08) (2014) 1495--1539.
\newblock \href {http://dx.doi.org/10.1142/S0218202514400028}
  {\path{doi:10.1142/S0218202514400028}}.

\bibitem{Cangiani14}
A.~Cangiani, E.~H. Georgoulis, P.~Houston, hp-version discontinuous {G}alerkin
  methods on polygonal and polyhedral meshes, Mathematical Models and Methods
  in Applied Sciences 24~(10) (2014) 2009--2041.
\newblock \href {http://dx.doi.org/10.1142/S0218202514500146}
  {\path{doi:10.1142/S0218202514500146}}.

\bibitem{Giani14}
S.~Giani, P.~Houston, {hp-}adaptive composite discontinuous {G}alerkin methods
  for elliptic problems on complicated domains, Numerical Methods for Partial
  Differential Equations 30~(4) (2014) 1342--1367.
\newblock \href {http://dx.doi.org/10.1002/num.21872}
  {\path{doi:10.1002/num.21872}}.

\bibitem{Cangiani16}
A.~Cangiani, Z.~Dong, E.~H. Georgoulis, P.~Houston, hp-version discontinuous
  {G}alerkin methods for advection-diffusion-reaction problems on polytopic
  meshes, Mathematical Modelling and Numerical Analysis 50~(3) (2016) 699--725.
\newblock \href {http://dx.doi.org/10.1051/m2an/2015059}
  {\path{doi:10.1051/m2an/2015059}}.

\bibitem{Bassi.Rebay.ea:1997}
F.~Bassi, S.~Rebay, G.~Mariotti, S.~Pedinotti, M.~Savini, {A high-order
  accurate discontinuous finite element method for inviscid and viscous
  turbomachinery flows}, in: R.~Decuypere, G.~Dibelius (Eds.), Proceedings of
  the 2nd European Conference on Turbomachinery Fluid Dynamics and
  Thermodynamics, Technologisch Instituut, Antwerpen, Belgium, 1997, pp.
  99--108.

\bibitem{Moulitsas.Karypis.ea:2001}
I.~Moulitsas, G.~Karypis, {MGridGen/ParmGridGen,} {Serial/Parallel} library for
  generating coase meshes for multigrid methods, Technical Report Version 1.0,
  University of Minnesota, Department of Computer Science/Army HPC Research
  Center, http://www-users.cs.umn.edu/$\sim$moulitsa/software.html (2001).

\bibitem{appPoly}
L.~Botti, Influence of reference-to-physical frame mappings on approximation
  properties of discontinuous piecewise polynomial spaces, Journal of
  Scientific Computing 52~(3) (2012) 675--703.

\bibitem{Polygons-cf:2011}
F.~Bassi, L.~Botti, A.~Colombo, S.~Rebay, Agglomeration based discontinuous
  {G}alerkin discretization of the {E}uler and {N}avier-{S}tokes equations,
  Computers \& Fluids 61 (2012) 77--85.
\newblock \href {http://dx.doi.org/10.1016/j.compfluid.2011.11.002}
  {\path{doi:10.1016/j.compfluid.2011.11.002}}.

\bibitem{BrennerScott:2008}
S.~C. Brenner, L.~R. Scott, The Mathematical Theory of Finite Element Methods,
  3rd Edition, Springer-Verlag, New York--Berlin--Heidelberg, 2008.

\bibitem{DiPiErn11}
D.~A. {Di~Pietro}, A.~Ern, Mathematical Aspects of Discontinuous {G}alerkin
  Methods, Vol.~69 of Maths \& Applications, Springer-Verlag, 2011.

\bibitem{Brezzi.Manzini.ea:2000}
F.~Brezzi, G.~Manzini, D.~Marini, P.~Pietra, A.~Russo, {Discontinuous
  {G}alerkin approximations for elliptic problems}, Numer. Methods Partial
  Differential Equations 16 (2000) 365--378.

\bibitem{Arnold.Brezzi.ea:2002}
D.~N. Arnold, F.~Brezzi, B.~Cockburn, D.~Marini, Unified analysis of
  discontinuous {G}alerkin methods for elliptic problems, SIAM J. Numer. Anal.
  39~(5) (2002) 1749--1779.

\bibitem{Bassi.Crivellini.ea:2006}
F.~Bassi, A.~Crivellini, D.~A. {Di~Pietro}, S.~Rebay, {An artificial
  compressibility flux for the discontinuous {G}alerkin solution of the
  incompressible {N}avier-{S}tokes equations}, J. Comput. Phys. 218 (2006)
  794--815.

\bibitem{DiPiStokes07}
D.~A. Di~Pietro, Analysis of a discontinuous {G}alerkin approximation of the
  {S}tokes problem based on an artificial compressibility flux, International
  Journal for Numerical Methods in Fluids 55~(8) (2007) 793--813.
\newblock \href {http://dx.doi.org/10.1002/fld.1495}
  {\path{doi:10.1002/fld.1495}}.

\bibitem{BriggsMultigridBook:2000}
W.~L. Briggs, V.~E. Henson, S.~F. McCormick, A multigrid tutorial (2nd ed.),
  Society for Industrial and Applied Mathematics, Philadelphia, PA, USA, 2000.

\bibitem{SmithBookDPM:2004}
B.~F. Smith, P.~E. Bj{\o}rstad, W.~Gropp, Domain Decomposition: Parallel
  Multilevel Methods for Elliptic Partial Differential Equations, Cambridge
  University Press, 2004.

\bibitem{TrottenbergMultigridBook:2001}
U.~Trottenberg, C.~W. Oosterlee, A.~Sch\"{u}ller, Multigrid, Academic Press,
  Inc., 2001.

\bibitem{AntoniettiDiosBrenner}
P.~F. Antonietti, B.~A. de~Dios, S.~C. Brenner, L.~yeng Sung, {S}chwarz methods
  for a preconditioned {W}{O}{P}{S}{I}{P} method for elliptic problems,
  Computational Methods in Applied Mathematics Comput. Methods Appl. Math.
  12~(3) (2012) 241--272.
\newblock \href {http://dx.doi.org/10.2478/cmam-2012-0021}
  {\path{doi:10.2478/cmam-2012-0021}}.

\bibitem{Toselli03}
D.~Sch\"{o}tzau, C.~Schwab, A.~Toselli, Mixed hp-{DGFEM} for incompressible
  flows, SIAM Journal on Numerical Analysis 40~(6) (2002) 2171--2194.
\newblock \href {http://dx.doi.org/10.1137/S0036142901399124}
  {\path{doi:10.1137/S0036142901399124}}.

\bibitem{ethier07}
K.~Shahbazi, P.~F. Fischer, C.~R. Ethier, A high-order {D}iscontinuous
  {G}alerkin method for the unsteady incompressible {N}avier-{S}tokes
  equations, J. Comput. Phys. 222~(1) (2007) 391--407.
\newblock \href {http://dx.doi.org/10.1016/j.jcp.2006.07.029}
  {\path{doi:10.1016/j.jcp.2006.07.029}}.

\bibitem{petsc-user-ref}
S.~Balay, S.~Abhyankar, M.~F. Adams, J.~Brown, P.~Brune, K.~Buschelman,
  L.~Dalcin, V.~Eijkhout, W.~D. Gropp, D.~Kaushik, M.~G. Knepley, L.~C.
  McInnes, K.~Rupp, B.~F. Smith, S.~Zampini, H.~Zhang, H.~Zhang,
  \href{http://www.mcs.anl.gov/petsc}{{PETS}c users manual}, Tech. Rep.
  ANL-95/11 - Revision 3.7, Argonne National Laboratory (2016).
\newline\urlprefix\url{http://www.mcs.anl.gov/petsc}

\bibitem{Toulorge20138}
T.~Toulorge, C.~Geuzaine, J.-F. Remacle, J.~Lambrechts, Robust untangling of
  curvilinear meshes, Journal of Computational Physics 254 (2013) 8 -- 26.
\newblock \href {http://dx.doi.org/10.1016/j.jcp.2013.07.022}
  {\path{doi:10.1016/j.jcp.2013.07.022}}.

\bibitem{SaadFlexy93}
Y.~Saad, A flexible inner-outer preconditioned {GMRES} algorithm, SIAM Journal
  on Scientific Computing 14~(2) (1993) 461--469.
\newblock \href {http://dx.doi.org/10.1137/0914028}
  {\path{doi:10.1137/0914028}}.

\bibitem{petsc-web-page}
S.~Balay, S.~Abhyankar, M.~F. Adams, J.~Brown, P.~Brune, K.~Buschelman,
  L.~Dalcin, V.~Eijkhout, W.~D. Gropp, D.~Kaushik, M.~G. Knepley, L.~C.
  McInnes, K.~Rupp, B.~F. Smith, S.~Zampini, H.~Zhang, H.~Zhang,
  \href{http://www.mcs.anl.gov/petsc}{{PETS}c {W}eb page},
  \url{http://www.mcs.anl.gov/petsc} (2016).
\newline\urlprefix\url{http://www.mcs.anl.gov/petsc}

\bibitem{petsc-efficient}
S.~Balay, W.~D. Gropp, L.~C. McInnes, B.~F. Smith, Efficient management of
  parallelism in object oriented numerical software libraries, in: E.~Arge,
  A.~M. Bruaset, H.~P. Langtangen (Eds.), Modern Software Tools in Scientific
  Computing, Birkh{\"{a}}user Press, 1997, pp. 163--202.

\bibitem{MOAB}
T.~J. Tautges, R.~Meyers, K.~Merkley, C.~Stimpson, C.~Ernst, {MOAB:} a
  mesh-oriented database, {SAND2004-1592}, Sandia National Laboratories, report
  (Apr. 2004).

\bibitem{Metis}
G.~Karypis, V.~Kumar, {METIS}, a software package for partitioning unstructured
  graphs, partitioning meshes, and computing fill-reducing orderings of sparse
  matrices, Technical Report Version 4.0, University of Minnesota, Department
  of Computer Science/Army HPC Research Center (1998).

\bibitem{KK_PsiTC_98}
C.~Kelley, D.~Keyes, Convergence analysis of pseudo-transient continuation,
  SIAM Journal on Numerical Analysis 35~(2) (1998) 508--523.
\newblock \href {http://dx.doi.org/10.1137/S0036142996304796}
  {\path{doi:10.1137/S0036142996304796}}.

\bibitem{mulder1985}
W.~A. Mulder, B.~Van~Leer, Experiments with implicit upwind methods for the
  {E}uler equations, Journal of Computational Physics 59~(2) (1985) 232--246.

\bibitem{Kovasznay48}
L.~I.~G. Kovasznay, Laminar flow behind a two-dimensional grid, Mathematical
  Proceedings of the Cambridge Philosophical Society 44 (1948) 58--62.
\newblock \href {http://dx.doi.org/10.1017/S0305004100023999}
  {\path{doi:10.1017/S0305004100023999}}.

\bibitem{botti2015713}
L.~Botti, A choice of forcing terms in inexact {N}ewton iterations with
  application to pseudo-transient continuation for incompressible fluid flow
  computations, Applied Mathematics and Computation 266 (2015) 713 -- 737.
\newblock \href {http://dx.doi.org/10.1016/j.amc.2015.05.136}
  {\path{doi:10.1016/j.amc.2015.05.136}}.

\bibitem{Womersley55}
J.~R. Womersley, Method for the calculation of velocity, rate of flow and
  viscous drag in arteries when the pressure gradient is known, The Journal of
  Physiology 127~(3) (1955) 553--563.

\bibitem{Cezeaux97}
J.~L. Cezeaux, A.~van Grondelle, Accuracy of the inverse {W}omersley method for
  the calculation of hemodynamic variable, Annals of Biomedical Engineering
  25~(3) (1997) 536--546.

\bibitem{Antiga2008}
L.~Antiga, M.~Piccinelli, L.~Botti, B.~Ene-Iordache, A.~Remuzzi, D.~A.
  Steinman, An image-based modeling framework for patient-specific
  computational hemodynamics, Medical {\&} Biological Engineering {\&}
  Computing 46~(11) (2008) 1097--1112.
\newblock \href {http://dx.doi.org/10.1007/s11517-008-0420-1}
  {\path{doi:10.1007/s11517-008-0420-1}}.

\bibitem{Cebral10}
F.~Mut, R.~Aubry, R.~Löhner, J.~R. Cebral, Fast numerical solutions of
  patient-specific blood flows in 3d arterial systems, International Journal
  for Numerical Methods in Biomedical Engineering 26~(1) (2010) 73--85.
\newblock \href {http://dx.doi.org/10.1002/cnm.1235}
  {\path{doi:10.1002/cnm.1235}}.

\end{thebibliography}
\end{document}